\documentclass[twocolumn,english,pra]{revtex4-2}
\usepackage[T1]{fontenc}
\usepackage[latin9]{inputenc}
\usepackage{color}
\usepackage{babel}
\usepackage{amsmath}
\usepackage{amssymb}
\usepackage{esint}
\hypersetup{
           breaklinks=true,   
           colorlinks=true,   
           pdfusetitle=true,  
}
\usepackage{hyperref}

\makeatletter

\providecommand{\tabularnewline}{\\}

\usepackage{braket}
\usepackage{babel}

\makeatother

\begin{document}
\title{Quantum hydrodynamics of coupled electron-nuclear systems}
\author{Rocco Martinazzo$^{1,2,*}$, Irene Burghardt$^{3}$}
\affiliation{$^{1}$Department of Chemistry, Università degli Studi di Milano,
Via Golgi 19, 20133 Milano, Italy}
\email{rocco.martinazzo@unimi.it}

\affiliation{$^{2}$Istituto di Scienze e Tecnologie Molecolari, CNR, via Golgi
19, 20133 Milano, Italy}
\affiliation{$^{3}$Institute of Physical and Theoretical Chemistry, Goethe University
Frankfurt, Max-von-Laue-Str. 7, D-60438 Frankfurt/Main, Germany}
\begin{abstract}
The quantum dynamics of electron-nuclear systems is analyzed from
the perspective of the exact factorization of the wavefunction, with
the aim of defining \emph{gauge} invariant equations of motion for
both the nuclei and the electrons. For pure states this is accomplished
with a quantum hydrodynamical description of the nuclear dynamics
and electronic density operators tied to the fluid elements. For
statistical mixtures of states the exact factorization approach is
extended to two limiting situations that we call ``type-\emph{n}''
and ``type-\emph{e}'' mixtures, depending on whether the nuclei
or the electrons are, respectively, in an intrinsically mixed state.
In both cases a fully \emph{gauge} invariant formulation of the dynamics
is obtained again in hydrodynamic form with the help of mechanical
momentum moments (MMMs). Nuclear MMMs extend in a \emph{gauge} invariant
way the ordinary momentum moments of the Wigner distribution associated
with a density matrix of positional variables, electron MMMs are operator-valued
and represent a generalization of the (conditional) density operators
used for pure states. The theory presented here bridges exact quantum
dynamics with several mixed quantum-classical approaches currently
in use to tackle non-adiabatic molecular problems, offering a foundation
for systematic improvements. It further connects to non-adiabatic
theories in condensed-phase systems. As an example, we re-derive the
finite-temperature theory of electronic friction of Dou, Miao \& Subotnik
(\emph{Phys. Rev. Lett.} \textbf{119}, 046001 (2017)) from the dynamics
of ``type-\emph{e}'' mixtures and discuss possible improvements.
\end{abstract}
\maketitle
\tableofcontents{}

\section{Introduction }

\textbf{} Describing the correlated behavior of electrons and nuclei
beyond the Born-Oppenheimer (BO) or adiabatic approximations poses
a formidable challenge with significant implications in various fields,
ranging from the photochemistry and photophysics of molecular systems,
aggregates, and functional materials \citep{Polli2010,Silva2012,Crespo-Otero2018,Jang2018b,Nelson2020a,Gutzler2021,Hosseinizadeh2021,Yang2022}
to chemical kinetics at metallic surfaces \citep{Wodtke2008,Shenvi2009,Bunermann2015,Kandratsenka2018a,Auerbach2021,Kruger2022}
and the transport of charge through molecular junctions, mesoscopic
devices, and bulk materials \citep{Repp2010,Hartle2011,Giustino2017,Evers2020,Zhou2021,Li2022c,Panhans2023,Hou2023,Li2023a}.
The BO (adiabatic) approximation significantly simplifies calculations
by assuming that only one electronic state is populated. This simplification
enables the complete separation of the electronic and nuclear problems.
However, when the underlying assumptions no longer hold the brute-force
approach of incorporating the relevant electronic states in a Born-Huang-like
expansion of the total wavefunction \citep{Bohm2001} is impractical
for all but small-sized systems. The limitation mainly arises because
the density of electronic states increases with system size, and the
electronic spectrum eventually coalesces into a continuum in an extended
system. Furthermore, solving the nuclear quantum dynamical problem
becomes soon impractical when increasing the system size, even for
single-surface (i.e., BO) problems. Therefore, only numerical methods
based on quantum-classical or quantum-semiclassical approximations,
where electrons are treated quantum mechanically while nuclei are
treated classically or quasi-classically, can offer promise for handling
large systems. This is reasonable for many applications since the
atomic nuclei considered are often heavy enough to behave as classical
objects, exhibiting only weak nuclear quantum effects. The latter
can thus be addressed separately, for example through a fully quantum
approach within either a reduced dimensionality description or a surrogate
model.

This classical treatment of the nuclear dynamics is suitable for single-surface
dynamics, especially when investigating 'global' observables for which
self-averaging occurs and quantum features are washed out. However,
it remains questionable for non-adiabatic dynamics, even for nuclei
that would behave as classical particles in a single-surface dynamics
under similar conditions. The challenge lies in determining the appropriate
forces acting on the nuclei and understanding the correct nuclear
influence on the quantum dynamics of the electrons. These challenges
become increasingly more intricate when a larger number of electronic
states are involved in the coupled dynamics. To make progress, it
is essential to depart from the BO framework and consider the electronic
problem in the time domain rather than the energy domain. However,
defining the appropriate equations of motion is not straightforward,
and different approaches emerge depending on when the classical limit
for the nuclei is introduced \citep{Nelson2020a}.

In this context, the exact factorization (EF) of the total electron-nuclear
wavefunction, as introduced by Abedi \emph{et al.} \citep{Abedi2010,Abedi2012},
represents a groundbreaking approach. It offers an exact scheme with
a built-in separation of variables that can be harnessed to account
for their distinct dynamics. Nevertheless, one must handle the EF
equations of motion with care when devising approximations, as the
nuclear and electronic wavefunctions are \emph{not} unique, and it
is crucial to ensure that any derived scheme is independent of this
arbitrariness. The objective of this paper is precisely to remove
this arbitrariness and to reformulate the EF dynamical equations in
a manner that eliminates this issue entirely. 

The proposed approach is ideally suited for introducing a classical
description for the nuclei while maintaining a quantum treatment of
the electronic motion. It can be viewed as an extension of De Broglie-Bohm-Madelung
hydrodynamics to multi-state dynamics, which is \emph{exact} for any
number of electronic states. Additionally, this approach allows extensions
of the EF method to investigate dynamical processes involving statistical
mixtures of states, thereby enabling the study of more realistic scenarios
like those appearing at finite temperatures.

The EF is an ``intermediate'' representation that consists in projecting
the total wavefunction of a molecular system $\ket{\Psi}$ in the
position basis of nuclear variables, henceforth denoted as $\{\ket{\mathbf{x}}$\}.
This defines a local (i.e., position dependent) electronic state $\ket{u(\mathbf{x})}$
and a nuclear wavefunction $\psi(\mathbf{x})$, upon requiring that
$\ket{u(\mathbf{x})}$ remains normalized at any time, namely
\[
\braket{\mathbf{x}|\Psi}_{X}=\psi(\mathbf{x})\ket{u(\mathbf{x})}
\]
where the subscript $X$ reminds that integration is performed over
nuclear variables only. In other words, in EF the total wavefunction
is exactly expanded as
\begin{equation}
\ket{\Psi_{t}}=\int d\mathbf{x}\psi_{t}(\mathbf{x})\ket{\mathbf{x}}\ket{u_{t}(\mathbf{x})}\label{eq:EF wavefunction}
\end{equation}
This form closely resembles the adiabatic \emph{ansatz} -- where
the $\ket{u_{t}}$'s of Eq. \ref{eq:EF wavefunction} are replaced
by eigenstates of the local electronic Hamiltonians -- and it is
thus an ideal starting point for addressing a non-adiabatic dynamics.
The equations of motion for $\psi(\mathbf{x})$ and $\ket{u(\mathbf{x})}$
can be obtained from the total Hamiltonian in coordinate representation
for the nuclear variables which, in rather general terms, can be taken
of the form 
\begin{equation}
\hat{H}=\hat{T}+H_{\text{el}}(\mathbf{x})\label{eq:total Hamiltonian}
\end{equation}
where $\hat{T}$ is the nuclear kinetic energy operator 
\[
\hat{T}=\frac{1}{2}\sum_{ij}\xi^{ij}\hat{p}_{i}\hat{p}_{j}\ \ \ \text{with}\ \hat{p}_{j}=-i\hbar\partial_{j}
\]
$\xi^{ij}$ is an inverse mass tensor for the nuclei and $H_{\text{el}}(\mathbf{x})$
is the electronic operator with the nuclei clamped at a configuration
$\mathbf{x}$. In this way, one obtains an effective Schr\"{o}dinger
equation for the nuclear wavefunction
\begin{equation}
H^{\text{eff}}\psi=i\hbar\frac{\partial\psi}{\partial t}\label{eq:EF nuclear equation}
\end{equation}
and an equation of motion for the electronic wavefunction
\begin{equation}
i\hbar(\partial_{t}+iA_{0})\ket{u}=(H_{\text{el}}-E)\ket{u}+K[\psi]\ket{u}\label{eq:EF electronic equation}
\end{equation}
In the above equations the effective Hamiltonian $H^{\text{eff}}$
for the nuclear dynamics closely resembles the adiabatic Hamiltonian
and is specified below (Eq. \ref{eq:Hamiltonian}), $E=\braket{u|H_{\text{el}}|u}$
is the average electronic energy and $A_{0}=i\braket{u|\partial_{t}u}$
is an arbitrary (but real) term that guarantees norm conservation
to the electronic wavefunction. 

The detailed derivation of the equations of motion along these lines
can be found in Ref. \citep{Martinazzo2022,Martinazzo2022a}. Here,
we focus on the expression of $K[\psi]$, the functional of $\psi$
that subsumes the electron-nuclear coupling and that reads as

\begin{equation}
K[\psi]\ket{u}=-i\hbar\sum_{j}V^{j}Q\ket{\partial_{j}u}-\hbar R\ket{u}\label{eq:e-n functional}
\end{equation}
where $Q=1-\ket{u}\bra{u}$, $V^{j}=(\hat{v}^{j}\psi)/\psi$ is a
complex-valued nuclear velocity field (to be discussed below, in Sec.
\ref{subsec:Momentum-and-velocity-fields}), $R\ket{u}=\frac{\hbar}{2}\sum_{ij}\xi^{ij}D_{ij}\ket{u}$,
and 
\[
D_{ij}\ket{u}=iA_{i}Q\ket{\partial_{j}u}+iA_{j}Q\ket{\partial_{i}u}+Q\ket{\partial_{i}\partial_{j}u}
\]

One key feature of the exact factorization approach is, as mentioned
above, some freedom in defining the electron and the nuclear wavefunctions
appearing in Eq. \ref{eq:EF wavefunction}, namely in the choice of
their \emph{phase}. This \emph{gauge} freedom introduces spurious
terms in the corresponding equations of motion and hampers the physical
interpretation of the dynamics, for which only \emph{gauge} invariant
quantities can be relevant. The equations above, Eqs. \ref{eq:EF nuclear equation}
and \ref{eq:EF electronic equation}, are ``microscopically'' \emph{gauge}
\emph{covariant}, meaning that they have separate, \emph{gauge} covariant
contributions, as discussed at length in Refs. \citep{Martinazzo2022,Martinazzo2022a}.
We shall show below how to recast them in a form that is fully \emph{gauge}
\emph{invariant} and that can thus display the true, physical couplings
between nuclei and electrons. 

The paper is organized as follows. In Section \ref{sec:QHD-with-gauge}
we review the configuration-space hydrodynamical formulation of quantum
mechanics developed by De Broglie, Bohm and Madelung \citep{Bialynicki-Birula,RobertE.Wyatt2005}---
henceforth, quantum hydrodynamics (QHD) -- with particular emphasis
on \emph{gauge} fields and on the issue of \emph{gauge} invariance.
We address both pure and mixed states and derive a \emph{gauge} invariant
formulation of the dynamics for either cases. This Section is instrumental
to Section \ref{sec:Quantum-Hydrodynamics-of} where we consider the
specific problem of the nuclear - electron dynamics, in the exact
factorization framework. There, the main focus is on the electron
dynamics since the description of the nuclear dynamics follows from
Section \ref{sec:QHD-with-gauge}. Again, we address both the pure-
and the mixed-state cases and extend the exact factorization to two
limiting situations where either the nuclei or the electrons are in
intrinsically mixed states. Next, in Section \ref{sec:Applications}
we present applications of the theory to electron-nuclear dynamics
beyond the Born-Oppenheimer approximation. The purpose of this Section
is to show that the present theory forms a unifying framework for
non-adiabatic problems, encompassing molecular and condensed-phase
problems, from few electronic states to dense manifolds. The emphasis
is on the mixed quantum-classical limit, where electrons are treated
quantally and nuclei are described classically. Couplings between
the two variables are here introduced from the exact equations of
motion, with no \emph{ad hoc} assumptions, and commonly adopted approaches
are framed in one and the same picture. Finally, in Section \ref{sec:Conclusions}
we summarize the main findings and draw some conclusions. 

\section{QHD with gauge fields\label{sec:QHD-with-gauge}}

As mentioned above, in the exact-factorization approach a key ingredient
is represented by the (\emph{gauge} invariant) velocity fields of
the nuclei appearing in the functional of Eq. \ref{eq:e-n functional}.
They subsume the effect that the nuclear dynamics has on the evolution
of the (local) electronic states, beyond the ordinary potential coupling
term built in the electronic Hamiltonian $H_{\text{el}}$ (see Eq.
\ref{eq:total Hamiltonian}). Here, these fields are described in
their natural setting, which is that of the hydrodynamical view of
quantum mechanics. We refer in particular to the configuration space
hydrodynamical picture set by the (closely related) De Broglie - Bohm
and Madelung views \citep{Holland1993}. Specifically, we shall focus
on the latter view, as it offers a complete \emph{gauge}-invariant
formulation of quantum mechanics, which is a central concern for our
aims.

In this Section we shall consider a generic \emph{pseudo}-electromagnetic
Hamiltonian in Schr\"{o}dinger representation for a number $D$ of
degrees of freedom $\mathbf{x}=(x^{1},x^{2},..x^{D})^{t}$ subjected
to \emph{gauge} fields. At a later stage it will play the role of
effective Hamiltonian for the marginal probability amplitude describing
the nuclei in the exact factorization approach (Eq. \ref{eq:EF nuclear equation}).
This Hamiltonian takes the form
\begin{equation}
H=\sum_{ij}\frac{\xi^{ij}}{2}\hat{\pi}_{i}\hat{\pi}_{j}+\mathcal{V}-\hbar A_{0}\label{eq:Hamiltonian}
\end{equation}
where $\xi^{ij}$ is a (coordinate-independent) inverse mass tensor,
$\hat{\pi}_{j}=-i\hbar\partial_{j}-\hbar A_{j}$ and $A_{j}$, $\mathcal{V}$
are local operators depending on $\mathbf{x}$ and eventually on time
$t$. Here, $\mathcal{V}$ is taken to be \emph{gauge} invariant,
while the \emph{gauge} potentials $A_{k}$ and $A_{0}$ are required
to transform according to
\[
A_{k}\rightarrow A_{k}+\partial_{k}\varphi\ \ \ k=0,1,..D
\]
under the $U(1)$ \emph{gauge} transformation $\psi\rightarrow\psi'=e^{i\varphi}\psi$
appropriate for a spinless wavefunction (here $k=0$ denotes the time
coordinate, $\partial_{0}\equiv\partial_{t}$). In the exact factorization
approach $\mathcal{V}=\braket{u|H_{\text{el}}|u}+\phi_{\text{FS}}$,
where the first term is the instantaneous Born-Oppenheimer potential
and $\phi_{\text{FS}}$ is its ``diagonal correction'' 
\[
\phi_{\text{FS}}=\frac{\hbar^{2}}{2}\sum_{ij}\xi^{ij}g_{ij}
\]
where $g_{ij}$ is the Fubini-Study metric tensor \citep{Provost1980},
the real part of the (instantaneous) quantum geometric tensor \citep{Berry1984}
\[
q_{ij}=\braket{\partial_{i}u|Q|\partial_{j}u}
\]
The latter is a key object since it subsumes the \emph{gauge} fields
introduced by the electronic wavefunction, in a \emph{gauge} invariant
way. 

The Hamiltonian of Eq. \ref{eq:Hamiltonian} is \emph{gauge} invariant
\emph{in form}, meaning that the transformed Hamiltonian
\[
H\rightarrow H'\equiv-\hbar\partial_{t}\varphi+e^{i\varphi}He^{-i\varphi}
\]
is the generator of the dynamics in the new \emph{gauge,} as can
be easily verified with a direct calculation
\[
i\hbar\partial_{t}\left(e^{i\varphi}\psi\right)=-\hbar(\partial_{t}\varphi)\psi'+e^{i\varphi}H\psi\equiv H'\psi'
\]
In practice, the $A_{k}$'s can have different physical meaning. They
can represent the coupling to a true electromagnetic field, and in
this case the replacement 
\[
A_{k}\rightarrow\frac{q}{\hbar c}A_{k}^{\text{em}}\ \ A_{0}\rightarrow-\frac{q}{\hbar}A_{0}^{\text{em}}
\]
($q_{k}$ being the particle charge) connects to the standard, non-relativistic,
semiclassical theory of the dynamics of a spinless charge in an electromagnetic
field. However, the $A_{k}$'s can also represent the coupling to
a \emph{pseudo}-electromagnetic field, for instance that arising from
the electronic subsystem in the exact factorization approach. In this
case, they read explicitly as
\[
A_{k}=i\braket{u|\partial_{k}u}\ \ k=0,1,..N
\]
and are known collectively as Berry connection (for $k=1,2,..$).
Here, the $\ket{u}\equiv\ket{u(\mathbf{x})}$'s are local electronic
states and the transformation properties of the connection reflect
the (conjugate) $U(1)$ action on these states that is needed to preserve
the invariance of the total wavefunction, i.e., $\ket{u}\rightarrow\ket{u'}=e^{-i\varphi}\ket{u}$. 

In either case, it is expedient to introduce the differential $1-$form
 
\[
\omega=\sum_{k}\hbar A_{k}dx^{k}
\]
 and its \emph{gauge}-invariant exterior derivative $d\omega=\sum_{kj}B_{kj}dx^{k}dx^{j}\equiv\mathcal{B}$.
A second \emph{gauge}-invariant quantity is  
\[
\mathcal{E}=\hbar dA_{0}-\partial_{t}\omega=\hbar\sum_{k}(\partial_{k}A_{0}-\partial_{t}A_{k})dx^{k}
\]
to which we can add $-d\mathcal{V}$ if desired, without altering
its \emph{gauge} invariance. The magnetic and electric fields (either
real or \emph{pseudo}) are then described by the components of the
above forms in the canonical basis $\{dx^{k}\}$, i.e., 
\begin{equation}
B_{kj}=\hbar(\partial_{k}A_{j}-\partial_{j}A_{k})\ \ \ E_{k}=\hbar(\partial_{k}A_{0}-\partial_{t}A_{k})\label{eq:E,B fields}
\end{equation}
and represent the ``physical'' (i.e., \emph{gauge} invariant) effects
that the \emph{gauge} fields have on the particle dynamics. They satisfy
two differential identities
\begin{equation}
d\mathcal{B}=0\ \ \ d\mathcal{E}=-\partial_{t}\mathcal{B}\label{eq:pseudo Maxwell laws}
\end{equation}
that are, respectively, the magnetic Gauss and the Maxwell-Faraday
induction law. They hold for \emph{both} the real and the \emph{pseudo}-electromagnetic
fields and, in terms of components, read as 
\[
\partial_{i}B_{jk}+\partial_{j}B_{ki}+\partial_{k}B_{ij}=0\ \ \ \partial_{k}E_{j}-\partial_{j}E_{k}=-\partial_{t}B_{kj}
\]

In the following, we first introduce the key momentum and velocity
fields, using their most general setting established by the above
Hamiltonian. Then, we follow the historical route to quantum hydrodynamics
by considering the problem of a single particle in three dimensional
space, here subjected to a real or \emph{pseudo}- electromagnetic
field. The general result, which follows next, is a straightforward
extension that could be directly derived from the Schr\"{o}dinger
equation, but the single particle problem is pivotal to identify its
key physical ingredients. The established setting is instrumental
to the analysis of the combined electron-nuclear dynamics in the exact
factorization picture that is performed in Section \ref{sec:Quantum-Hydrodynamics-of}. 

\subsection{Momentum and velocity fields\label{subsec:Momentum-and-velocity-fields}}

The operator $\hat{\pi}_{k}=-i\hbar\partial_{k}-\hbar A_{k}$ entering
Eq. \ref{eq:Hamiltonian} represents the $k^{\text{th}}$ component
of the \emph{mechanical} momentum, which connects to the velocity
components $\hat{v}^{k}$ \emph{via} the mass tensor, namely through
$\hat{\pi}_{k}=\sum_{j}\xi_{kj}\hat{v}^{j}$ where $\xi_{kj}$ is
the mass tensor, $\sum_{j}\xi_{kj}\xi^{ji}=\delta_{k}^{i}$, and $\hat{v}^{k}=\frac{i}{\hbar}[H,x^{k}]$
is the $k^{\text{th}}$ component of the velocity. Both $\hat{\pi}_{k}$
and $\hat{v}_{k}$ are \emph{gauge} covariant, meaning that 
\[
\psi\rightarrow\psi'=e^{i\varphi}\psi\ \ \ (\hat{G}\psi)\rightarrow(\hat{G}'\psi')=e^{i\varphi}(\hat{G}\psi)
\]
hence the expressions
\[
\Pi_{k}=\frac{\hat{\pi}_{k}\psi}{\psi}\ \ \ V^{k}=\frac{\hat{v}^{k}\psi}{\psi}
\]
define two kinds of complex-valued, \emph{gauge}-invariant fields
in any point of configuration space but the ones where the wavefunction
vanishes. They are, respectively, the momentum and the velocity\textbf{
}fields and comprise qualitatively different contributions in their
real and imaginary parts, henceforth denoted as $\pi_{k}$, $w_{k}$
for the first and $v_{k}$, $u_{k}$ for the second. By definition
we have, for instance,
\begin{equation}
\Pi_{k}(\mathbf{x},t)=-i\hbar\partial_{k}\ln\psi(\mathbf{x},t)-\hbar A_{k}(\mathbf{x},t)\label{eq:complex momentum field}
\end{equation}
and, using the polar representation of the wavefunction, 
\begin{equation}
\pi_{k}(\mathbf{x},t)=\Re\Pi_{k}=\hbar\left(\partial_{k}\theta(\mathbf{x},t)-A_{k}(\mathbf{x},t)\right)\label{eq:real momentum field}
\end{equation}
\begin{equation}
w_{k}(\mathbf{x},t)=\Im\Pi_{k}=-\hbar\partial_{k}\ln|\psi(\mathbf{x},t)|=-\frac{\hbar}{2}\partial_{k}\ln n(\mathbf{x},t)\label{eq:imaginary momentum field}
\end{equation}
where $\theta=\text{arg}\psi$ and $n=|\psi|^{2}$ is the particle
density in configuration space. Note the \emph{both} $\pi_{k}$ and
$w_{k}$ are ill-defined (singular) when $\psi=0$ \footnote{One can also use the complex-valued momentum densities $N_{k}=n\Pi_{k}=\psi^{*}\hat{\pi}_{k}\psi$,
which are yet \emph{gauge} invariant and free of singularities. Likewise,
for the velocity fields: one can introduce the complex-valued current
densities $Y^{k}=nV^{k}=\psi^{*}\hat{v}^{k}\psi$, which vanish at
the wavefunction nodes, rather than being singular. This shows that
the singularities are harmless for the evolution of the probability
fluid discussed below. }. Similarly, we have $v^{k}(\mathbf{x},t)=\Re V^{k}=\sum_{j}\xi^{kj}\pi_{j}(\mathbf{x},t)$
and $u^{k}(\mathbf{x},t)=\Im V^{k}=\sum_{j}\xi^{kj}w_{j}(\mathbf{x},t)$. 

The thus defined \emph{gauge}-invariant, complex-valued fields $\Pi_{k}$
and $V^{k}$ (or the real-valued pairs $(\pi_{k},w_{k})$ and $(v^{k},u^{k})$),
one for each $k=1,2,..N$, replace the complex-valued wavefunction
$\psi$ (or its polar pair $(|\psi|,\theta)$) in a redundant way,
hence they satisfy some ``consistency'' conditions. These are more
easily formulated for the momentum fields, because of their direct
connection to $\psi$. A first relation concerns the imaginary parts
$w_{k}$'s which are longitudinal by construction, hence conservative
in any connected domain,
\[
\sum_{k}\int_{\gamma(a,b)}w_{k}dx^{k}=-\frac{\hbar}{2}\ln\frac{n(\mathbf{x}_{b})}{n(\mathbf{x}_{a})}
\]
for any path joining the point $\mathbf{x}_{a}$ with $\mathbf{x}_{b}$.
This can be better expressed as 
\[
n(\mathbf{x}_{b})=\exp\left(-\frac{2}{\hbar}\int_{\gamma(a,b)}\mathcal{W}\right)n(\mathbf{x}_{a})
\]
where $\mathcal{W}$ is an\emph{ }exact 1-form $\mathcal{W}=\sum_{k}w_{k}dx^{k}=-\hbar/2d\ln\,n$
and the path is arbitrary, subjected only to the constraint that it
does not go through the wavefunction nodes or excluded region (if
any). The velocity counterpart of this relation is similar but employs
a different ``metric''. For a single particle in 3D we can write
\[
n\mathbf{u}=-\frac{\hbar}{2m}\mathbf{\nabla}n
\]
which can be viewed as a sort of Fick's law relating the density gradient
to a kind of (number) density current on the left, with $D=\hbar/2m$
playing the role of diffusion coefficient. For this reason $\mathbf{u}$
is termed osmotic\textbf{ }velocity, a term that we shall adopt for
$u^{k}$, irrespective of the system dimensions (and nature of the
mass tensor). 

The real part of the momentum fields, $\pi_{k}$, are trickier for
at least two reasons. First of all, the definition of the fields includes
the potentials $A_{k}$ which are essential for the \emph{gauge} invariance
and are usually non-longitudinal, thereby providing a ``diamagnetic''
contribution to the current. Secondly, but equally important, even
when $A_{k}=0$ the momentum $\pi_{k}=\hbar(\partial_{k}\theta(\mathbf{x},t)-A_{k}(\mathbf{x},t))$
is only \emph{apparently} longitudinal: the wavefunction phase $\theta$
is defined modulo $2\pi$ only, meaning that $\theta=\text{arg\ensuremath{\psi}}$
is \emph{not} a smooth function on the whole complex plane, rather
it presents a branch cut (typically placed on the real negative axis,
in such a way that $\text{arg}(\mathbb{C})=(-\pi,\pi]$). That is,
when selecting \emph{smoothly} the phase $\theta$ along a loop there
is no guarantee that it gets back to its original value, it may well
change by an integer multiple of $2\pi$. Accounting for this aspect
we must have a quantization condition on the ``circulation'', namely
\begin{equation}
\Gamma_{\gamma}=\sum_{k}\oint_{\gamma}(\pi_{k}+\hbar A_{k})dx^{k}=2\pi\hbar\,n\ \ n\in\mathbb{Z}\label{eq:circulation}
\end{equation}
for any closed path $\gamma$ in configuration space. Here, $n$ is
the topological value that describes the way a loop winds around a
singularity (or an excluded volume). In fact, $n$ can be nonzero
only in a multiply connected domain, since otherwise we could exploit
Stokes' theorem to write
\[
\Gamma_{\gamma}=\int_{S}d\Pi+\int_{S}d\omega
\]
where $S$ is any open surface having $\gamma$ as its frontier, $d\Pi=d(\sum_{k}\pi_{k}dx^{k})$
is the vorticity of the (mechanical) momentum field and the second
integral is the $B-$flux through $S$ (i.e., the Berry flux or the
flux of the magnetic field, depending on the context). Stokes' theorem
guarantees that  the surface integral (hence, the circulation of
Eq. \ref{eq:circulation}) changes smoothly down to zero value when
contracting the loop to a single point, thereby implying that $n=0$
must hold for the finite loop as well. We thus see that in the typical
situation where there is no excluded volume the mechanical momentum
field $\pi_{k}$ has a non-trivial vorticity only for a non-vanishing
magnetic field, \emph{except} at the singularities, i.e. at the wavefunction
nodes. 

Thus, a third issue worth mentioning here is the behavior of the momentum
field around the wavefunction nodes \citep{RobertE.Wyatt2005}. Suppose
$\mathbf{x}_{0}$ is a node for the time evolving wavefunction $\psi(\mathbf{x},t$)
at $t=0$, i.e., $\psi(\mathbf{x}_{0},0)=0$, and consider $\psi(\mathbf{x}_{0},\delta t)$
for infinitesimal times $\delta t=\pm|\delta t|$. Unless $\partial_{t}\psi(\mathbf{x}_{0},0)=0$
(a case that can be considered accidental) the wavefunction takes
opposite values at opposite times
\[
\psi(\mathbf{x}_{0},-|\delta t|)=-\psi(\mathbf{x}_{0},+|\delta t|)
\]
and, by construction, passes through the origin of the complex plane
at $t=0$. Hence, its phase changes \emph{abruptly} by $\pi$
\[
\theta(\mathbf{x}_{0},|\delta t|)=\theta(\mathbf{x}_{0},-|\delta t|)+\pi\ \ \ (\text{mod}2\pi)
\]
Cleary, at $t=0$ and $\mathbf{x}=\mathbf{x}_{0}$ the phase $\theta$
is completely undefined and the partial derivatives $\partial_{t}(\pi_{k}+\hbar A_{k})$
are singular. Note that, in general, the wavefunction nodes are objects
of codimension 2 as they are found at the intersection of two surfaces,
i.e. those defined by the conditions $\text{\ensuremath{\Re\psi(\mathbf{x})=\Im\psi(\mathbf{x})=0}}$. 

\subsection{QHD for a particle in 3D space with \emph{gauge} fields\label{subsec:QHD-for-a-particle-in-3D}}

We start with the simple example of a particle of mass $m$ in 3D
space, namely with the Hamiltonian
\[
H=\frac{(-i\hbar\boldsymbol{\nabla}-\hbar\mathbf{A})}{2m}^{2}+\mathcal{V}-\hbar A_{0}
\]
that may represent a charge of mass $m$ subjected to a true electromagnetic
field, provided the \emph{gauge} potentials are correctly scaled for
the particle charge.  As mentioned above, there is no real difficulty
in dealing directly with the general case of Eq. \ref{eq:Hamiltonian},
but this single-particle example allows us to single out the key physical
ingredients that are otherwise hard to identify in the general setting.
We only sketch the derivations and present the main (textbook) results
\citep{Bialynicki-Birula,RobertE.Wyatt2005}, though we put some emphasis
on the \emph{gauge} invariance of the theory which is crucial for
the issues discussed in the next Section. 

In this problem, the momentum and velocity fields are simply related
to each other, and we choose to work with the velocity field $\mathbf{V}=\mathbf{v}+i\mathbf{u}$
where $\mathbf{v}=\frac{\hbar}{m}\left(\boldsymbol{\nabla}\theta-\mathbf{A}\right)$
and $\mathbf{u}=-\frac{\hbar}{2m}\boldsymbol{\nabla}\ln\,n=-\frac{\hbar}{m}\boldsymbol{\nabla}\ln R$,
with $R=\sqrt{n}=|\psi|$. From the Schr\"{o}dinger equation in ``local''
form 
\[
i\hbar\partial_{t}\ln\psi=\frac{\psi^{*}H\psi}{n}
\]
(which is valid everywhere in configuration space but the wavefunction
nodes) one obtains the set of coupled (real) equations for the pair
of variables ($n,\theta$) that represent the wavefunction in polar
form. These equations require a complex-valued ``energy density''
that, for $H$ above, reads as
\begin{equation}
\psi^{*}H\psi\equiv\frac{m}{2}n\,\mathbf{V}^{*}\mathbf{V}+n(\mathcal{V}-\hbar A_{0})-i\frac{\hbar}{2}\boldsymbol{\nabla}(n\mathbf{V})\label{eq:energy density}
\end{equation}
They are, respectively, the celebrated continuity equation for the
particle density \footnote{This equation shows that $\mathbf{v}$ indeed plays the role of a
classical velocity field, determining the particle current density
$\mathbf{j}$ through $\mathbf{j}=n\mathbf{v}$.}
\begin{equation}
\partial_{t}n+\boldsymbol{\nabla}(n\mathbf{v})=0\label{eq:continuity equation}
\end{equation}
and a kind of Hamilton-Jacobi equation for the ``action'' \footnote{More precisely, this is a double function $S(\mathbf{x},t,\mathbf{x}_{0},t_{0}$)
and it is the action functional evaluated along the stationary path,
as a function of the endpoints of the action integral.} $S=\hbar\theta\equiv S(\mathbf{x},t)$
\begin{equation}
-\frac{\partial S}{\partial t}=\frac{\left(\boldsymbol{\nabla}S-\hbar\mathbf{A}\right)^{2}}{2m}+(\mathcal{V}-\hbar A_{0})+Q\label{eq:Hamilton-Jacobi}
\end{equation}
which involves the quantum potential 
\begin{equation}
Q=\frac{m}{2}\mathbf{u}\mathbf{u}+\frac{\hbar}{2n}\boldsymbol{\nabla}(n\mathbf{u})=-\frac{\hbar^{2}}{2m}\frac{\nabla^{2}R}{R}\label{eq:quantum potential}
\end{equation}
Eq. \ref{eq:Hamilton-Jacobi} represents the key dynamical equation
of the De Broglie - Bohm approach, corresponding \footnote{We exploit here the similarity with the classical problem. Later we
shall give a \emph{direct} proof of these equations of motion.} to the particle equation of motion 
\[
\mathbf{x}(t_{0})=\mathbf{x}_{0}\ \ \ \mathbf{v}(t_{0})=\mathbf{v}_{0}=\frac{1}{m}\left(\boldsymbol{\nabla}S(\mathbf{x}_{0},t_{0})-\mathbf{\hbar A}(\mathbf{x}_{0},t_{0})\right)
\]
\[
m\dot{\mathbf{v}}=\mathbf{E}+\mathbf{v}\times\mathbf{B}-\boldsymbol{\nabla}(\mathcal{V}+Q)
\]
where $\mathbf{v}=\dot{\mathbf{x}}$, $\mathbf{E}=\hbar(\boldsymbol{\nabla}A_{0}-\partial_{t}\mathbf{A})$
is the electric field and $\mathbf{B}=\boldsymbol{\nabla}\times(\hbar\mathbf{A})$
is the pseudo-vector describing the magnetic field. Solution of this
equation for any initial position $\mathbf{x}_{0}$ (and corresponding
$\mathbf{v}_{0}$, as determined by the initial wavefunction) provides
in principle a solution of the quantum dynamical problem --- i.e.
the wavefunction at later times. However, note that on comparing with
the \emph{classical} Hamilton-Jacobi equation, the added term $Q=Q(n(\mathbf{x},t))$
is a potential that depends on the \emph{particle density}, and this
is the crucial factor that makes the system quantum. The trajectory
equation does determine (as in classical mechanics) the evolution
of a classical-like system but now the trajectories are \emph{quantum}
trajectories whose (changing) density affects the equation of motion
itself. This implies that it is \emph{not} possible to solve for a
single or a handful of trajectories, unless we know \emph{in advance}
the evolving density $n(\mathbf{x},t$). 

Were it not for the classical analogy, the phase equation of the De
Broglie - Bohm approach, Eq. \ref{eq:Hamilton-Jacobi}, would be an
ordinary partial differential equation, not easier to solve than the
Schr\"{o}dinger equation we started from. Importantly, this equation
involves yet the potentials, that are \emph{gauge} dependent, and
thus it is not in a \emph{gauge} invariant form. In fact, $\theta(\mathbf{x},t)=S(\mathbf{x},t)/\hbar$
itself is \emph{gauge} dependent. In order to fix this problem and
to find the (\emph{gauge} invariant) \emph{particle} equations of
motion one can take the spatial derivative of the Schr\"{o}dinger
equation, to write
\begin{equation}
m\frac{\partial\mathbf{V}}{\partial t}\mathbf{=-\hbar}\frac{\partial\mathbf{A}}{\partial t}\mathbf{-\boldsymbol{\nabla}}\left(\frac{H\psi}{\psi}\right)\label{eq:momentum EOM}
\end{equation}
where the r.h.s. can be obtained from the energy density of Eq. \ref{eq:energy density}.
The result is the ``Lagrangian-frame'' expression
\begin{equation}
m\frac{d\mathbf{v}}{dt}=\mathbf{E}\mathbf{+}\mathbf{v}\times\mathbf{B}\mathbf{-\boldsymbol{\nabla}}\left(\mathcal{V}+Q\right)\label{eq:Madelung I}
\end{equation}
where 
\begin{equation}
\frac{d}{dt}:=\frac{\partial}{\partial t}+\mathbf{v}\boldsymbol{\nabla}\label{eq:material derivative}
\end{equation}
is the material derivative \footnote{Occasionally, when no confusion arises, we shall also use the overdot
to denote such a derivative.} and $\mathbf{E}$ and $\mathbf{B}$ are, respectively, the \emph{pseudo}-electric
and the \emph{pseudo}-magnetic fields introduced above. This equation
needs to be complemented with continuity equation, 
\begin{equation}
\frac{dn}{dt}+n\boldsymbol{\nabla}\mathbf{v}=0\label{eq:Madelung II}
\end{equation}
(here written in Lagrangian form, too) and, together, the two represent
Madelung's hydrodynamical equations of motion. The solutions of these
equations solve the original quantum dynamical problem if one enforces
the quantization condition on the circulation, namely Eq. \ref{eq:circulation}.
This is necessary to guarantee the existence of a scalar function
$S\equiv\hbar\theta$ that is independent of the path (modulo $2\pi\hbar$).
This quantization condition (a boundary condition for the velocity
field) is required at initial time only, since Kelvin's circulation
theorem \citep{Fetter2003,Saffman1992} -- i.e., $\dot{\Gamma}_{\gamma}=0$
for any closed curve $\gamma$ in motion with the fluid elements ---
also applies to the fluid dynamics described by Eqs. \ref{eq:Madelung I}
and \ref{eq:Madelung II} (see Appendix \ref{app:Circulation}). 

Eqs. \ref{eq:Madelung I} and \ref{eq:Madelung II} can be further
presented as a set of coupled equations for the real (\textbf{v)}
and imaginary (\textbf{u}) velocity fields, namely 
\begin{align}
m\frac{d\mathbf{u}}{dt} & =\frac{\hbar}{2}\boldsymbol{\nabla}(\boldsymbol{\nabla}\mathbf{v})-m(\mathbf{u}\boldsymbol{\nabla})\mathbf{v}+\mathbf{u}\times\mathbf{B}\nonumber \\
m\frac{d\mathbf{v}}{dt} & =\mathbf{F}+m(\mathbf{u}\boldsymbol{\nabla})\mathbf{u}-\frac{\hbar}{2}\nabla^{2}\mathbf{u}\label{eq: u,v equations-1}
\end{align}
where $\mathbf{F}=\mathbf{E}\mathbf{+}\mathbf{v}\times\mathbf{B}-\boldsymbol{\nabla}\mathcal{V}$
represents the total (\emph{classical}) force acting on the particle.
This set of equations describes a particle dynamics subjected to the
classical force $\mathbf{F}$ and ``an internal force'' (second
and third term on the r.h.s. of the last equation) that depends on
the \emph{auxiliary field }$\mathbf{u}$. The latter evolves along
the flow as dictated by the first equation, and replaces the quantum
potential appearing in the De Broglie - Bohm formulation \footnote{It should be noted that the coupling between the $\mathbf{v}$ and
$\mathbf{u}$ fields occurs through their spatial derivatives. It
is with this mechanism that non-locality enters the theory: one needs
the infinitesimal evolution of a (large) set of particles in order
to build the \emph{space behavior} of the fields and be able to proceed
with the next step.}. Noteworthy, Eq.s \ref{eq: u,v equations-1} can be combined into
a single equation for the complex-valued fields, 
\begin{equation}
m\frac{\tilde{d}\mathbf{V}}{dt}=\tilde{\mathbf{F}}-i\frac{\hbar}{2m}\boldsymbol{\nabla}\times\mathbf{B}+i\frac{\hbar}{2}\nabla^{2}\mathbf{V}\label{eq:complex Madelung}
\end{equation}
where now $\tilde{\mathbf{F}}=\mathbf{E}\mathbf{+}\mathbf{V}\times\mathbf{B}-\boldsymbol{\nabla}\mathcal{V}$
is a \emph{complex-valued} classical force and 
\begin{equation}
\frac{\tilde{d}}{dt}=\left(\frac{\partial}{\partial t}+\mathbf{V}\boldsymbol{\nabla}\right)\label{eq:complex material derivative}
\end{equation}
is the complex-valued material derivative, both involving the complex-valued
velocity field $\mathbf{V}$ in place of the real one. This result
follows more easily by introducing the complex-valued action $F(\mathbf{x})$
according to 
\[
\psi=\exp\left(i\frac{F(\mathbf{x})}{\hbar}\right)
\]
(i.e., setting $F(\mathbf{x})=S(\mathbf{x})-\frac{i}{2}\ln\,n(\mathbf{x})$)
and writing a complex-valued Hamiltonian-Jacobi equation for it. However,
in practice, the first (vectorial) equation of Eq. \ref{eq: u,v equations-1}
is conveniently replaced by the continuity equation for the scalar
$n$, at the expense of breaking the symmetry between velocity fields,
hence Eq. \ref{eq:complex Madelung} becomes just a compact way of
presenting the QHD equations of motion.

Finally, we notice that the key equation of motion, Eq. \ref{eq:Madelung I},
can be recast as a momentum balance equation --- i.e. as a quantum
Navier-Stokes equation ---
\begin{equation}
m\,n\frac{d\mathbf{v}}{dt}=n\,\mathbf{F}_{b}+\boldsymbol{\nabla}\boldsymbol{\sigma}\label{eq:momentum balance equation}
\end{equation}
where the bulk force $\mathbf{F}_{b}$ is classical ($\mathbf{F}_{b}\equiv\mathbf{F}$)
and the surface force is quantum. The latter reads component-wise
as $(\boldsymbol{\nabla}\boldsymbol{\sigma})_{k}=\sum_{j}\partial_{j}\sigma_{jk}$,
and is determined by the quantum stress tensor 
\begin{equation}
\sigma_{jk}=-\delta_{jk}P_{q}-m\,nu_{j}u_{k}\label{eq:quantum stress tensor}
\end{equation}
where $P_{q}=-\frac{\hbar^{2}}{4m}\nabla^{2}n$ is the (positive definite)
quantum pressure. The analogy with the classical Navier-Stokes, though,
can \emph{only} be formal. The fluid of quantum probability addressed
here does not experience any frictional effect since this would require
an interaction between fluid molecules, one that can transfer momentum
when the molecules travel at different speeds. Here the interaction
is a manifestation of the non-locality of quantum mechanics, it depends
on the density gradient and it is same irrespective of whether or
not the ``particles'' travel at the same velocity.

\subsection{Many-particle QHD with \emph{gauge} fields\label{subsec:Many-particle-QHD-with}}

We are now ready to undertake the more complicated task of developing
the hydrodynamic picture for the many-body Hamiltonian of Eq. \ref{eq:Hamiltonian}.
We shall use the complex valued momentum fields introduced in Sec.
\ref{subsec:Momentum-and-velocity-fields}, Eq. \ref{eq:complex momentum field},
and the result
\begin{equation}
\frac{H\psi}{\psi}=\sum_{ij}\frac{\xi^{ij}}{2}\Pi_{i}^{*}\Pi_{j}-i\frac{\hbar}{2}\sum_{j}\frac{\partial_{j}(nV^{j})}{n}+\mathcal{V}-\hbar A_{0}\label{eq:complex potential of Euler force}
\end{equation}
Eq. \ref{eq:complex potential of Euler force} defines the potential
of the complex Eulerian force 
\begin{equation}
\frac{\partial\Pi_{k}}{\partial t}=-\hbar\frac{\partial A_{k}}{\partial t}-\partial_{k}\left(\frac{H\psi}{\psi}\right)\label{eq:fundamental momentum equation}
\end{equation}
that shows that non-longitudinal contributions to the momentum field
can \emph{only} come from the time-dependence of the vector potential
$A_{k}$,  consistently with the generalized Maxwell-Faraday induction
law of Eq. \ref{eq:pseudo Maxwell laws}. There remains to calculate
the spatial derivative of Eq. \ref{eq:complex potential of Euler force}.
A simple calculation gives for the non-trivial kinetic contribution
\begin{equation}
\partial_{k}\left(\frac{T\psi}{\psi}\right)=-i\frac{\hbar}{2}\sum_{j}\partial_{k}\partial_{j}V^{j}+\sum_{j}V^{j}(\partial_{k}\Pi_{j})\label{eq:derivative of complex kinetic potential}
\end{equation}
that can be further rearranged as 
\begin{align}
\partial_{k}\left(\frac{T\psi}{\psi}\right) & =-i\frac{\hbar}{2}\sum_{ij}\xi^{ij}\partial_{j}\partial_{i}\Pi_{k}+i\frac{\hbar}{2}\sum_{ij}\xi^{ij}\partial_{j}B_{ki}\label{eq:derivative of complex kinetic potential II}\\
 & +\sum_{j}V^{j}(\partial_{j}\Pi_{k})-\sum_{j}V^{j}B_{kj}
\end{align}
where we have used $\partial_{k}\Pi_{j}-\partial_{j}\Pi_{k}=-\hbar(\partial_{k}A_{j}-\partial_{j}A_{k})=-B_{kj}$.
Adding the potential contribution and rearranging, we finally find
\begin{equation}
\frac{\tilde{d}\,\Pi_{k}}{dt}=\tilde{F}_{k}-i\frac{\hbar}{2}\sum_{ij}\xi^{ij}\partial_{j}B_{ki}+i\frac{\hbar}{2}\sum_{ij}\xi^{ij}\partial_{i}\partial_{j}\Pi_{k}\label{eq:complex Madelung (general)}
\end{equation}
where we have introduced the complex material derivative 
\begin{equation}
\frac{\tilde{d}}{dt}=\frac{\partial}{\partial t}+\sum_{j}V^{j}\partial_{j}\label{eq:complex material derivative 2}
\end{equation}
 and the complex-valued total classical force, 
\begin{equation}
\tilde{F}_{k}=E_{k}+\sum_{j}V^{j}B_{kj}-\partial_{k}\mathcal{V}\label{eq:complex valued total classical force}
\end{equation}
Eq. \ref{eq:complex Madelung (general)} generalizes Eq. \ref{eq:complex Madelung}
for the single particle in 3D space, here re-written in terms of the
momentum field, 
\[
\frac{\tilde{d}\,\mathbf{\boldsymbol{\Pi}}}{dt}=\tilde{\mathbf{F}}-i\frac{\hbar}{2m}\boldsymbol{\nabla}\times\mathbf{B}+i\frac{\hbar}{2m}\nabla^{2}\boldsymbol{\Pi}
\]
where in 3D the pseudo-vector field $\mathbf{B}$ is related to the
tensor $B_{kj}$ by $\mathbf{B}_{i}=\frac{1}{2}\sum_{kj}\epsilon_{ikj}B_{kj}$
(here $\epsilon_{ijk}$ is the Levi-Civita symbol that stands for
the fully antisymmetric unit tensor, $\epsilon_{123}=1$). 

Eq. \ref{eq:complex Madelung (general)} is the complex-valued form
of Madelung's hydrodynamic equations. It can be easily split into
a pair of real-valued equations by taking its real and imaginary parts.
Upon using $\frac{\hbar}{2}\sum_{ij}\xi^{ij}\partial_{j}\left(\partial_{i}\pi_{k}-B_{ki}\right)=\frac{\hbar}{2}\sum_{ij}\xi^{ij}\partial_{j}\partial_{k}\pi_{i}\equiv\frac{\hbar}{2}\sum_{j}\partial_{k}\partial_{j}v^{j}$,
we find
\begin{equation}
\frac{d\pi_{k}}{dt}=F_{k}+\sum_{j}u^{j}\partial_{j}w_{k}-\frac{\hbar}{2}\sum_{ij}\xi^{ij}\partial_{i}\partial_{j}w_{k}\label{eq:real Madelung (general)}
\end{equation}
and
\begin{equation}
\frac{dw_{k}}{dt}=\frac{\hbar}{2}\sum_{j}\partial_{k}\partial_{j}v^{j}-\sum_{j}u^{j}\partial_{j}\pi_{k}+\sum_{j}u^{j}B_{kj}\label{eq:imaginary Madelung (general)}
\end{equation}
which generalize Eqs. \ref{eq: u,v equations-1}, here re-written
in terms of momentum fields,
\begin{align*}
\frac{d\boldsymbol{\pi}}{dt} & =\mathbf{F}+(\mathbf{u}\boldsymbol{\nabla})\mathbf{w}-\frac{\hbar}{2m}\nabla^{2}\mathbf{w}\\
\frac{d\mathbf{w}}{dt} & =\frac{\hbar}{2}\boldsymbol{\nabla}(\boldsymbol{\nabla}\mathbf{v})-(\mathbf{u}\boldsymbol{\nabla})\boldsymbol{\pi}+\mathbf{u}\times\mathbf{B}
\end{align*}
Eq. \ref{eq:imaginary Madelung (general)} is, of course, continuity
equation in disguise. Hence, at the expense of breaking the symmetry
between real and imaginary momentum fields, it is conveniently replaced
by
\begin{equation}
\partial_{t}n+\sum_{j}\partial_{j}(v^{j}n)=0\label{eq:continuity equation Euler (general)}
\end{equation}
or, equivalently, by its Lagrangian form
\begin{equation}
\frac{dn}{dt}+n\sum_{j}\partial_{j}v^{j}=0\label{eq:continuity equation Lagrange (general)}
\end{equation}
Eqs. \ref{eq:continuity equation Lagrange (general)} and \ref{eq:real Madelung (general)}
form the basis for the trajectory approach to quantum dynamics underlying
the Madelung hydrodynamic picture. Eq. \ref{eq:real Madelung (general)}
shows that, in analogy with the single particle case, the dynamics
is governed by the classical force (here comprising both ``external''
and \emph{gauge} fields) and by a quantum force
\begin{equation}
(F_{q})_{k}=\sum_{j}u^{j}\partial_{j}w_{k}-\frac{\hbar}{2}\sum_{ij}\xi^{ij}\partial_{i}\partial_{j}w_{k}\label{eq:quantum force}
\end{equation}
depending \emph{only} on the density. The latter can be rearranged
to identify the quantum potential $Q$ 
\begin{equation}
Q=-\frac{\hbar^{2}}{2}\sum_{ij}\xi^{ij}\frac{\partial_{i}\partial_{j}R}{R}\label{eq:quantum potential (general)}
\end{equation}
an expression that closely parallels that found above for the single
particle in 3D space, Eq. \ref{eq:quantum potential}. The analogy
can be tightened if we introduce a mass-weighted\textbf{ }Laplacian
according to
\begin{equation}
\Delta_{M}:=\sum_{ij}\xi^{ij}\partial_{i}\partial_{j}\label{eq:mass-weighted Laplacian}
\end{equation}
and rewrite Eqs. \ref{eq:quantum potential (general)} and \ref{eq:quantum potential}
in the suggestive form
\begin{equation}
Q=-\frac{\hbar^{2}}{2R}\Delta_{M}R\label{eq:quantum potential (general) II}
\end{equation}
\begin{equation}
(F_{q})_{k}=-\partial_{k}Q\label{eq:quantum force 2}
\end{equation}
The mass-weighted Laplacian is further in accordance with other analogies
provided above in the equations of motion, e.g. it can be used to
write the momentum equation as
\begin{equation}
\frac{d\pi_{k}}{dt}=F_{k}+\sum_{j}u^{j}\partial_{j}w_{k}-\frac{\hbar}{2}\Delta_{M}w_{k}\label{eq:real Madelung (final)}
\end{equation}
in close analogy with Eq. \ref{eq: u,v equations-1}, or that for
the complex-valued field
\begin{equation}
\frac{\tilde{d}\,\Pi_{k}}{dt}=\tilde{F}_{k}-i\frac{\hbar}{2}\sum_{ij}\xi^{ij}\partial_{j}B_{ki}+i\frac{\hbar}{2}\Delta_{M}\Pi_{k}\label{eq:complex Madelung (final)}
\end{equation}
(cf. with Eq. \ref{eq:complex Madelung}). 

The resulting quantum potential can also be used to obtain the density
of quantum force which enters the momentum balance equation (i.e.
the Navier-Stokes equation for the momentum) in the form of divergence
of a stress tensor
\begin{equation}
n\frac{d\pi_{k}}{dt}=nF_{k}+\sum_{j}\partial_{j}\sigma_{k}^{j}\label{eq:quantum Navier-Stokes}
\end{equation}
with the quantum\textbf{ }stress\textbf{ }tensor given by 
\begin{equation}
\sigma_{k}^{j}=-\delta_{k}^{j}P_{q}-nu^{j}w_{k}\label{eq:quantum stress tensor (general)}
\end{equation}
and the quantum pressure $P_{q}=-\frac{\hbar^{2}}{4}\Delta_{M}n$.
Eq. \ref{eq:quantum stress tensor (general)} parallels (generalizes)
Eq. \ref{eq:quantum stress tensor} found for the particle in 3D space.
Notice that $u^{j}$ is the $j^{\text{th}}$ imaginary component of
the velocity, while $w_{k}$ is the $k^{\text{th}}$ imaginary component
of the mechanical momentum. 

\subsection{Statistical mixtures\label{subsec:Statistical-mixtures}}

The theory presented so far has only addressed the case of a \emph{pure}
\emph{state}, for which a set of two hydrodynamic equations suffices.
These can be formulated as the momentum particle equation augmented
with the continuity equation or as two Euler equations, one for the
density (continuity equation) and one for the momentum density. For
(proper) \emph{mixtures of} \emph{states} an analogous and yet very
elegant description is possible in terms of momentum moments of the
underlying phase-space (pseudo) distribution \citep{Moyal1949,Irving1951,Johansen1998},
which generalize the density and the momentum density appearing in
the pure state problem to higher order quantities. The main difference
between a mixed and a pure state is that the coupled equations of
motion in general do \emph{not} close at the first order level (i.e.,
at the level of the momentum density), rather they form an infinite
set, arranged in a hierarchical form, that in practice must be closed
somehow. It is only for a pure state that the second (and higher)
order moment can be expressed in terms of the zero-th and the first
moments only, although in a non-trivial way \citep{Burghardt2001}\footnote{This is due to the fact the a pure state is fully characterized by
a wavefunction and the latter only requires an amplitude and a phase.}. 

The standard construction of momentum moments and of the related hierarchy
of equations of motion is well known. It is however of limited help
here since the presence of \emph{gauge} potentials in our Hamiltonian,
Eq. \ref{eq:Hamiltonian}, makes the ordinary momentum moments \emph{gauge}
dependent, hence unsuited for a \emph{gauge} invariant hydrodynamic
description of the quantum dynamics. Thus, we introduce here a new
type of \emph{gauge} invariant moments --- which we call mechanical\textbf{
}momentum\textbf{ }moments (MMMs) because of their relationship with
the mechanical momentum operator --- and show that their equations
of motion take a simple, physically sound form. In the following we
provide the main results of these developments, the details of the
calculations can be found in Appendix \ref{app:momentum moments}.

The ordinary momentum moments are the moments of the partial ``distributions''
that are obtained when fixing $\mathbf{q}$ in the Wigner function
$\rho_{W}(\mathbf{q},\mathbf{p}$), namely
\[
S_{i_{1}i_{2}..i_{D}}^{(n)}(\mathbf{q})=\int d^{D}\mathbf{p}\,p_{1}^{i_{1}}p_{2}^{i_{2}}..p_{D}^{i_{D}}\rho_{W}(\mathbf{q},\mathbf{p})
\]
defines the $i_{1}i_{2}..i_{D}$ component of the moment of order
$n=\sum_{k}^{D}i_{k}$, $D$ is the number of degrees of freedom and
$\rho_{W}(\mathbf{q},\mathbf{p})$ is the Wigner function
\[
\rho_{W}(\mathbf{q},\mathbf{p})=\int\frac{d^{D}\mathbf{r}}{(2\pi\hbar)^{D}}\sigma\left(\mathbf{\mathbf{q}}-\frac{\mathbf{r}}{2},\mathbf{q}+\frac{\mathbf{r}}{2}\right)e^{i\frac{\mathbf{p}\mathbf{r}}{\hbar}}
\]
associated to the system density matrix $\sigma(\mathbf{x},\mathbf{x}')$.
For instance, in this notation, $S_{00..0}^{(0)}(\mathbf{q})$ is
just the density in configuration space, $S_{01..0}^{(1)}(\mathbf{q})$
is the second component of the first order moment, etc. 

There exist several equivalent ways to express these moments, the
one that is most useful for our purposes is by means of 
\[
\mathcal{P}_{ijk..}^{(n)}(\mathbf{q})=\frac{\braket{\mathbf{q}|[\hat{p}_{i},[\hat{p}_{j},[\hat{p}_{k},..\hat{\sigma}]_{+}]_{+}]_{+}|\mathbf{q}}}{2^{n}}
\]
where $[A,B]_{+}=AB+BA$ is the anti-commutator and now the hat denotes
abstract operators, e.g, $\hat{p}_{k}$ and $\hat{\sigma}$ for the
$k^{\text{th}}$ momentum and for the density operator, respectively.
In the last expression, the subscript labels the coordinates and specifies
the order of the products --- the momentum operators appearing in
the nested products are labeled with indexes from right to left, starting
from the innermost product. However, since the $\hat{p}_{k}$'s commute
with each other, the $\mathcal{P}_{ijk..}^{(n)}(\mathbf{q})$ 's so
defined are fully symmetric in their indexes and coincide with the
moments above. That is, 
\begin{align*}
\mathcal{P}_{ijk..}^{(n)}(\mathbf{q}) & =\mathcal{P}_{ikj..}^{(n)}(\mathbf{q})=\mathcal{P}_{kij..}^{(n)}(\mathbf{q})..\equiv S_{i_{1}i_{2}..i_{D}}^{(n)}\\
 & \ \ \ i_{m}=\text{count}([ijk..],m)
\end{align*}
where $\text{count}([ijk..],m)$ counts the occurrences of $m$ in
the list of indexes $ijk..$. This representation of the moments is
redundant but it is very useful if we intend to generalize the moment
construction to \emph{gauge} covariant, non-commuting operators. Specifically,
in connection with the dynamics provided by our Hamiltonian, Eq. \ref{eq:Hamiltonian},
we define mechanical\textbf{ }momentum\textbf{ }moments according
to
\begin{align}
\mathcal{M}^{(0)}(\mathbf{q}) & =\braket{\mathbf{q}|\hat{\sigma}|\mathbf{q}}\equiv n(\mathbf{q})\nonumber \\
\mathcal{M}_{k}^{(1)}(\mathbf{q}) & =\frac{\braket{\mathbf{q}|[\hat{\pi}_{k},\hat{\sigma}]_{+}|\mathbf{q}}}{2}\nonumber \\
\mathcal{M}_{jk}^{(1)}(\mathbf{q}) & =\frac{\braket{\mathbf{q}|[\hat{\pi}_{j}[\hat{\pi}_{k},\hat{\sigma}]_{+}]_{+}|\mathbf{q}}}{2^{2}}\nonumber \\
.. & ..\nonumber \\
\mathcal{M}_{ijk..}^{(n)}(\mathbf{q}) & =\frac{\braket{\mathbf{q}|[\hat{\pi}_{i},[\hat{\pi}_{j},[\hat{\pi}_{k},..\hat{\sigma}]_{+}]_{+}]_{+}|\mathbf{q}}}{2^{n}}\label{eq:mechanical momentum moments - 1}
\end{align}
This definition can be equivalently expressed in a form that is rather
useful for calculations, namely
\begin{equation}
\begin{array}{cc}
\mathcal{M}_{ijk..}^{(n)}(\mathbf{q}) & =\left(\frac{\pi_{i}+\pi_{i}^{'*}}{2}\right)\left(\frac{\pi_{j}+\pi_{j}^{'*}}{2}\right)\left(\frac{\pi_{k}+\pi_{k}^{'*}}{2}\right)\\
 & ..\,\sigma(\mathbf{x},\mathbf{x}')\vert_{\mathbf{x}=\mathbf{x}'=\mathbf{q}}
\end{array}\label{eq:eq:mechanical momentum moments - 2}
\end{equation}
where now $\pi_{k}=-i\hbar\partial_{k}-\hbar A_{k}$, $\pi'_{k}=-i\hbar\partial'_{k}-\hbar A'_{k}$
and $A_{k}\equiv A_{k}(\mathbf{x})$, $A_{k}'\equiv A_{k}(\mathbf{x}')$
(see Appendix \ref{app:momentum moments} for details). Note that
here the symbol $\pi_{k}$, differently from the previous Sections,
is used to mean a differential operator. 

The moments defined in this way are, by construction, \emph{gauge}
invariant and they thus fulfill our requirement. The price to be paid
is that they are no longer symmetric in their indexes since 
\[
[\pi_{j}+\pi_{j}^{'*},\pi_{k}+\pi_{k}^{'*}]=[\pi_{j},\pi_{k}]+[\pi_{j}^{'*},\pi_{k}^{'*}]\equiv i\hbar(B_{jk}-B_{jk}^{'})
\]
where again we have primed (unprimed) quantities for functions of
$\mathbf{x}'$ ($\mathbf{x}$). For instance, a direct calculation
shows that 
\[
\mathcal{M}_{ijk}^{(3)}-\mathcal{M}_{ikj}^{(3)}=\left(-\frac{i\hbar}{2}\right)^{2}\partial_{i}B_{jk}(\mathbf{q})\,\mathcal{M}^{(0)}(\mathbf{q})
\]
\begin{align*}
\mathcal{M}_{mijk}^{(4)}-\mathcal{M}_{mikj}^{(4)} & =\left(-\frac{i\hbar}{2}\right)^{2}\left(\partial_{i}B_{jk}(\mathbf{q})\,\mathcal{M}_{m}^{(1)}(\mathbf{q})+\right.\\
 & \left.+\partial_{m}B_{jk}(\mathbf{q})\,\mathcal{M}_{i}^{(1)}\right)
\end{align*}
The exception is the symmetry in the two leftmost indexes 
\[
\mathcal{M}_{ijk..}^{(n)}(\mathbf{q})=\mathcal{M}_{jik..}^{(n)}(\mathbf{q})
\]
which holds for any moment of order $n\ge2$. 

Note that we \emph{could} define moments which are yet \emph{gauge}
invariant and symmetric in their indexes, for instance 
\begin{align*}
\mathcal{N}_{ijk..}^{(n)}(\mathbf{q}) & =\left(\frac{p_{i}+p_{i}^{'*}}{2}-\hbar A_{i}(\mathbf{q})\right)\left(\frac{p_{j}+p_{j}^{'*}}{2}-\hbar A_{j}(\mathbf{q})\right)\\
 & \times\left(\frac{p_{k}+p_{k}^{'*}}{2}-\hbar A_{k}(\mathbf{q})\right)..\,\sigma(\mathbf{x},\mathbf{x}')\vert_{\mathbf{x}=\mathbf{x}'=\mathbf{q}}
\end{align*}
However, only for the $\mathcal{M}_{ijk..}^{(n)}$'s defined above
the equations of motion take a simple form. 

The moment equations of motion follow indeed from the Liouville -
von Neumann (LvN) equation and for the first three moments take the
form (see Appendix \ref{app:momentum moments})
\begin{equation}
\frac{\partial\mathcal{M}^{(0)}}{\partial t}(\mathbf{q})=-\sum_{ij}\xi^{ij}\partial_{i}\mathcal{M}_{j}^{(1)}(\mathbf{q})\label{eq:Zero-th moment EOM}
\end{equation}
\begin{align}
\frac{\partial\mathcal{M}_{k}^{(1)}}{\partial t}(\mathbf{q}) & =-\sum_{ij}\xi^{ij}\partial_{i}\mathcal{M}_{kj}^{(2)}(\mathbf{q})+\nonumber \\
 & +\left(-\partial_{k}\mathcal{V}(\mathbf{q})+E_{k}\right)\mathcal{M}^{(0)}(\mathbf{q})+\label{eq:First moment EOM}\\
 & +\sum_{ij}\xi^{ij}B_{ki}(\mathbf{q})\mathcal{M}_{j}^{(1)}(\mathbf{q})\nonumber 
\end{align}
\begin{align}
\frac{\partial\mathcal{M}_{km}^{(2)}}{\partial t}(\mathbf{q}) & =-\sum_{ij}\xi^{ij}\partial_{i}\mathcal{M}_{kmj}^{(3)}(\mathbf{q})+\nonumber \\
 & +\left(-\partial_{k}\mathcal{V}+E_{k}\right)\mathcal{M}_{m}^{(1)}(\mathbf{q})+\nonumber \\
 & +\left(-\partial_{m}\mathcal{V}+E_{m}\right)\mathcal{M}_{k}^{(1)}(\mathbf{q})+\nonumber \\
 & +\sum_{ij}\xi^{ij}\left(B_{ki}(\mathbf{q})\mathcal{M}_{mj}^{(2)}(\mathbf{q})+B_{mi}(\mathbf{q})\mathcal{M}_{kj}^{(2)}(\mathbf{q})\right)\label{eq:Second moment EOM}
\end{align}

In general, for the $n^{\text{th}}$ moment we have an ``upward''
coupling to the $(n+1)^{\text{th}}$ layer of moments that occurs
through a surface term containing the spatial derivatives of the $(n+1)^{\text{th}}$
moments, and both ``downward'' and ``horizontal'' couplings that
occur through ``classical'' forces. The rank-1 \emph{pseudo}-electric
field $E_{k}$ and the intrinsic force $-\partial_{k}\mathcal{V}$
link the $n^{\text{th}}$ layer of moments to the layer \emph{below}
it. The rank-2 \emph{pseudo-magnetic} field $B_{kj}$ increases the
order by one and thus describes ``horizontal'' coupling to moments
of the same order $n$ (starting from $n=1$). 

In the above hydrodynamic set of equations the first, Eq. \ref{eq:Zero-th moment EOM},
is continuity\textbf{ }equation, with $\mathcal{J}^{i}(\mathbf{q})=\sum_{j}\xi^{ij}\mathcal{M}_{j}^{(1)}(\mathbf{q})$
playing the role of density current. The second equation, Eq. \ref{eq:First moment EOM},
extends the quantum\textbf{ }Navier-Stokes\textbf{ }equation for a
pure state, Eq. \ref{eq:quantum Navier-Stokes}, 
\[
n\frac{d\pi_{:k}}{dt}=nF_{k}+\sum_{j}\partial_{j}\sigma_{k}^{j}
\]
here written in the form 
\[
\partial_{t}(n\pi_{k})=nF_{k}-\sum_{j}\partial_{j}(nv^{j}\pi_{k})+\sum_{j}\partial_{j}\sigma_{k}^{j}
\]
where use has been made of the identity $\partial_{t}(n\pi_{k})+\sum_{j}\partial_{j}(nv^{j}\pi_{k})\equiv n\,d\pi_{k}/dt$.
 The comparison with the pure state is insightful, since it suggests
to single out the ``uncorrelated'' term from the second moment by
defining the second cumulant as
\begin{equation}
\mathcal{C}_{kj}^{(2)}(\mathbf{q})=\mathcal{M}_{kj}^{(2)}(\mathbf{q})-\frac{\mathcal{M}_{k}^{(1)}(\mathbf{q})\mathcal{M}_{j}^{(1)}(\mathbf{q})}{\mathcal{M}^{(0)}(\mathbf{q})}\label{eq:second moment correlation}
\end{equation}
For a pure state we have 
\[
-\sum_{ij}\xi^{ij}\partial_{i}\mathcal{C}_{kj}=\sum_{i}\partial_{i}\sigma_{k}^{i}\equiv\mathcal{F}_{k}^{S}
\]
and the cumulant is seen to be the source of the density of surface
forces, $\mathcal{F}_{k}^{S}$ (the only ones having a quantum origin).
In general, i.e. for an arbitrary mixed state, we can always introduce
the rank-2 tensor $\mathcal{C}_{kl}^{(2)}(\mathbf{q})$ according
to Eq. \ref{eq:second moment correlation} and re-write Eq. \ref{eq:First moment EOM}
in terms of it. The result is 
\begin{align}
\frac{d\pi_{k}}{dt}(\mathbf{q}) & =-\frac{1}{\mathcal{M}^{(0)}(\mathbf{q})}\sum_{ij}\xi^{ij}\partial_{i}\mathcal{C}_{kj}^{(2)}(\mathbf{q})+\nonumber \\
 & +(-\partial_{k}\mathcal{V}(\mathbf{q})+E_{k})+\sum_{j}B_{kj}(\mathbf{q})v^{j}(\mathbf{q})\label{eq:First moment EOM - Lagrange}
\end{align}
where $\pi_{k}=\mathcal{M}_{k}^{(1)}/\mathcal{M}^{(0)}$ and $v^{j}=\sum_{i}\xi^{ji}\pi_{i}$
are the usual momentum and velocity fields. We thus see that, in the
form of Eq. \ref{eq:First moment EOM - Lagrange}, the first moment
equation describes the Lagrangian-frame evolution of momentum in terms
of classical, bulk forces and a surface term which is responsible
for quantum effects. 

\section{Quantum Hydrodynamics of exactly factorized electrons-nuclei\label{sec:Quantum-Hydrodynamics-of}}

We are now ready to address the main issue of this paper, that is
replacing the EF equations with \emph{gauge} invariant equations of
motion for both electrons and nuclei. We shall start with the pure
state case where this program can be accomplished in an exact way,
and later we shall explore extensions of the theory to special statistical
mixtures that require some assumptions. 

\subsection{Pure states\label{subsec:Pure-states}}

We recall that the generic Hamiltonian of Eq. \ref{eq:Hamiltonian},
is also the Hamiltonian governing the time evolution of the marginal
probability amplitude describing the nuclei in the exact factorization
approach to the electron-nuclear problem. Hence, as shown in the previous
Section this immediately leads to a \emph{gauge} invariant formulation
of the nuclear dynamics in hydrodynamical form, for instance that
provided in Lagrangian frame by the complex momentum equation presented
in Eq. \ref{eq:complex Madelung (final)}, with the complex momentum
fields $\Pi_{k}$ given by Eq. \ref{eq:complex material derivative 2}.
In that case, the dynamical problem needs to be augmented by the coupled
equation of motion for the local electronic states, Eq. \ref{eq:EF electronic equation},
\begin{equation}
i\hbar Q\left(\frac{\partial}{\partial t}+\sum_{j}V^{j}\partial_{j}\right)\ket{u}=QH_{\text{el}}\ket{u}-\hbar R\ket{u}\label{eq:electronic equation of motion}
\end{equation}
here written for the only relevant component of $\partial_{t}\ket{u}$
--- the \emph{gauge} covariant projection $Q\partial_{t}\ket{u}$
which is ``free'' of the \emph{gauge} choice --- and rearranged
to make explicit on the l.h.s. the complex material derivative of
Eq. \ref{eq:complex material derivative}. This equation can be presented
in complex hydrodynamical form the closely parallels that momentum
equation for the nuclei (Eq. \ref{eq:complex Madelung (final)}),
i.e., in the form
\[
i\hbar Q\frac{\tilde{d}}{dt}\ket{u}=Q\left(H_{\text{el}}\ket{u}-\frac{\hbar^{2}}{2}\bar{\Delta}_{M}\right)\ket{u}
\]
where $\bar{\Delta}_{M}:=\sum_{ij}\xi^{ij}D_{i}D_{j}$ is a mass-weighted
\emph{gauge} covariant Laplacian (cf. with Eq. \ref{eq:mass-weighted Laplacian}),
defined with the help of the \emph{gauge} covariant derivative $D_{j}=\partial_{j}+iA_{j}$.
Alternatively, one can work with real-valued nuclear fields and consider
\begin{equation}
i\hbar Q\frac{d\ket{u}}{dt}=QH_{\text{el}}\ket{u}+\hbar\sum_{j}u^{j}Q\partial_{j}\ket{u}-\hbar R\ket{u}\label{eq:electronic equation of motion II}
\end{equation}
in parallel to the real-valued momentum equation for the nuclei, Eq.
\ref{eq:real Madelung (final)}, and jointly with the continuity equation
(Eq. \ref{eq:continuity equation Lagrange (general)}). Upon making
explicit the role of the density gradients, one can present the dynamical
equations in a suggestive form, namely
\begin{equation}
iQ\frac{d\ket{\varphi}}{dt}=Q\left(\ket{\Phi}-\frac{\hbar}{2}\sum_{ij}\xi^{ij}(\partial_{i}\ln\,n)\partial_{j}\ket{\varphi}-\frac{\hbar}{2}\bar{\Delta}_{M}\ket{\varphi}\right)\label{eq:HEF trajectory equation I}
\end{equation}
for the electrons and
\begin{equation}
\frac{d\pi_{k}}{dt}=\left(F_{k}-\frac{\hbar}{2}\sum_{ij}\xi^{ij}(\partial_{i}\ln\,n)\partial_{j}w_{k}-\frac{\hbar}{2}\Delta_{M}w_{k}\right)\label{eq:HEF trajectory equation II}
\end{equation}
for the nuclei, where now $\ket{\varphi}=\hbar\ket{u}$ and $\ket{\Phi}=H_{\text{el}}\ket{u}/\hbar$.
For $k=1,2,..D$ Eq. \ref{eq:HEF trajectory equation II} represents
trajectory equations similar to those of the De Broglie - Bohm - Madelung
theory, and defines the evolution of the elementary volumes of the
quantum probability fluid in nuclear configuration space\textbf{ }(noting
that the second and third terms in Eq. \ref{eq:HEF trajectory equation II}
are just the quantum force, see Eq. \ref{eq:quantum force})\textbf{.
}However, they are now augmented by an equation for the electronic
state, Eq. \ref{eq:HEF trajectory equation I}, that \emph{follows}
each fluid element
\[
\ket{\varphi}\equiv\hbar\ket{u(\mathbf{x}_{t},t)}
\]
similarly to the quantum hydrodynamical description of a particle
with a spin \citep{Bialynicki-Birula}. This ``quantum vector field''
associated to the nuclear fluid evolves according to the local electronic
Hamiltonian (first term on the r.h.s. of Eq. \ref{eq:HEF trajectory equation I})
and the coupling to the nuclear motion (second and third term on the
r.h.s. of Eq. \ref{eq:HEF trajectory equation I}). The latter takes
a form which is remarkably similar to the quantum force contribution
appearing in the quantum trajectory of motion for the nuclei: they
share precisely the same coupling factor (density dependent) and differ
\emph{only} in the auxiliary field they couple to, which is $\ket{\varphi}$
for the electronic variables and the set of $w_{k}$'s for the nuclear
variables \footnote{It may be worth noticing that in the \emph{realified} Hilbert electronic
space $i\ket{\varphi}$ and $\ket{\varphi}$ are linearly independent
from each other. In the dynamical equations they are ``paired''
to each other precisely as $\mathbf{\pi}_{k}=\Re\Pi_{k}$ and $w_{k}=\Im\Pi_{k}$.}. 

In the above equations the effect of the electron dynamics on the
nuclear motion is condensed in the ``classical'' force $F_{k}$
(the real part of Eq. \ref{eq:complex valued total classical force}),
containing both the averaged electronic forces and the \emph{gauge}
fields,
\begin{equation}
F_{k}=-\partial_{k}\mathcal{V}+\sum_{j}v^{j}B_{kj}+\hbar(\partial_{k}A_{0}-\partial_{t}A_{k})\label{eq:total classical force}
\end{equation}
Here, three distinct contributions are evident. The first term is
the scalar potential known from the Born-Oppenheimer approximation,
though now involving a time dependent electronic state. It explicitly
reads as
\[
\mathcal{V}=E_{\text{el}}+\phi_{\text{FS}}=\braket{u|H_{\text{el}}|u}+\hbar^{2}\sum_{ij}\frac{\xi^{ij}}{2}g_{ij}
\]
where $E_{\text{el}}=E_{\text{el}}(\mathbf{x})$ is the average (local)
electronic energy, $g_{ij}=\Re q_{kj}$ is the Fubini-Study metric
tensor and $q_{ij}=\braket{\partial_{i}u|Q|\partial_{j}u}$ the instantaneous
quantum geometric tensor. This is the term that was dubbed ``\emph{gauge}-invariant
part of the time-dependent potential energy surface'' in the seminal
work by Gross and coworkers on exact factorization \footnote{In Appendix C of our previous work, Ref. \citep{Martinazzo2022a},
the original electronic equation of the exact factorization approach
was written as $i\hbar\partial_{t}\ket{u}=(H_{\text{el}}-\varepsilon)\ket{u}+F\ket{u}$,
and it was proved that $F=K[\psi]+\phi_{\text{FS}}$, where $K[\psi]$
is zero-averaged, i.e. $\braket{u|K[\psi]|u}\equiv0$. Gross and coworkers
defined the time-dependent potential energy surface (TDPES) as $\braket{u|H_{\text{el}}+F-i\hbar\partial_{t}|u}\equiv\braket{u|H_{\text{el}}|u}+\phi_{\text{FS}}-\hbar A_{0}$
where only the first two terms are \emph{gauge} invariant. The last
term must be combined with the Berry connection to form the electron-dynamical
force $F_{k}^{\text{ED}}$.}. The second term of Eq. \ref{eq:total classical force} represents
the usual \emph{pseudo}-magnetic force driven by the Berry curvature
$B_{kj}=-2\hbar\Im q_{kj}$ while the last term of Eq. \ref{eq:total classical force}
is the ``electron dynamical'' force, the only one containing the
time-derivative of the electronic state, $F_{k}^{\text{ED}}\equiv-2\hbar\Im\braket{\partial_{k}u|Q\partial_{t}u}$.
This last contribution precisely corresponds to the \emph{pseudo}-electric
field $E_{k}$, as defined in Eq. \ref{eq:E,B fields}. However, in
the context of the exact factorization of the wavefunction, it would
be misleading to label it as ``electric'' due to the presence of
a velocity-dependent (magnetic) contribution. Indeed, upon making
apparent the material derivative of the electronic state, we find
 
\begin{equation}
F_{k}^{\text{ED}}=-2\hbar\Im\braket{\partial_{k}u|Q|\dot{u}}-\sum_{j}v^{j}B_{kj}=G_{k}^{\text{ED}}-\sum_{j}v^{j}B_{kj}\label{eq:ED force in Euler and Lagrangian "frame"}
\end{equation}
where $G_{k}^{\text{ED}}$ --- defined by this equation --- is the
``Lagrangian-frame'' electron-dynamical force. $G_{k}^{\text{ED}}$
is free of velocity contributions, as a simple inspection of Eq. \ref{eq:electronic equation of motion II}
reveals.  Hence, the total classical force of Eq. \ref{eq:total classical force}
does \emph{not} explicitly depend on the nuclear velocity,
\[
F_{k}=-\partial_{k}\mathcal{V}+G^{\text{ED}}
\]
Furthermore, upon using Eq. \ref{eq:electronic equation of motion II},
$G_{k}^{\text{ED}}$ can be split as 
\begin{align}
G_{k}^{\text{ED}}=\delta F_{k}^{\text{BO}}+\delta F_{k}^{\text{el}}\label{eq:G force: BO and electric correction}
\end{align}
where, $\delta F_{k}^{\text{BO}}=2\Re\braket{\partial_{k}u|QH_{\text{el}}|u}$
is a dynamical correction (due to the electronic response) to the
Born-Oppenheimer force $F_{k}^{\text{BO}}=-\partial_{k}\braket{u|H_{\text{el}}|u}$,
and 
\begin{equation}
\delta F_{k}^{\text{el}}=2\hbar\sum_{j}u^{j}g_{kj}-2\hbar\Re\braket{\partial_{k}u|R|u}\label{eq:dynamical correction to electric force}
\end{equation}
is the corresponding correction to the \emph{pseudo}-electric force
$-\partial_{k}\phi_{\text{FS}}$ appearing already in the adiabatic
approximation. Since the \emph{dynamically corrected} BO and electric
forces read as, respectively,
\begin{equation}
\mathcal{F}_{k}^{\text{BO}}:=F_{k}^{\text{BO}}+\delta F_{k}^{\text{BO}}\equiv-\braket{u|\partial_{k}H_{\text{el}}|u}\label{eq:dynamically corrected BO force}
\end{equation}
and \footnote{To obtain this equation we exploited the geometrical identity $\Re\braket{\partial_{k}u|QD_{i}D_{j}u}=\frac{1}{2}(\partial_{j}g_{ik}+\partial_{i}g_{kj}-\partial_{k}g_{ij}),$see
Ref. \citep{Martinazzo2022a}.}
\begin{equation}
\mathcal{F}_{k}^{\text{el}}:=-\partial_{k}\phi_{\text{FS}}+\delta F_{k}^{\text{el}}\equiv-\frac{\hbar^{2}}{n}\sum_{ij}\xi^{ij}\partial_{i}(ng_{kj})\label{eq:dynamically corrected E force}
\end{equation}
we arrive at
\begin{equation}
F_{k}=-\braket{u|\partial_{k}H_{\text{el}}|u}-\frac{\hbar^{2}}{n}\sum_{ij}\xi^{ij}\partial_{i}(ng_{kj})\label{eq:dynamically corrected total force}
\end{equation}
This is the same result as obtained in our previous work \citep{Martinazzo2022,Martinazzo2022a},
if it were not for the fact that here the\textbf{ }\emph{pseudo}-magnetic\textbf{
}force\textbf{ }vanishes\textbf{ }\emph{precisely}\textbf{ (}and\textbf{
}not\textbf{ }on\textbf{ }average\textbf{)} \footnote{In our previous work, we called $\delta F_{k}^{\text{BO}}$ a ``genuine
non-Born-Oppenheimer force'' and used the symbol $F_{k}^{\text{NBO}}$
for it.}.\textbf{ }The\textbf{ }reason\textbf{ }is\textbf{ }that\textbf{ }we\textbf{
}are\textbf{ }now\textbf{ }dealing\textbf{ }with\textbf{ }the\textbf{
}classical\textbf{ }force (as needed in the hydrodynamic picture)
and not with the quantum one used in Refs. \citep{Martinazzo2022,Martinazzo2022a},
i.e., that in the Heisenberg picture \footnote{In other words, this magnetic force only appears in the Euler frame
where $\mathbf{x}$ is fixed and the fluid elements move with speed
$v^{i}(\mathbf{x},t)$. In the Lagrangian frame $\mathbf{x}_{t}$
is at rest with the moving fluid, and cannot couple to the magnetic
field.}. For comparison, Table \ref{tab:Quantum-mechanical-forces} gives
the full set of Heisenberg-picture force components for both an adiabatic
and an exact dynamics. 

A few comments are in order at this point. First of all, $\mathcal{F}_{k}^{\text{el}}$
(Eq. \ref{eq:dynamically corrected E force}) is manifestly zero-averaged
when interpreted as (local) \emph{operator} acting on the nuclear
wavefunction
\[
\int d\mathbf{x}\psi^{*}F_{k}^{\text{el},c}\psi=-\hbar^{2}\sum_{ij}\xi^{ij}\int d\mathbf{x}\partial_{i}(ng_{kj})\equiv0
\]
but it is generally non-zero as a \emph{classical force, }i.e.\emph{
}as required by the trajectory equation, Eq. \ref{eq:HEF trajectory equation II}.
Furthermore, as already observed in Refs. \citep{Martinazzo2022,Martinazzo2022a},
the dynamical correction to the BO force turns the force of the electronically-averaged
potential into a \emph{mean electronic force}: this is a key difference
since the first is manifestly \emph{conservative} while the latter
is \emph{not }(unless the electronic state is stationary)\emph{. }The
same occurs for the \emph{pseudo}-electric (Fubini-Study) force, which
is turned non-conservative when dynamically corrected.\emph{ }Hence,
in a sense, the electron dynamics works as an electromotive force
for the nuclear dynamics. 
\begin{table*}
\begin{centering}
\begin{tabular}{c|c|c|c}
\hline 
 & Adiabatic & Exact & Exact\tabularnewline
\hline 
\hline 
BO\textcolor{white}{\huge{}I} & $-\partial_{k}\braket{u_{0}|H_{\text{el}}|u_{0}}$ & $-\partial_{k}\braket{u|H_{\text{el}}|u}$ & $-\braket{u|\partial_{k}H_{\text{el}}|u}$\tabularnewline
\hline 
el\textcolor{white}{\huge{}I} & $-\frac{\hbar^{2}}{2}\sum\xi^{ij}\partial_{k}g_{ij}^{0}$ & $-\frac{\hbar^{2}}{2}\sum\xi^{ij}\partial_{k}g_{ij}$ & $-\frac{\hbar^{2}}{n}\sum\xi^{ij}\partial_{i}(ng_{jk})$\tabularnewline
\hline 
mag\textcolor{white}{\huge{}I} & $\frac{\hbar}{2}\sum_{j}(\hat{v}_{j}^{0}B_{kj}^{0}+B_{kj}\hat{v}_{j}^{0})$ & $\frac{\hbar}{2}\sum_{j}(\hat{v}_{j}B_{kj}+B_{kj}\hat{v}_{j})$ & $\frac{\hbar}{2}\sum_{j}(\Delta\hat{v}_{j}B_{kj}^{0}+B_{kj}\Delta\hat{v}_{j})$\tabularnewline
\hline 
ED\textcolor{white}{\huge{}I} &  & $-2\hbar\Im\braket{\partial_{k}u|Q|\partial_{t}u}$ & \tabularnewline
\hline 
\end{tabular}
\par\end{centering}
\caption{\label{tab:Quantum-mechanical-forces}Quantum mechanical (Heisenberg)
forces {[}BO = Born-Oppenheimer, el = \emph{pseudo}-electric, mag
= \emph{pseudo}-magnetic, ED = electrodynamical{]} acting on the nuclei
in both the adiabatic approximation (second column) and the exact
dynamics, with either the electron dynamical force (third column)
or the dynamical corrections (fourth columns). The superscript $^{0}$
indicates the use of a stationary-state connection and metric, $n=|\psi|^{2}$
is the nuclear density and $\Delta\hat{v}_{j}=\hat{v}_{j}-\Re V^{j}$.
The forces are all \emph{local} except for the \emph{pseudo}-magnetic
one. As a consequence, only the latter needs to be \textquotedblleft classicalized\textquotedblright{}
in the hydrodynamical picture, with the replacement $\hat{v}^{j}\rightarrow v^{j}$.
This procedure makes the dynamically corrected \emph{pseudo}-magnetic
force identically vanishing in the hydrodynamic picture.}
\end{table*}

The equations of motion (Eqs. \ref{eq:HEF trajectory equation I},
\ref{eq:HEF trajectory equation II}) are not yet in the desired form,
since the electronic equation (\ref{eq:HEF trajectory equation I})
is \emph{gauge} covariant rather than \emph{gauge} invariant. To bring
it in the required form we focus on the conditional\textbf{ }electronic\textbf{
}density\textbf{ }operator $\rho_{\text{el}}=\ket{u}\bra{u}\equiv\rho_{\text{el}}(\mathbf{x})$
and on its dynamics, as it follows from Eq. \ref{eq:electronic equation of motion II}.
As shown in detail in Appendix \ref{app:gauge invariant formulation of electron dynamics},
this is a \emph{closed} equation for the density operator, involving
$\rho_{\text{el}}$ itself and its spatial first and second derivatives.
We emphasize that the $\rho_{\text{el}}(\mathbf{x})$'s are \emph{conditional}
operators, and that they are only loosely connected to the true electronic
density operator $\text{\ensuremath{\gamma}}_{\text{el}}$, hence
the existence and knowledge of an electronic equation for the $\rho_{\text{el}}(\mathbf{x})$'s
are not obvious. In particular, we remind that $\text{\ensuremath{\gamma}}_{\text{el}}$
is the convex sum of the $\rho_{\text{el}}$'s over nuclear configuration
space, weighted by the nuclear density (see Ref. \citep{Martinazzo2022a},
Sec. IIIC), 
\begin{equation}
\text{\ensuremath{\gamma}}_{\text{el}}=\int d\mathbf{x}n(\mathbf{x})\rho_{\text{el}}(\mathbf{x})\label{eq:total electronic density operator (pure state)}
\end{equation}
The Liouville-von Neumann-like equation of motion governing the evolution
of $\rho_{\text{el}}$ reads as (see Appendix \ref{app:gauge invariant formulation of electron dynamics})
\begin{equation}
i\hbar\dot{\rho}_{\text{el}}=[H_{\text{el}}+\delta H_{\text{en}},\rho_{\text{el}}]\label{eq:electronic equation final}
\end{equation}
where the dot stands for the material derivative and $\delta H_{\text{en}}$
is the electron-nuclear correction to the electronic Hamiltonian defining
the Liouvillian
\begin{equation}
\delta H_{\text{en}}=-\frac{\hbar^{2}}{2n}\sum_{ij}\xi^{ij}\partial_{i}(n\partial_{j}\rho_{\text{el}}),\label{eq:electron-.nuclear correction to the Lioviallian}
\end{equation}
Eq. \ref{eq:electronic equation final} is both trace- and purity-
conserving, i.e., 
\[
\text{Tr}_{\text{el}}\dot{\rho}_{\text{el}}=0\ \ \ \text{Tr}_{\text{el}}\left(\rho_{\text{el}}\dot{\rho}_{\text{el}}\right)=0
\]
see Appendix \ref{app:gauge invariant formulation of electron dynamics}
for a proof \footnote{These properties guarantee that $\rho_{\text{el}}$ represents a pure-state
density at any time, provided it does at initial time. As a corollary,
the instantaneous electronic ket $\ket{u_{t}(\mathbf{x})}$ can be
retrieved from $\rho_{\text{el}}(\mathbf{x})$ by projection, using
an (almost) arbitrary $\ket{\psi}\in\mathcal{H}_{\text{el}}$. In
this picture, it is the (\emph{quasi}) arbitraness of $\ket{\psi}$
that provides the \emph{gauge} freedom.}. Jointly with the continuity equation for the nuclear density and
with the momentum equation for the nuclei, 
\begin{align}
\frac{d\pi_{k}}{dt} & =-\braket{u|\partial_{k}H_{\text{el}}|u}-\frac{\hbar^{2}}{n}\sum_{ij}\xi^{ij}\partial_{i}(ng_{kj})+\nonumber \\
 & +\frac{\hbar^{2}}{4n}\sum_{ij}\xi^{ij}\partial_{i}(n\partial_{j}\partial_{k}\ln\,n)\label{eq:nuclear equation final}
\end{align}
it forms a full set of \emph{gauge} invariant equations for both the
electrons and the nuclei. Here, in re-writing Eq. \ref{eq:HEF trajectory equation II}
we have made explicit the role of the electronic Hamiltonian and that
of the \emph{gauge} fields, and we have combined the quantum contributions
in a single term. 

Eqs. \ref{eq:electronic equation final} and \ref{eq:nuclear equation final}
represent the first main finding of this work: they provide an exact,
\emph{gauge} invariant description of the electronic and nuclear dynamics,
in the Lagrangian frame defined by the nuclear quantum fluid. Alternatively,
the nuclear hydrodynamical equations of motion can be given in the
Eulerian frame. Using the moment notation and the results of Section
\ref{subsec:Statistical-mixtures}, they read as 
\begin{equation}
\frac{\partial\mathcal{M}^{(0)}}{\partial t}(\mathbf{q})=-\sum_{ij}\xi^{ij}\partial_{i}\mathcal{M}_{j}^{(1)}(\mathbf{q})\label{eq:moment nuclear equation 1}
\end{equation}
\begin{equation}
\frac{\partial\mathcal{M}_{k}^{(1)}}{\partial t}(\mathbf{q})=-\sum_{ij}\xi^{ij}\partial_{i}\mathcal{M}_{kj}^{(2)}(\mathbf{q})+F_{k}\mathcal{M}^{(0)}(\mathbf{q})\label{eq:moment nuclear equation 2 first}
\end{equation}
for the zero-th and the first moment, respectively, and replace Eqs.
\ref{eq:nuclear equation final} and \ref{eq:continuity equation Lagrange (general)}
formulated in Lagrangian frame. Here, the net force $F_{k}$ is (see
Eq. \ref{eq:dynamically corrected total force})
\begin{equation}
F_{k}=-\text{Tr}_{\text{el}}(\rho_{\text{el}}\partial_{k}H_{\text{el}})-\frac{\hbar^{2}}{\mathcal{M}^{(0)}}\sum_{ij}\xi^{ij}\partial_{i}(\mathcal{M}^{(0)}g_{kj})\label{eq:hydrodynamical force in EF}
\end{equation}
and the second moment is subjected to the pure-state closure 
\begin{align}
\mathcal{M}_{kj}^{(2)}(\mathbf{q}) & =\frac{\mathcal{M}_{k}^{(1)}(\mathbf{q})\mathcal{M}_{j}^{(1)}(\mathbf{q})}{\mathcal{M}^{(0)}(\mathbf{q})}+\nonumber \\
 & -\frac{\hbar^{2}}{4}\left(\partial_{k}\partial_{j}\mathcal{M}^{(0)}(\mathbf{q})-\frac{\partial_{k}\mathcal{M}^{(0)}(\mathbf{q})\partial_{j}\mathcal{M}^{(0)}(\mathbf{q})}{\mathcal{M}^{(0)}(\mathbf{q})}\right)\label{eq:pure state closure}
\end{align}

Before closing this Section we emphasize that one main advantage of
a \emph{gauge} invariant formulation of the dynamics is that it removes
the need of obtaining the electronic wavefunctions with a smooth phase
factor, when the latter is computed on-the-fly along the nuclear trajectories.
This issue can pose significant difficulties, especially in systems
with broken time-reversal invariance--- e.g. in the presence of a
magnetic field. In such cases, the overlap between adjacent electronic
states, $\braket{u(\mathbf{x}_{t})|u(\mathbf{x}_{t+\Delta t})}$,
takes arbitrary phases and it becomes challenging to make it smooth\emph{
}by selecting an appropriate phase from the random values generated
by the electronic solver. In the above formulation, on the other hand,
the only information required consists of the \emph{gauge} invariant
\emph{fields}, and these can be directly obtained from $\rho_{\text{el}}$
and its spatial derivatives. For instance,
\begin{equation}
q_{kj}=\text{Tr}_{\text{el}}\left(\rho_{\text{el}}\left(\partial_{k}\rho_{\text{el}}\right)\left(\partial_{j}\rho_{\text{el}}\right)\right)\label{eq:quantum geometric tensor (pure state)}
\end{equation}
provides the instantaneous quantum geometric tensor, hence the Fubini-Study
metric (which is needed for the dynamics, see the force of Eq. \ref{eq:hydrodynamical force in EF})
and the Berry curvature. 

Furthermore, we notice that upon multiplying Eq. \ref{eq:electronic equation final}
by $n(\mathbf{x})$ and integrating over all nuclear configuration
space one obtains
\begin{equation}
\frac{\partial\gamma_{\text{el}}}{\partial t}=-\frac{i}{\hbar}\int d\mathbf{x}\,n(\mathbf{x})\,[H_{\text{el}}(\mathbf{x}),\rho_{\text{el}}(\mathbf{x})]\label{eq:LvN for total rho_el}
\end{equation}
which gives the evolution of the ``true'' density operator of the
electronic subsystem --- that is, $\gamma_{\text{el}}$ of Eq. \ref{eq:total electronic density operator (pure state)}
--- in terms of local (i.e., conditional) contributions. This equation
is an \emph{exact} identity for globally pure states that can be generalized
to the statistical mixtures discussed below. It shows that out of
the two terms on the r.h.s. of Eq. \ref{eq:electronic equation final}
only the first one (the Liouvillian associated to $H_{\text{el}}$)
contributes directly to the change of the global electronic properties,
the second one (i.e. the electron-nuclear coupling $\delta H_{\text{en}}$)
affects the local properties (the $\rho_{\text{el}}$'s) that in turn
determine a change in $\gamma_{\text{el}}$. In a sense, Eq. \ref{eq:LvN for total rho_el}
proves that $H_{\text{el}}$ plays a primary role in the evolution
of the electronic properties. We should emphasize, though, that this
result is of limited help in general since most often the observables
of interest are electron-nuclear rather than purely electronic, and
thus require more detailed information than those contained in $\gamma_{\text{el}}$.
A notable example consists in the adiabatic\textbf{ }populations $p_{n}^{\text{ad}}$
(apparently, pure electronic observables) which are defined as
\[
p_{n}^{\text{ad}}=\int d\mathbf{x}n(\mathbf{x})\text{Tr}(\rho_{\text{el}}P_{n})
\]
where $P_{n}(\mathbf{x})$ is the eigenprojector on the $n^{\text{th}}$
eigenstate of $H_{\text{el}}$. These are, in fact, expectation values
of the following \emph{electron-nuclear} operator
\[
\hat{P}_{n}^{\text{ad}}=\int d\mathbf{x}\ket{\mathbf{x}}P_{n}(\mathbf{x})\bra{\mathbf{x}}
\]
and always require $n(\mathbf{x})$ and the conditional density operators
$\rho_{\text{el}}$, unless the relevant eigenspace is trivial and
$P_{n}(\mathbf{x})$ does not depend on $\mathbf{x}$. 

\subsection{Type-\emph{n} statistical mixtures\label{subsec:Type-n-statistical-mixtures}}

In Section \ref{subsec:Pure-states} we have provided an exact, \emph{gauge}-invariant
description of the dynamics using the exact factorization of the wavefunction
describing the globally pure state of the combined electronic-nuclear
system. We now consider extensions to this picture, starting from
the situation where the electron-nuclear system is in a simple, proper
statistical mixture of states where electrons can be yet described
by a \emph{local} pure electronic state. By this we mean that the
nuclear-coordinate representation of the \emph{total} density operator
changes according to \footnote{Please notice that the hat is used here to identify operators in the
electronic variables, the density is a \emph{matrix} in nuclear coordinates
and an \emph{operator} in the electronic ones.}
\begin{align*}
\hat{\rho}(\mathbf{x},\mathbf{x}') & =\psi(\mathbf{x})\psi^{*}(\mathbf{x}')\ket{u(\mathbf{x})}\bra{u(\mathbf{x}')}\\
 & \rightarrow\hat{\rho}(\mathbf{x},\mathbf{x}')=\sigma(\mathbf{x},\mathbf{x}')\ket{u(\mathbf{x})}\bra{u(\mathbf{x}')}
\end{align*}
where $\sigma(\mathbf{x},\mathbf{x}')$ --- the \emph{apparent} nuclear
density matrix --- is no longer constrained to have a pure state
form. We call this special mixture of states a type-\textbf{$n$ }statistical\textbf{
}mixture. 

In this case, the equations of motion for the nuclei and for the electrons
follow easily as a simple generalization of the equations given in
the Section \ref{subsec:Statistical-mixtures} and \ref{subsec:Pure-states}.
Specifically, electrons are governed by the same equations we have
seen above, namely Eq. \ref{eq:electronic equation final} for the
Lagrangian-frame dynamics of the conditional electronic density operators.
We only need to perform an ensemble average of the complex-valued
nuclear velocity or momentum fields appearing there, specifically
in the definition of the material derivative on the l.h.s. of Eq.
\ref{eq:electronic equation final} and in the \emph{e-n} correction
term on its r.h.s. (Eq. \ref{eq:electron-.nuclear correction to the Lioviallian}).
With nuclei described by the mixed-state apparent density matrix $\sigma(\mathbf{x},\mathbf{x}')$
the real component of the momentum field takes the form of a ratio
of mechanical momentum moments 
\[
\pi_{k}(\mathbf{q})=\frac{\mathcal{M}_{k}^{(1)}(\mathbf{q})}{\mathcal{M}^{(0)}(\mathbf{q})}
\]
since $\mathcal{J}^{i}(\mathbf{q})=\sum_{j}\xi^{ij}\mathcal{M}_{j}^{(1)}(\mathbf{q})$
is the correct expression for the nuclear current density. As for
the imaginary component we need a similar construction, that translates
into a logarithmic average $n^{\text{eff}}$ of the nuclear density
entering the electronic equation, namely
\[
\ln n^{\text{eff}}(\mathbf{q})=\sum_{\alpha}\omega_{\alpha}\ln\,n_{\alpha}(\mathbf{q})
\]
where $\omega_{\alpha}$ are statistical weights and $n_{\alpha}(\mathbf{q})=\psi_{\alpha}^{*}(\mathbf{q})\psi_{\alpha}(\mathbf{q})$
is the density of the $\alpha^{\text{th}}$ representative of the
ensemble. Here, we can replace $n^{\text{eff}}$ with $\mathcal{M}^{(0)}$,
which is logarithmically correct up to second order in the (relative)
density fluctuations in the ensemble
\[
n^{\text{eff}}\approx\mathcal{M}^{(0)}\exp\left(-\frac{\braket{\delta n^{2}}}{(\mathcal{M}^{(0)})^{2}}\right)
\]
to write the electron-nuclear correction term as in the pure state
case, 
\begin{align*}
\delta\bar{H}_{\text{en}} & =-\frac{\hbar^{2}}{2\mathcal{M}^{(0)}}\sum_{ij}\xi^{ij}\partial_{i}(\mathcal{M}^{(0)}\partial_{j}\rho_{\text{el}})
\end{align*}
with the ensemble average nuclear density $\mathcal{M}^{(0)}$ in
place of $\psi^{*}\psi$ and $\rho_{\text{el}}$ the usual conditional
density operator for the electrons.

As for the nuclei, the dynamical behavior of any realization of the
ensemble is governed by one and the same Hamiltonian $H^{\text{eff}}$,
with the \emph{gauge} fields coming from the evolution of the (local)
pure electronic state. As a result, the nuclear dynamics can be described
by a Liouville - von Neumann equation for the density matrix $\sigma(\mathbf{x},\mathbf{x}')$
of the kind already mentioned in Sec. \ref{subsec:Statistical-mixtures},
meaning that it can be given as a hierarchy of momentum moment equations
\[
\frac{\partial\mathcal{M}^{(0)}}{\partial t}(\mathbf{q})=-\sum_{ij}\xi^{ij}\partial_{i}\mathcal{M}_{j}^{(1)}(\mathbf{q})
\]
\begin{align*}
\frac{\partial\mathcal{M}_{k}^{(1)}}{\partial t}(\mathbf{q}) & =-\sum_{ij}\xi^{ij}\partial_{i}\mathcal{M}_{kj}^{(2)}(\mathbf{q})+F_{k}\mathcal{M}^{(0)}(\mathbf{q})
\end{align*}
\begin{align*}
\frac{\partial\mathcal{M}_{km}^{(2)}}{\partial t}(\mathbf{q}) & =-\sum_{ij}\xi^{ij}\partial_{i}\mathcal{M}_{kmj}^{(3)}(\mathbf{q})+\\
 & +F_{k}\mathcal{M}_{m}^{(1)}(\mathbf{q})+F_{m}\mathcal{M}_{k}^{(1)}(\mathbf{q})+\\
 & +\sum_{ij}\xi^{ij}\left(B_{ki}\mathcal{C}_{mj}^{(2)}(\mathbf{q})+B_{mi}\mathcal{C}_{kj}^{(2)}(\mathbf{q})\right)
\end{align*}
\[
...
\]
where again (in the approximation $n^{\text{eff}}\approx\mathcal{M}^{(0)}$)
the net force $F_{k}$ is given by Eq. \ref{eq:hydrodynamical force in EF}
and $\mathcal{C}_{kj}^{(2)}(\mathbf{q})$ was defined in Eq. \ref{eq:second moment correlation}. 

Hence, in a type-$n$ statistical mixture the electronic equation
of motion takes precisely the same form of the pure state case, Eq.
\ref{eq:electronic equation final}, and the hierarchy of equations
for the nuclear moments generalizes the pure state result of Eqs.
\ref{eq:moment nuclear equation 1} and \ref{eq:moment nuclear equation 2 first},
following the general scheme presented in Section \ref{subsec:Statistical-mixtures}.
We notice that in the latter set of equations cancellation of the
\emph{gauge} fields is incomplete beyond the first moment: coupling
to the magnetic field occurs through the first-moment correlations,
which are generally non-vanishing beyond the first moment.

\subsection{Type-\emph{e} statistical mixtures\label{subsec:Type-e-statistical-mixtures}}

We now consider the opposite limit of a statistical mixture where
the nuclei can be yet represented by a wavefunction $\psi(\mathbf{x})$
and the electrons are in an mixed state, meaning that even locally
in nuclear coordinates they are not constrained to have a pure state
form. In other words, in this case the total density operator takes
the form
\begin{equation}
\hat{\rho}(\mathbf{x},\mathbf{x}')=\psi(\mathbf{x})\psi^{*}(\mathbf{x}')\hat{\mathcal{W}}(\mathbf{x},\mathbf{x}')\label{eq:type-e mixture}
\end{equation}
where $\hat{\mathcal{W}}(\mathbf{x},\mathbf{x}')$ is the ensemble
average of the pure-state dyad $\hat{\mathcal{W}}_{\alpha}(\mathbf{x},\mathbf{x}')=\ket{u_{\alpha}(\mathbf{x})}\bra{u_{\alpha}(\mathbf{x}')}$,
\[
\hat{\mathcal{W}}(\mathbf{x},\mathbf{x}')=\sum_{\alpha}\omega_{\alpha}\hat{\mathcal{W}}_{\alpha}(\mathbf{x},\mathbf{x}')
\]
with $\omega_{\alpha}$ statistical weights, $\alpha=1,2,..n_{\alpha}$.
We call the special kind of mixture of Eq. \ref{eq:type-e mixture}
a type-\emph{e }statistical mixture, to emphasize that now the electrons
are in a intrinsically mixed state. 

In this situation the fact that a \emph{unique} nuclear wavefunction
can be singled out from the ensemble means  that the electronic states
$\ket{u_{\alpha}(\mathbf{x})}$'s have only an overall \emph{gauge}
degree of freedom, i.e. their phases cannot be changed independently
of each other. Rather, a \emph{single} set of \emph{gauge} potentials
exist, $A_{k}$ for $k=0,1,2,..$ (in keeping with the previous notation
we use $k=0$ for the time coordinate) and this defines completely
the dynamical problem for the nuclei and makes the theory yet relatively
simple. We notice that this does not imply that the $\ket{u_{\alpha}}$'s
define the \emph{same} Berry connection, only that the \emph{gauge}
potentials are just the ensemble average 
\begin{equation}
A_{k}=\sum_{\alpha}w_{\alpha}A_{k}^{\alpha}\equiv\sum_{\alpha}\omega_{\alpha}\left(i\braket{u_{\alpha}|\partial_{k}u_{\alpha}}\right)\label{eq:ensemble connection}
\end{equation}
in such a way that the same \emph{gauge} transformation applied to
the $\ket{u_{\alpha}}$'s, i.e. $\ket{u_{\alpha}}\rightarrow e^{-i\varphi}\ket{u_{\alpha}}$,
gives rise to the desired transformation law for the \emph{gauge}
potentials
\[
A_{k}\rightarrow A_{k}+\partial_{k}\varphi
\]
In turn, this also implies that both the \emph{pseudo}-magnetic and
\emph{pseudo}-electric fields are ensemble averages of the pure-state
expressions (see Eq. \ref{eq:E,B fields})
\begin{align}
B_{kj} & =\sum_{\alpha}\omega_{\alpha}\hbar(\partial_{k}A_{j}^{\alpha}-\partial_{j}A_{k}^{\alpha})\equiv\sum_{\alpha}\omega_{\alpha}B_{kj}^{\alpha}\nonumber \\
E_{k} & =\sum_{\alpha}\omega_{\alpha}\hbar(\partial_{k}A_{0}^{\alpha}-\partial_{t}A_{k}^{\alpha})\equiv\sum_{\alpha}\omega_{\alpha}E_{k}^{\alpha}\label{eq:E,B ensemble averages}
\end{align}
However, in order to completely define the nuclear problem, we shall
also need  an ensemble average for the metric tensor, $g_{ij}^{\text{av}}$.
 This requires separate considerations (see below, Eq. \ref{eq:average quantum geometric tensor})
since for mixed states the tensor cannot be represented in terms of
derivatives of the conditional electronic density operator. Once this
issue is settled, the nuclear dynamics follows straightforwardly from
the pure-state case, and can be subsumed in the \emph{gauge}-invariant
quantum hydrodynamical equations, e.g. continuity equation for the
nuclear density (Eq. \ref{eq:continuity equation Lagrange (general)})
and the momentum equation of Eq. \ref{eq:HEF trajectory equation II},
\begin{equation}
\frac{d\pi_{k}}{dt}=-\text{Tr}_{e}\left(\rho_{\text{el}}\partial_{k}H_{\text{el}}\right)-\frac{\hbar^{2}}{n}\sum_{ij}\xi^{ij}\partial_{i}(ng_{kj}^{\text{av}})+(F_{q})_{k}\label{eq:momentum equation type-e}
\end{equation}
with the quantum force $(F_{q})_{k}$, as usual, given by Eq. \ref{eq:quantum force}.
Thus, we are left with the problem of seeking a \emph{gauge} invariant
formulation of the electron dynamics that can account for the mixed
state character of the local electronic states. 

Formally, given a total density operator $\hat{\rho}(\mathbf{x},\mathbf{x}')$
\emph{and} a nuclear wavefunction $\psi(\mathbf{x})$ (hence, the
pure-state apparent nuclear density matrix $\sigma(\mathbf{x},\mathbf{x}')=\psi(\mathbf{x})\psi^{*}(\mathbf{x}')$)
we can define a ``doubly conditional electronic operator'' according
to
\[
\hat{\mathcal{W}}(\mathbf{x},\mathbf{x}')=\frac{\hat{\rho}(\mathbf{x},\mathbf{x}')}{\sigma(\mathbf{x},\mathbf{x}')}
\]
for any $\mathbf{x}$ and $\mathbf{x}'$ where the wavefunction does
not vanish. This operator determines the conditional density operator
for electrons through its diagonal elements
\[
\rho_{\text{el}}(\mathbf{q})=\hat{\mathcal{W}}(\mathbf{q},\mathbf{q})
\]
and, in general, determines the true nuclear density matrix through
its off diagonal elements
\[
\rho_{n}(\mathbf{x},\mathbf{x}')=\sigma(\mathbf{x},\mathbf{x}')\text{Tr}_{\text{el}}\hat{\mathcal{W}}(\mathbf{x},\mathbf{x}')
\]
These ``doubly conditional'' electronic operators $\hat{\mathcal{W}}(\mathbf{x},\mathbf{x}')$
are the key elements of the theory of type-$e$ statistical mixtures.
In the following we shall describe them by means of a set of moments,
which we call mechanical\textbf{ }electronic\textbf{ }moments, in
full analogy with the mechanical momentum moments of Sections \ref{subsec:Statistical-mixtures}
and \ref{subsec:Type-n-statistical-mixtures}. We report only the
main findings of these developments and leave the details of the calculations
in Appendix \ref{app:electronic momentum moments}. 

We start defining the ``mechanical'' electronic moments 
\[
\hat{\mathcal{M}}^{(0)}(\mathbf{q})=\hat{\mathcal{W}}(\mathbf{q},\mathbf{q})\equiv\rho_{\text{el}}(\mathbf{q})
\]
\[
\hat{\mathcal{M}}_{k}^{(1)}(\mathbf{q})=\frac{\mu_{k}+\mu_{k}^{'*}}{2}\hat{\mathcal{W}}(\mathbf{x},\mathbf{x}')\vert_{\mathbf{x}=\mathbf{x}'=\mathbf{q}}
\]
\[
\hat{\mathcal{M}}_{kj}^{(2)}(\mathbf{q})=\frac{\mu_{k}+\mu_{k}^{'*}}{2}\frac{\mu_{j}+\mu_{j}^{'*}}{2}\hat{\mathcal{W}}(\mathbf{x},\mathbf{x}')\vert_{\mathbf{x}=\mathbf{x}'=\mathbf{q}}
\]
\[
...
\]
where now $\mu_{k}=-i\hbar\partial_{k}+\hbar A_{k}\equiv-i\hbar D_{k}$
and $D_{k}=\partial_{k}+iA_{k}$ is the \emph{gauge} covariant derivative.
By definition, these moments are \emph{gauge} invariant with respect
to the \emph{gauge} transformations mentioned above (note that sign
change in the \emph{gauge} potentials when compared to the mechanical
momenta of the nuclei, i.e., $\mu_{k}\equiv-\pi_{k}^{*}$ ). For a
pure state, for instance, we find 
\begin{equation}
\hat{\mathcal{M}}^{(0)}(\mathbf{q})=\ket{u(\mathbf{q})}\bra{u(\mathbf{q})}=\rho_{\text{el}}(\mathbf{q})\label{eq:M_0 (pure-state)}
\end{equation}
\begin{equation}
\hat{\mathcal{M}}_{k}^{(1)}(\mathbf{q})=-\frac{i\hbar}{2}\left(\ket{D_{k}u(\mathbf{q})}\bra{u(\mathbf{q})}-\ket{u(\mathbf{q})}\bra{D_{k}u(\mathbf{q})}\right)\label{eq:;M_1 (pure-sate)}
\end{equation}
\begin{align}
\hat{\mathcal{M}}_{kj}^{(2)}(\mathbf{q}) & =\frac{\hbar^{2}}{4}\left(\ket{D_{k}u(\mathbf{q})}\bra{D_{j}u(\mathbf{q})}+\ket{D_{j}u(\mathbf{q})}\bra{D_{k}u(\mathbf{q})}\right)+\nonumber \\
 & -\frac{\hbar^{2}}{4}\left(\ket{D_{k}D_{j}u(\mathbf{q})}\bra{u(\mathbf{q})}+\ket{u(\mathbf{q})}\bra{D_{k}D_{j}u(\mathbf{q})}\right)\label{eq:M_2 (pure-state)}
\end{align}

The equations of motion for these moments can be derived from an analysis
of the pure-state case, provided the results are cast in a form that
survives the ensemble average (see Appendix \ref{app:electronic momentum moments}).
The appropriate form of the electronic equation for the state vectors
involves only\emph{ }the\emph{ gauge} covariant derivatives $D_{k}$'s,
since these are fixed in our setting by the presence of a single set
of \emph{gauge} potentials $A_{k}$. This is Eq. \ref{eq:electronic equation of motion}
in the form
\begin{align}
i\hbar\ket{D_{0}u} & =(H_{\text{el}}-\mathcal{V})\ket{u}-i\hbar\sum_{j}V^{j}\ket{D_{j}u}+\nonumber \\
 & -\frac{\hbar^{2}}{2}\sum_{ij}\xi^{ij}\ket{D_{i}D_{j}u}\label{eq:electronic equation (gauge covariant derivatives)}
\end{align}
where we have set $D_{0}\equiv\partial_{t}+iA_{0}$ and removed the
projector $Q=1-P$, $P=\ket{u}\bra{u}$, which is annoying for the
ensemble average. This could be done at the price of adding the following
term in the space projected by $P$ (``$P$-space'') 
\[
\mathcal{V}=E_{\text{el}}+\frac{\hbar^{2}}{2}\sum_{ij}\xi^{ij}g_{ij}\equiv\braket{u|H_{\text{el}}|u}+\frac{\hbar^{2}}{2}\sum_{ij}\xi^{ij}\braket{D_{i}u|D_{j}u}
\]
which is precisely the potential energy for the nuclear motion (noting
that $\braket{D_{i}u|D_{j}u}\equiv\braket{\partial_{i}u|Q|\partial_{j}u}$),
and which guarantees $\braket{u|D_{0}u}\equiv0$. Indeed, we have
\[
i\hbar\braket{u|D_{0}u}=(E_{\text{el}}-\mathcal{V})-\frac{\hbar^{2}}{2}\sum_{ij}\xi^{ij}\braket{u|D_{i}D_{j}u}\equiv0
\]
since $\braket{u|D_{j}u}=0$ implies $\braket{u|D_{i}D_{j}u}\equiv-\braket{D_{i}u|D_{j}u}$.
When extending the above equation to a statistical mixture of the
type considered here, i.e. with the replacement $\ket{u}\rightarrow\ket{u_{\alpha}}$
for $\alpha=1,2,..n_{\alpha}$ in Eq. \ref{eq:electronic equation (gauge covariant derivatives)},
orthogonality no longer holds for each representative 
\[
\braket{u_{\alpha}|D_{0}u_{\alpha}}=i(A_{0}-A_{0}^{\alpha})\ \ \ \ A_{0}^{\alpha}:=i\braket{u_{\alpha}|\partial_{t}u_{\alpha}}
\]
but holds \emph{on average }if we set 
\begin{equation}
\mathcal{V}=\sum_{\alpha}\omega_{\alpha}(\,\braket{u_{\alpha}|H_{\text{el}}|u_{\alpha}}+\frac{\hbar^{2}}{2}\sum_{ij}\xi^{ij}\braket{D_{i}u_{\alpha}|D_{j}u_{\alpha}}\,)\label{eq:effective V}
\end{equation}
since this implies
\begin{align*}
\hbar\overline{(A_{0}^{\alpha}-A_{0})} & =\hbar\sum_{j}V^{j}\overline{(A_{j}-A_{j}^{\alpha})}+\\
 & -i\frac{\hbar^{2}}{2}\sum_{ij}\xi^{ij}\partial_{i}\overline{(A_{j}-A_{j}^{\alpha})}\equiv0
\end{align*}
Here, we have used the overbar to denote the ensemble average and
exploited the definition of the ensemble connection, Eq. \ref{eq:ensemble connection}. 

It is worth stressing at this point that the electronic equation of
motion for the ensemble realizations (Eq. \ref{eq:electronic equation (gauge covariant derivatives)}
with $\ket{u_{\alpha}}$ in place of $\ket{u}$) remains \emph{gauge}
covariant, even if the $A_{k}$'s are not defined by any Berry connection.
Indeed, for that property to hold one only needs that the potentials
transform as $A_{k}\rightarrow A_{k}+\partial_{k}\varphi$ under the
\emph{gauge} transformation $\ket{u_{\alpha}}\rightarrow e^{-i\varphi}\ket{u_{\alpha}}$.
This is the essence of the \emph{gauge} covariant derivative and it
is the key for Eq. \ref{eq:electronic equation (gauge covariant derivatives)}
to work properly for the members of our ensemble \footnote{What is specific of the Berry connection is that it guarantees the
orthogonality condition mentioned above.}. 

Incidentally, the above arguments also show that the metric tensor
entering the nuclear potential $\mathcal{V}$ (Eq. \ref{eq:effective V})
is $g_{ij}^{\text{av}}\equiv\Re q_{ij}^{\text{av}}$ where
\begin{equation}
q_{ij}^{\text{av}}=\overline{\braket{D_{i}u_{\alpha}|D_{j}u_{\alpha}}}\label{eq:average quantum geometric tensor}
\end{equation}
is the appropriate expression for the averaged quantum geometric tensor.
This is consistent with the expression for the average \emph{gauge}
fields given in Eq. \ref{eq:E,B ensemble averages} and formally settles
any issue concerning the nuclear dynamics. 

Now, as shown in Appendix \ref{app:electronic momentum moments},
when using Eq. \ref{eq:electronic equation (gauge covariant derivatives)}
as effective electronic equation for our type-\emph{e} ensemble, the
zero-th moment equation takes the form 
\begin{align}
\frac{d\hat{\mathcal{M}}^{(0)}}{dt}(\mathbf{q}) & =-\frac{i}{\hbar}[H_{\text{el}}(\mathbf{q}),\hat{\mathcal{M}}^{(0)}(\mathbf{q})]+\nonumber \\
 & -\frac{1}{n(\mathbf{q})}\sum_{ij}\xi^{ij}\partial_{i}\left(n(\mathbf{q})\hat{\mathcal{M}}_{j}^{(1)}(\mathbf{q})\right)\label{eq:zero-moment electronic equation}
\end{align}
where $d/dt$ denotes the material derivative and $n(\mathbf{q})=|\psi(\mathbf{x})|^{2}$
is the usual nuclear density. This is actually one of the several
possible ways of presenting the equation of motion for the conditional
density of a pure state (see Appendix \ref{app:gauge invariant formulation of electron dynamics}),
the one appropriate for the kind of ensemble averages we are interested
here. The key difference between a type-\emph{e} mixed state and a
pure state is that only for the latter the first order moment appearing
on the r.h.s. can be expressed in terms of the zero-th order one and
its spatial derivative, 
\[
\hat{\mathcal{M}}_{j}^{(1)}=i\frac{\hbar}{2}[\hat{\mathcal{M}}^{(0)},\partial_{j}\hat{\mathcal{M}}^{(0)}]\ \ \ \text{(pure state)}
\]
Eq. \ref{eq:zero-moment electronic equation} is a Liouville-von Neumann-like
equation with a surface term (second term on the r.h.s.) that resembles
the flux term of the continuity equation. Upon multiplying by $n(\mathbf{q})$
and integrating over all nuclear configuration space we obtain
\begin{equation}
\frac{\partial\gamma_{\text{el}}}{\partial t}=-\frac{i}{\hbar}\int d\mathbf{q}\,n(\mathbf{q})\,[H_{\text{el}}(\mathbf{q}),\hat{\mathcal{M}}^{(0)}(\mathbf{q})]\label{eq:mxed-state LvN for total rho_el}
\end{equation}
where $\gamma_{\text{el}}$ is the ``true'' density operator of
the electronic subsystem, appropriate to the present statistical mixture
(see Eq. \ref{eq:total electronic density operator (pure state)}),
\begin{equation}
\gamma_{\text{el}}=\int d\mathbf{q}\,n(\mathbf{q})\,\hat{\mathcal{M}}^{(0)}(\mathbf{q})\label{eq:mixed-state expansion for total rho_el}
\end{equation}
This equation generalizes Eq. \ref{eq:LvN for total rho_el} found
for pure states, and is subjected to the same caveats: most often
one needs the electronic moment $\hat{\mathcal{M}}^{(0)}(\mathbf{q})$
(and the nuclear density $n(\mathbf{q})$) for computing the observables
of interest, hence the global electronic density operator $\gamma_{\text{el}}$
is of little help.

Now, the main difference between the type-$e$ statistical mixture
addressed here and the pure-state case of Sec. \ref{subsec:Pure-states}
is that the evolution of the zero-th electronic moment requires the
first moment (last term on the r.h.s. of Eq. \ref{eq:zero-moment electronic equation}),
which in turn involves higher order moments. In other words, the Liouville
von Neumann like equation of Eq. \ref{eq:zero-moment electronic equation}
no longer suffices to describe the electron dynamics, and an infinite
hierarchy of equations is necessary. For instance, the dynamical equation
of the first order moments reads as 
\begin{align}
\frac{d\hat{\mathcal{M}}_{k}^{(1)}}{dt} & =-\frac{i}{\hbar}[H_{\text{el}},\hat{\mathcal{M}}_{k}^{(1)}]-\frac{1}{2}\left[\partial_{k}H_{\text{el}}-\overline{\partial_{k}H_{\text{el}}},\mathcal{M}^{(0)}\right]_{+}+\nonumber \\
 & +\frac{\hbar^{2}}{n}\sum_{ij}\xi^{ij}\partial_{i}(ng_{kj}^{\text{av}})\hat{\mathcal{M}}^{(0)}+\nonumber \\
 & -\sum_{ij}\xi^{ij}(\partial_{i}\ln\,n)\mathcal{M}_{kj}^{(2)}\label{eq:first-moment electronic equation}\\
 & -\sum_{ij}\xi^{ij}\partial_{i}\hat{\mathcal{M}}_{kj}^{(2)}-\sum_{ij}\xi^{ij}B_{ki}\hat{\mathcal{M}}_{j}^{(1)}+\nonumber \\
 & +\frac{i\hbar}{2}\sum_{ij}\xi^{ij}(\partial_{k}\partial_{i}\ln\,n)\hat{\mathcal{M}}_{j}^{(1)}-\sum_{j}(\partial_{k}v^{j})\hat{\mathcal{M}}_{j}^{(1)}\nonumber 
\end{align}
The evolution of these moments requires the second-order moments in
a flux-like term (fourth line), even if the \emph{e-n} coupling term
(third line) is neglected. This is the general way the couplings to
higher order moments occurs. In fact, higher order moment equations
become increasingly complicated, but they appear unlikely to play
a key role in practical applications. 

Before closing this Section we notice that the electronic moments
have well defined traces,
\begin{equation}
\text{Tr}_{\text{el}}\hat{\mathcal{M}}^{(0)}=1\ \ \ \text{Tr}\mathcal{\hat{M}}_{k}^{(1)}=0\ \ \ \text{Tr}\mathcal{\hat{M}}_{kj}^{(2)}=\hbar^{2}g_{kj}^{\text{av}}\label{eq:trace of electronic moments}
\end{equation}
as can be checked with a direct calculation on the pure-state expressions
(Eqs. \ref{eq:M_0 (pure-state)}, \ref{eq:;M_1 (pure-sate)} and \ref{eq:M_2 (pure-state)}),
followed by an ensemble average. The zero-th moment equation, Eq.
\ref{eq:zero-moment electronic equation}, is clearly trace-conserving
since both the commutator and the first moment on the r.h.s. are traceless.
As for the first moment equation, on the other hand, the trace of
the r.h.s. of Eq. \ref{eq:first-moment electronic equation} reduces
to 
\begin{align*}
\frac{\hbar^{2}}{n}\sum_{ij}\xi^{ij}\partial_{i}(ng_{kj}^{\text{av}})-\hbar^{2}\sum_{ij}\xi^{ij}(\partial_{i}\ln\,n)g_{kj}^{\text{av}}+\\
-\hbar^{2}\sum_{ij}\xi^{ij}\partial_{i}g_{kj}^{\text{av}} & \equiv0
\end{align*}
thereby proving that this equation is trace-conserving too. 

As a final remark we notice that the type-\emph{e} statistical mixture
considered in this Section requires the instantaneous \emph{pseudo}-magnetic
field $B_{kj}=-2\hbar\Im q_{kj}^{\text{av}}$, as is evident from
the moment equation of first order given in Eq. \ref{eq:first-moment electronic equation}.
This cannot be obtained from Eq. \ref{eq:quantum geometric tensor (pure state)}
since the latter only holds for pure states, and cannot be generalized
to statistical mixtures. The situation differs from that of the metric
$g_{kj}^{\text{av}}$, which is directly provided by the second moments
through their trace (Eq. \ref{eq:trace of electronic moments}). The
required equation of motion for $B_{kj}$ can be obtained from Maxwell-Faraday
induction law (Eq. \ref{eq:pseudo Maxwell laws}) and is conveniently
re-expressed in terms of the forces acting on the nuclei since these
are in any case needed for the particle dynamics. It reads as
\begin{equation}
-\frac{\partial B_{kj}}{\partial t}=\partial_{k}F_{j}-\partial_{j}F_{k}+\sum_{m}\partial_{j}(v^{m}B_{km})-\partial_{k}(v^{m}B_{jm})\label{eq:vortex equation}
\end{equation}
 where $F_{k}$ can either be the classical force 
\[
F_{k}=-\text{Tr}_{e}\left(\rho_{\text{el}}\partial_{k}H_{\text{el}}\right)-\frac{\hbar^{2}}{n}\sum_{ij}\xi^{ij}\partial_{i}(ng_{kj}^{\text{av}})
\]
or the total force of Eq. \ref{eq:momentum equation type-e} (the
quantum component is longitudinal and does not contribute to the curl
appearing in Eq. \ref{eq:vortex equation}) . Eq. \ref{eq:vortex equation}
follows from Eq. \ref{eq:total classical force} upon singling out
the electric field components and performing the ensemble average.
It plays the role of a vortex equation for the dynamics of the quantum
probability fluid. 

\subsection{Remark on mechanical momentum moments\label{subsec:Remark-on-mechanical}}

The momentum moments introduced in the previous Sections replace in
a sense the ordinary moments and, as seen above, provide a \emph{gauge}
invariant description of the dynamics. However, they are not (separately)
fully equivalent to the ordinary moments: the latter allow, at least
in principle, to retrieve a density matrix, while this is not obviously
the case for the mechanical moments which, by definition, require
the \emph{gauge} potentials. 

The point, however, is that, after all, we are always interested in
the \emph{full} density matrix 
\[
\hat{\rho}(\mathbf{x},\mathbf{x}')=\sigma(\mathbf{x},\mathbf{x}')\hat{\mathcal{W}}(\mathbf{x},\mathbf{x}')
\]
The issue of the \emph{gauge} invariance only arises when looking
separately at the electronic and the nuclear problems, the total density
matrix above is in fact globally \emph{gauge} invariant by construction.
The mechanical moments allow one to split the problem into \emph{gauge}
invariant sub-problems. This becomes clear when focusing on the \emph{ordinary}
moments of this matrix, for instance
\[
\hat{\mathcal{Q}}_{k}^{(1)}(\mathbf{q})=-i\hbar\frac{\partial_{k}-\partial'_{k}}{2}\sigma(\mathbf{x},\mathbf{x}')\hat{\mathcal{W}}(\mathbf{x},\mathbf{x}')\vert_{\mathbf{x}=\mathbf{x}'=\mathbf{q}}
\]
Here, 
\begin{align*}
-ih\partial_{k}\left(\sigma(\mathbf{x},\mathbf{x}')\hat{\mathcal{W}}(\mathbf{x},\mathbf{x}')\right) & =\left(\pi_{k}\sigma(\mathbf{x},\mathbf{x}')\right)\hat{\mathcal{W}}(\mathbf{x},\mathbf{x}')+\\
 & +\sigma(\mathbf{x},\mathbf{x}')\left(\mu_{k}\hat{\mathcal{W}}(\mathbf{x},\mathbf{x}')\right)
\end{align*}
where $\pi_{k}=-i\hbar\partial_{k}-\hbar A_{k}$ and $\mu_{k}=-i\hbar\partial_{k}+\hbar A_{k}$
as defined above, hence 
\[
\hat{\mathcal{Q}}_{k}^{(1)}(\mathbf{q})=\mathcal{M}_{k}^{(1)}(\mathbf{q})\hat{\mathcal{M}}^{(0)}(\mathbf{q})+\mathcal{M}^{(0)}(\mathbf{q})\hat{\mathcal{M}}_{k}^{(1)}(\mathbf{q})
\]
This equation shows that, even though the nuclear (electronic) mechanical
moments are not sufficient to obtain the nuclear (electronic) density
matrix, the set of mechanical moments for both electrons and nuclei
\emph{does} provide a complete description of the system, i.e., the
full density operator. 

The ``weakness'' of the mechanical moments is only apparent, it
is actually a necessity: the marginal density matrix for the nuclei
and the conditional density matrix for the electrons --- as inspired
by the exact factorization of the total wavefunction -- are \emph{gauge}
dependent, hence require that a \emph{gauge} choice is made. 

\section{Applications\label{sec:Applications}}

\subsection{Mixed quantum-classical dynamics\label{subsec:Mixed-quantum-classical-dynamics}}

In this Section we make a connection between the exact electron-nuclear
quantum dynamics formulated for a pure state in Section \ref{subsec:Pure-states}
and commonly adopted mixed quantum-classical schemes in which nuclei
are treated as classical particles and electrons evolve quantum dynamically.
The purpose of this Section is to bridge the exact quantum theory
to practical approaches to the problem and, eventually, to identify
the key corrections to the latter that can bring results in better
agreement with the exact one.

We start discussing a little point that concerns the dynamical interpretation
behind Eq. \ref{eq:electronic equation of motion II} and the original
equation arising from the exact factorization of the electron-nuclear
wavefunction, namely Eq. \ref{eq:EF electronic equation} or the equivalent
\begin{align}
i\hbar Q\partial_{t}\ket{u(\mathbf{x},t)} & =QH_{\text{el}}(\mathbf{x})\ket{u(\mathbf{x},t)}-i\hbar\sum_{j}v^{j}Q\partial_{j}\ket{u(\mathbf{x},t)}\nonumber \\
 & +\hbar\sum_{j}u^{j}Q\partial_{j}\ket{u(\mathbf{x},t)}-\hbar R\ket{u(\mathbf{x},t)}\label{eq:electronic equation of motion (Euler)}
\end{align}
Eq. \ref{eq:electronic equation of motion (Euler)} is formulated
in the \emph{Eulerian frame}: for each $\mathbf{x}$ in the configuration
space of the system there exists an equation of motion of the kind
of Eq. \ref{eq:electronic equation of motion (Euler)}, and the local
electronic state fixed at $\mathbf{x}$ is seen to evolve under the
action of \emph{one and the same} electronic Hamiltonian $H_{\text{el}}(\mathbf{x})$
at any time, besides its coupling to the nuclear motion \footnote{Of course, a further dependence on $(\mathbf{x},t)$ arises on $Q,v^{j},u^{j}$
and $R$ but we omit it for clarity.}. Eq. \ref{eq:electronic equation of motion II}, on the other hand,
is formulated in the \emph{Lagrangian frame}, as appropriate for instance
for a ``direct'' dynamical approach: for each evolving representative
$\mathbf{x}_{t}$ there exists an electronic equation of the kind
of Eq. \ref{eq:electronic equation of motion II} and the electronic
state evolves under the action of the \emph{instantaneous} Hamiltonian
$H_{\text{el}}(\mathbf{x}_{t})$ along the particle trajectory describing
nuclear motion in the QHD setting, 
\begin{align}
i\hbar Q\frac{d}{dt}\ket{u(\mathbf{x}_{t})} & =QH_{\text{el}}(\mathbf{x}_{t})\ket{u(\mathbf{x}_{t})}+\nonumber \\
 & +\hbar\sum_{j}u^{j}Q\partial_{j}\ket{u(\mathbf{x}_{t})}-\hbar R\ket{u(\mathbf{x}_{t})}\label{eq:electronic equation of motion (Lagrange)}
\end{align}

Albeit equivalent in an exact framework, the two equations, Eqs. \ref{eq:electronic equation of motion (Euler)}
and \ref{eq:electronic equation of motion (Lagrange)}, represent
distinct starting points when introducing approximations. In the Lagrangian
picture the coupling to the nuclear motion appears \emph{reduced},
since the effect of the (real) velocity fields has been absorbed in
the total time derivative, with clear advantages in the \emph{diabatic}
(fast moving nuclei) limit where the electronic state has hardly time
to adjust to the ``evolving'' Hamiltonian. On the other hand, the
advection term becomes a complicating factor in the \emph{adiabatic}
limit where the electrons respond rapidly to the changing Hamiltonian,
on the time scale appropriate for a (slow) nuclear motion. 

This latter, adiabatic limit deserves some considerations since it
is rather instructive. It is easily formulated in the Eulerian frame,
where the adiabatic condition 
\begin{equation}
i\hbar Q\partial_{t}\ket{u(\mathbf{x},t)}\approx0\label{eq:adiabatic condition (Euler)}
\end{equation}
is a stationary condition for the local electronic states that, we
remind, are fixed in space in this picture. In the Lagrangian frame,
on the other hand, it reads as
\begin{equation}
i\hbar Q\frac{d}{dt}\ket{u(\mathbf{x}_{t})}\approx i\hbar\sum_{j}v^{j}Q\partial_{j}\ket{u}\label{eq:adiabatic condition (Lagrange)}
\end{equation}
and it is a bit less intuitive. It becomes clear when the adiabatic
dynamics is driven by classical parameters $x^{k}$, with velocity
$\dot{x}^{k}\equiv v^{k}$, since in that case the appropriate (variational)
equation of motion would be (see Eq. 4 in Ref. \citep{Martinazzo2022a})
\begin{equation}
i\hbar\frac{d}{dt}\ket{u}=EP\ket{u}+i\hbar\dot{P}\ket{u}\label{eq:adiabatic dynamics with classical parameters}
\end{equation}
where $P$ is the eigenprojector of interest, $E$ the corresponding
eigenvalue and $\dot{P}$ its directional derivative $\dot{P}=\sum_{j}v^{j}\partial_{j}P$.
Indeed, Eq. \ref{eq:adiabatic condition (Lagrange)} follows from
Eq. \ref{eq:adiabatic dynamics with classical parameters} upon projecting
on $Q$ space
\[
Q\frac{d}{dt}\ket{u}=\sum_{j}v^{j}Q\partial_{j}\left(\ket{u}\bra{u}\right)\,\ket{u}\equiv\sum_{j}v^{j}Q\partial_{j}\ket{u}
\]
and noticing that the dynamics in the $P$-space is just a \emph{gauge}
choice. Clearly, if the dynamics is close to be adiabatic the Euler
form is preferred and the stationary condition is conveniently turned
into a standard eigenvalue problem --- i.e., $QH_{\text{el}}\ket{u}=0$
--- in the limit where $v^{k}\rightarrow0$ and the \emph{e-n} coupling
terms on the r.h.s of Eq. \ref{eq:electronic equation of motion (Euler)}
can be neglected. Far from this dynamical limit, on the other hand,
the Lagrangian form has distinct advantages and represents the preferred
choice for describing the electron dynamics. As a matter of fact,
the \emph{gauge} invariant form of the electronic equation appears
natural in the Lagrangian frame (Eq \ref{eq:electronic equation final}),
in the Eulerian frame it would contain an ``odd'' advection term. 

Next, we introduce a classical approximation for the nuclei by naively
setting $\partial_{i}n\approx0$ in the equations of motions. This
turns the momentum equation (Eq. \ref{eq:HEF trajectory equation II})
into
\begin{equation}
\frac{d\pi_{k}}{dt}\approx-\braket{u|\partial_{k}H_{\text{el}}|u}-\hbar^{2}\sum_{ij}\xi^{ij}\partial_{i}g_{kj}\label{eq:force in classical limit}
\end{equation}
and amounts to neglect the quantum force of Eq. \ref{eq:quantum force}
\emph{and} to simplify the dynamically corrected electric force, Eq.
\ref{eq:dynamically corrected E force}, by neglecting a term
\[
-\frac{\hbar^{2}}{2}\sum_{ij}\xi^{ij}g_{kj}\partial_{i}\text{ln}\,n
\]
at the expense of breaking a global property of the corrected electric
force, i.e. the vanishing of its average over the nuclear quantum
state. 

Eq. \ref{eq:force in classical limit} immediately leads to classical
Born-Oppenheimer dynamics if the geometric term is neglected, since
$\braket{u|\partial_{k}H_{\text{el}}|u}=\partial_{k}\braket{u|H_{\text{el}}|u}$
when $\ket{u}$ is a (normalized) eigenstate of $H_{\text{el}}$.
However, in order to recover the full adiabatic limit once needs to
eliminate the electronic response from the above expression and enforce
the adiabatic condition with the vanishing of the dynamical response
$F_{k}^{\text{ED}}$ (third term in Eq. \ref{eq:total classical force})
\emph{before} setting $\ket{u}$ to be an eigenstate of $H_{\text{el}}$.
This gives the well known result 
\begin{equation}
\frac{d\pi_{k}}{dt}\approx F_{k}=-\partial_{k}\mathcal{V}+\sum_{j}v^{j}B_{kj}\label{eq:total classical force in the adiabatic approximation}
\end{equation}
where $\mathcal{V}=\braket{u|H_{\text{el}}|u}+\phi_{\text{FS}}$ is
the Born-Oppenheimer potential energy surface with the diagonal correction,
and the second term is the Berry force. This can be directly checked
with a calculation that starts from Eq. \ref{eq:force in classical limit},
upon noticing that 
\[
\braket{u|\partial_{k}H_{\text{el}}|u}=\partial_{k}\braket{u|H_{\text{el}}|u}-2\Re\braket{\partial_{k}u|QH_{\text{el}}|u}
\]
and 
\[
QH_{\text{el}}\ket{u}\approx i\hbar\sum_{j}v^{j}Q\partial_{j}\ket{u}+\frac{\hbar^{2}}{2}\sum_{ij}\xi^{ij}QD_{i}D_{j}\ket{u}
\]
hold in the classical, adiabatic limit, along with the geometric identity
\citep{Martinazzo2022a}
\begin{equation}
\Re\braket{\partial_{k}u|QD_{i}D_{j}u}=\frac{1}{2}(\partial_{j}g_{ik}+\partial_{i}g_{kj}-\partial_{k}g_{ij})\label{eq:Levi-Civita connection}
\end{equation}
These considerations will turn out to be useful in the next Section
where we address a \emph{quasi}-adiabatic regime. 

Beyond this adiabatic limit, the electron dynamics needs to be explicitly
taken into account and, in the classical approximation considered
here, it follows the equation 
\begin{equation}
i\hbar Q\frac{d}{dt}\ket{u(\mathbf{x}_{t})}\approx QH_{\text{el}}(\mathbf{x}_{t})\ket{u(\mathbf{x}_{t})}-\hbar R\ket{u(\mathbf{x}_{t})}\label{eq:ED in classical limit}
\end{equation}
here given in the traditional wavefunction picture. This equation
results from the condition $\partial_{i}n\approx0$ and the second
term on its r.h.s. is consistent with the geometric contribution appearing
in Eq. \ref{eq:force in classical limit}, meaning that the second
term on the r.h.s. of Eq. \ref{eq:force in classical limit} is precisely
the dynamically corrected electric force of Eq. \ref{eq:dynamically corrected E force}
at this level of approximation,
\[
\mathcal{F}^{\text{el}}\approx-\partial_{k}\phi_{FS}-2\hbar\Re\braket{\partial_{k}u|R|u}\equiv-\hbar^{2}\sum_{ij}\xi^{ij}\partial_{i}g_{kj}
\]
where we have exploited once again the geometric identity of Eq. \ref{eq:Levi-Civita connection}.
These geometrical contributions have nothing to do with the quantum
nature of the nuclear dynamics and thus survive the classical limit.
These are neglected in the popular Ehrenfest method,\textbf{ }whose
equations of motion
\begin{align}
\frac{d\pi_{k}}{dt} & \approx-\braket{u|\partial_{k}H_{\text{el}}|u}\label{eq:ehrenfest method mom}\\
i\hbar Q\frac{d}{dt}\ket{u(\mathbf{x}_{t})} & \approx QH_{\text{el}}(\mathbf{x}_{t})\ket{u(\mathbf{x}_{t})}\label{eq:ehrenfest method el}
\end{align}
result from Eqs. \ref{eq:force in classical limit} and \ref{eq:ED in classical limit},
upon disregarding the geometric terms. Eqs. \ref{eq:ehrenfest method mom}
and \ref{eq:ehrenfest method el} are \emph{precisely} the equations
of motion of the ``semiclassical Ehrenfest method'', here obtained
from \emph{first-principles} without any mean-field approximation
and \emph{ad-hoc} correction for energy conservation. The widely used
form of the Ehrenfest equations of motion is indeed easily recovered
from Eqs. \ref{eq:ehrenfest method mom} and \ref{eq:ehrenfest method el}
by fixing the \emph{gauge} of the electronic equation according to
$\hbar A_{0}=E_{\text{el}}$ and presenting the equations in an adiabatic
(time-independent) basis of eigenstates of the electronic Hamiltonian,
$\{\ket{u_{n}}\}$. They read as 
\begin{equation}
\dot{\pi}_{k}=-\sum_{n}\rho_{nn}\partial_{k}E_{n}+\sum_{n,m}^{n\neq m}\rho_{nm}(E_{m}-E_{n})d_{mn}\label{eq:Ehrenfest momentum - adiabatic}
\end{equation}
\begin{align}
\dot{\rho}_{nn} & =-2\mathbf{v}\sum_{k}\Re((\mathbf{d}_{nk}\rho_{kn})\label{eq:eq:Ehrenfest el diagonal - adiabatic}\\
\dot{\rho}_{nm} & =-\frac{i}{\hbar}(E_{n}-E_{m})\rho_{nm}+\nonumber \\
 & -\mathbf{v}\sum_{k}(\mathbf{d}_{nk}\rho_{km}+\rho_{nk}(\mathbf{d})_{mk}^{*})\ \ \ (n\neq m)\label{eq:eq:Ehrenfest el off-diagonal - adiabatic}
\end{align}
where the $\mathbf{d}_{nm}$'s are non-adiabatic coupling vectors,
$\mathbf{d}_{nm}=\braket{u_{n}|\boldsymbol{\nabla}u_{m}}$, $\boldsymbol{\nabla}=(\partial_{1},\partial_{2},..)$
and $\mathbf{v}=(v^{1},v^{2},..)$ are, respectively, the gradient
and nuclear velocity vectors in the nuclear configuration space, and
the $\rho_{nm}$'s are defined by $\rho_{nm}=\braket{u_{n}|\rho_{\text{el}}|u_{m}}$
and are known as (adiabatic) populations for $n=m$ and coherences
for $n\neq m$ (cf.. Eqs. 73, 74 and 78 in Ref. \citep{Shu2023}).
To arrive at Eqs. \ref{eq:Ehrenfest momentum - adiabatic}, \ref{eq:eq:Ehrenfest el diagonal - adiabatic}
and \ref{eq:eq:Ehrenfest el off-diagonal - adiabatic} we expressed
the off-diagonal matrix elements of the microscopic force in terms
of the derivative couplings through 
\[
\braket{u_{n}|\partial_{k}H_{\text{el}}|u_{m}}=(E_{m}-E_{n})\braket{u_{n}|\partial_{k}u_{m}}\,\,(n\neq m)
\]
and we exploited the identity
\begin{align}
\braket{u_{n}|\dot{\rho}|u_{m}} & =\dot{\rho}_{nm}-\braket{\dot{u}_{n}|\rho|u_{m}}-\braket{u_{n}|\rho|\dot{u}_{m}}\nonumber \\
 & \equiv\dot{\rho}_{nm}+\mathbf{v}\sum_{k}(\mathbf{d}_{nk}\rho_{km}-\rho_{nk}\mathbf{d}_{km})\label{eq:deivative in adiabatic basis}
\end{align}
which follows from $\ket{\dot{u}_{n}}\equiv\mathbf{v}\boldsymbol{\nabla}\ket{u_{n}}$
and $\mathbf{d}_{nk}=-\mathbf{d}_{kn}^{*}$.

Clearly, a first correction to the Ehrenfest dynamics would be the
inclusion of the geometric effects in both the nuclear and electron
dynamics, what we can call ``geometric Ehrenfest'' approximation.
Note that we refer here to effects that are of \emph{pseudo}-electric
(rather than \emph{pseudo}-magnetic) nature: Berrry-like forces are
absent when the electron dynamics is taken into account. 

However, it turns out that a main weakness of the quantum-classical
approach is \emph{not} in the classical approximation to the nuclear
dynamics but rather in the classical way the electron-nuclear coupling
is handled. By this we mean that one can perform a classical limit
for the nuclear dynamics by setting to zero the quantum force while
keeping the density dependence (which occurs through the imaginary
part of the nuclear momentum fields, $w_{k}=-\hbar/2\partial_{k}\ln n$)
in the rest. At this stage, if we neglect the geometric contributions
($\partial_{k}g_{ij}\approx0$) we obtain a pseudo-classical dynamics
governed by
\begin{equation}
\frac{d\pi_{k}}{dt}\approx-\braket{u|\partial_{k}H_{\text{el}}|u}+2\hbar\sum_{j}u^{j}\,g_{kj}\label{eq:force in CCMQCD}
\end{equation}
\begin{equation}
i\hbar Q\frac{d}{dt}\ket{u(\mathbf{x}_{t})}\approx QH_{\text{el}}(\mathbf{x}_{t})\ket{u(\mathbf{x}_{t})}+\hbar\sum_{j}u^{j}\ket{D_{j}u}\label{eq:ED in CCMCQCD}
\end{equation}
where $u^{j}=\sum_{i}\xi^{ij}w_{j}$ is the imaginary part of the
$j$ component of the velocity field, and the rightmost terms of the
two equations are again consistent with each other, 
\[
\mathcal{F}^{\text{el}}\approx\delta F_{k}^{\text{el}}\approx2\hbar\sum_{j}u^{j}g_{kj}
\]
This is \emph{pseudo}-classical because the nuclear trajectories are
not independent of each other, and essentially amounts to what is
known in the literature as coupled-trajectory mixed quantum-classical
dynamics \citep{Min2015,Agostini2016,Min2017b}, although in this
context could be better named ``Ehrenfest with quantum coupling''.
The method appears to correctly describe the wavepacket splitting
arising from the non-adiabatic couplings and to fix the ``overcoherence''
problem which plagues traditional mixed quantum-classical approaches
\citep{Tully1990,Bittner1995,Thachuk1998,Shu2023}. This is achieved
at the expense of introducing a coupling between trajectories through
the density dependent terms $u^{j}$'s that, in practice, are modeled
with a Gaussian smearing of the representative points \citep{Agostini2016}.
Clearly, a further improvement of the method consists in re-introducing
the geometric effects, hence neglecting only the quantum force from
the exact dynamical equations (``geometric Ehrenfest with quantum
coupling). 

\begin{table*}
\begin{centering}
\begin{tabular}{c|c|c}
\hline 
 & Electron dynamics ($\dot{\rho}_{\text{el}}$) & Classical forces ($\dot{\pi}_{k}$)\tabularnewline
\hline 
\hline 
Ehrenfest\textcolor{white}{\Huge{}I} & $(i\hbar)^{-1}[H_{\text{el}},\rho_{\text{el}}]$ & $-\text{Tr}_{\text{el}}(\rho_{\text{el}}(\partial_{k}H_{\text{el}}))$\tabularnewline
\hline 
geometric Ehrenfest\textcolor{white}{\Huge{}I} & $(i\hbar)^{-1}[H_{\text{el}}-\frac{\hbar^{2}}{2}\Delta_{M}\rho_{\text{el}},\rho_{\text{el}}]$ & $-\text{Tr}_{\text{el}}(\rho_{\text{el}}(\partial_{k}H_{\text{el}}))-\hbar^{2}\sum_{ij}\xi^{ij}\partial_{i}g_{kj}$\tabularnewline
\hline 
Ehrenfest with quantum coupling\textcolor{white}{\Huge{}I} & $(i\hbar)^{-1}[H_{\text{el}}+\hbar\sum_{j}u^{j}\partial_{j}\rho_{\text{el}},\rho_{\text{el}}]$ & $-\text{Tr}_{\text{el}}(\rho_{\text{el}}(\partial_{k}H_{\text{el}}))+2\hbar\sum_{j}u^{j}\,g_{kj}$\tabularnewline
\hline 
geometric Ehrenfest with quantum coupling\textcolor{white}{\Huge{}I} & $(i\hbar)^{-1}[H_{\text{el}}-\frac{\hbar^{2}}{2n}\sum_{ij}\xi^{ij}\partial_{i}(n\partial_{j}\rho_{\text{el}}),\rho_{\text{el}}]$ & $-\text{Tr}_{\text{el}}(\rho_{\text{el}}(\partial_{k}H_{\text{el}}))-\frac{\hbar^{2}}{n}\sum_{ij}\xi^{ij}\partial_{i}(ng_{kj})$\tabularnewline
\hline 
\end{tabular}
\par\end{centering}
\caption{\label{tab:Mixed quantum-classical dynamics}Summary of the mixed
quantum-classical methods that originate from the exact factorization
of the wavefunction, as described in the main text. The second column
defines the electron dynamics in \emph{gauge} invariant form by providing
the expression for the Lagrangian rate of change of the conditional
density operator $\rho_{\text{el}}$. The third column defines correspondingly
the nuclear dynamics, by providing the $k^{\text{th}}$ component
of the force acting on the (classically) moving nuclei, in terms of
\emph{guage} invariant electronic quantities. In the equations $\Delta_{M}$
is the mass-weighted Laplacian of Eq. \ref{eq:mass-weighted Laplacian},
$u^{j}$ is the imaginary part of the velocity field, $u^{j}=-\frac{\hbar}{2}\sum_{i}\xi^{ji}\partial_{i}\ln\,n$
(Sec. \ref{subsec:Momentum-and-velocity-fields}), $n$ is the nuclear
density and $g_{kj}$ is the Fubini-Study metric tensor, $g_{kj}=\Re\text{Tr}_{\text{el}}(\rho_{\text{el}}(\partial_{k}\rho_{\text{el}})(\partial_{j}\rho_{\text{el}}))$.}
\end{table*}
 All these approximations are best compared when the electron dynamics
is described in \emph{gauge}-invariant form, i.e. by means of the
Liouville-von Neumann-like equation of Eq. \ref{eq:electronic equation final}.
The approximate electronic-nuclear coupling vanishes in the Ehrenfest
method of Eqs. \ref{eq:ehrenfest method mom} and \ref{eq:ehrenfest method mom},
while it reads as 
\[
\delta H_{\text{en}}\approx-\frac{\hbar^{2}}{2}\sum_{ij}\xi^{ij}\partial_{i}\partial_{j}\rho_{\text{el}}
\]
 in the geometric limit described by Eq. \ref{eq:ED in classical limit},
and as 
\[
\delta H_{\text{en}}\approx\hbar\sum_{j}u^{j}\partial_{j}\rho_{\text{el}}
\]
in the pseudo-classical limit addressed by the coupled-trajectory
mixed quantum-classical method of Eqs. \ref{eq:force in CCMQCD},
\ref{eq:ED in CCMCQCD}. Finally, in the ``geometric Ehrenfest with
quantum coupling'' scheme the term takes its exact form, Eq. \ref{eq:electron-.nuclear correction to the Lioviallian},
\[
\delta H_{\text{en}}=-\frac{\hbar^{2}}{2n}\sum_{ij}\xi^{ij}\partial_{i}(n\partial_{j}\rho_{\text{el}})
\]
and the only approximation is the neglect of the quantum force in
the nuclear equation of motion. Table \ref{tab:Mixed quantum-classical dynamics}
summarizes the situation and shows the key equations of motion for
these mixed quantum-classical approximations that directly follow
from the exact factorization of the wavefunction. 

We have intentionally left trajectory surface hopping (TSH) \citep{Tully1990}
out of the discussion, since this latter method would require a separate
analysis that goes well beyond the purposes of the present manuscript.
TSH is a popular quantum-classical method in which electrons evolve
according to the same electronic equation of the Ehrenfest method
(Eq. \ref{eq:ehrenfest method el}) but, differently from the Ehrenfest
method, the nuclei sample \emph{stochastically} the energy states,
according to their probabilities $p_{n}=\braket{u_{n}|\rho_{\text{el}}|u_{n}}$.
In practice, this is realized by letting the nuclei move along an
energy surface (the ``active'' state) and hop to different states
with a rate that is usually determined by the electronic equation.
Since (see Eq. \ref{eq:deivative in adiabatic basis})
\[
-\dot{p}_{n}=-\braket{u_{n}|\dot{\rho}_{\text{el}}|u_{n}}+2\sum_{j}\sum_{m}\Re(\braket{u_{n}|\partial_{j}u_{m}}\rho_{mn})v^{j}
\]
and the first term vanishes in Ehrenfest dynamics,
\[
\braket{u_{n}|\dot{\rho}_{\text{el}}|u_{n}}=-\frac{i}{\hbar}\braket{u_{n}|[H_{\text{el}},\rho_{\text{el}}]|u_{n}}\equiv0,
\]
we have that 
\begin{align*}
\gamma_{n\rightarrow m}^{\text{hop}} & =\frac{2}{p_{n}}\sum_{j}\Re(\braket{u_{n}|\partial_{j}u_{m}}\rho_{mn})v^{j}
\end{align*}
(if positive) represents the correct transition probability from $n$
to $m$, depending on the nuclear velocity, on the derivative couplings
and on the coherences $\rho_{nm}$. 

Because of its stochastic nature, deriving TSH from deterministic
equations requires additional assumptions, a topic that warrants more
extensive exploration in a separate publication. Here we just notice
that TSH has been \emph{almost} derived \citep{Subotnik2013} from
the mixed quantum-classical Liouville dynamics (QCLD) developed by
Martens, Kapral and others \citep{Martens1997,Donoso1998,Kapral1999a,Nielsen2000a,Horenko2002},
and that the latter is closely related to the theory developed here.
This connection becomes apparent when examining the momentum moment
formulation of Eq. \ref{eq:moment nuclear equation 1} and \ref{eq:moment nuclear equation 2 first}
which --- similarly to QCLD --- was derived from a partial Wigner
transform, albeit \emph{without} approximations. 

In closing this Section we notice that one main problem of mixed quantum-classical
methods relying on Ehrenfest evolution of the electron dynamics is
the absence of any feedback from the actual \emph{distribution} of
the nuclei in configuration space. In the Lagrangian frame where the
nuclei are at rest the electrons experience only a strictly local,
potential-like coupling with the nuclei, through the electronic Hamiltonian
$H_{\text{el}}(\mathbf{x})$. In the exact dynamical equation, on
the other hand, the electron-nuclear coupling $\delta H_{\text{el}}$
of Eq. \ref{eq:electron-.nuclear correction to the Lioviallian},
provides a coupling which depends on (the gradient of) $n(\mathbf{x})$
and which is active when $\rho_{\text{el}}$ happens to change in
the neighborhoods of $\mathbf{x}$ (since it depends on $\partial_{j}\rho_{\text{el}}$).
In terms of hopping probabilities, adding this \emph{e-n }coupling
contribution introduces a correction to the transition probability.
Indeed, differently from Ehrenfest dynamics, we now have
\[
\braket{u_{n}|\dot{\rho}_{\text{el}}|u_{n}}=\frac{2}{\hbar}\Im\braket{u_{n}|\delta H_{\text{en}}\rho_{\text{el}}|u_{n}}
\]
hence the hopping probabilities get modified into
\[
\gamma_{n\rightarrow m}^{\text{hop}}\rightarrow\bar{\gamma}_{n\rightarrow m}^{\text{hop}}=\gamma_{n\rightarrow m}^{\text{hop}}-\frac{2}{p_{n}}\Im\left(X_{nm}\rho_{mn}\right)
\]
where 
\begin{align*}
X_{nm} & =\frac{1}{\hbar}\braket{u_{n}|\delta H_{\text{en}}|u_{m}}=\\
 & -\frac{\hbar}{2}\braket{u_{n}|\Delta_{M}\rho_{\text{el}}|u_{m}}+\sum_{j}u^{j}\braket{u_{n}|\partial_{j}\rho_{\text{el}}|u_{m}}
\end{align*}
Here, the first term is geometric, while the second is dynamical and
depends on the nuclear density gradient through the imaginary component
of the velocity field, $u^{j}$. Neglecting the first term one obtains
a corrected transition probability in the form
\begin{align*}
\bar{\gamma}_{n\rightarrow m}^{\text{hop}} & \approx-\frac{2}{p_{n}}\sum_{j}\Re((\partial_{j}P_{n})_{nm}\rho_{mn})v^{j}+\\
 & -\frac{2}{p_{n}}\sum_{j}\Im\left((\partial_{j}\rho_{\text{el}})_{nm}\rho_{mn}\right)u^{j}
\end{align*}
where $\braket{u_{n}|\partial_{j}u_{m}}=-\braket{u_{n}|\partial_{j}P_{n}|u_{m}}$
has been used for $m\neq n$. This correction turns TSH into a \emph{coupled}
trajectory surface hopping method (where coupling occurs through the
hopping probabilities only) and might be important when $\rho_{\text{el}}$
deviates from a pure adiabatic state (otherwise $\braket{u_{n}|\dot{\rho}_{\text{el}}|u_{n}}\equiv0$,
as in the ordinary TSH, and the correction vanishes). 

Of course, if the electron-nuclear term $\delta H_{\text{en}}$ were
also considered for the electron dynamics (even at an approximate
level) it would provide a further source of coupling for the trajectories
and it would directly affect the evolution of populations and coherences.
The specific form of electronic decoherence introduced by this electron-nuclear
quantum coupling will be addressed in a separate publication. Here,
we just notice that, as observed in Sec. \ref{subsec:Pure-states},
the exact electronic equation, Eq. \ref{eq:LvN for total rho_el},
is purity conserving, as it must be since it is just the electronic
equation for the pure-state, local electronic wavefunction defined
by the exact factorization approach (i.e., Eq. \ref{eq:EF electronic equation}),
recast in a density form. Hence, if decoherence happens to occur in
the energy basis between regions of strong non-adiabatic coupling
it \emph{must} be accompanied by population decay to a single adiabatic
state, i.e., it must be of the ``decay of mixing'' type, according
to the classification of Ref. \citep{Shu2023}. 

\subsection{Electronic friction at finite temperature\label{subsec:Electronic-friction-at}}

Quoting Dou \emph{et al.} \citep{Dou2018}, ``electronic friction
is a correction to the Born-Oppenheimer approximation, whereby nuclei
in motion experience a drag in the presence of a manifold of electronic
states''. It is one of the many non-adiabatic phenomena that may
occur when atoms or molecules scatter, vibrate or react at or close
to a metal surface \citep{Wodtke2008,Dou2018,Jiang2019b,Auerbach2021}.
It is the simplest one, since it is the first correction to the \emph{adiabatic}
picture: during their motion, nuclei induce $e-h$ pair excitations
into the metal thereby transferring energy to the substrate, but the
excitation is of such limited extent that it can be quickly ``moved
away'', on the time-scale of the nuclear motion. As a result, the
electronic state is always found close to a local adiabatic state
(the ground-state for $T=0$ K), and the nuclear dynamics is \emph{quasi}-adiabatic
with an added friction. 

The $T=0$ K electronic frictional regime was derived by the present
authors \citep{Martinazzo2022,Martinazzo2022a}, under suitable assumptions,
from the exact factorization of the electron-nuclear wavefunction.
The derivation was relatively simple --- an application of linear
response theory to the electronic motion described by Eq. \ref{eq:electronic equation of motion}---
and it should be now clear \emph{why}: we adopted the Eulerian frame
which, as discussed in the previous Section, is the most appropriate
frame for handling a \emph{quasi}-adiabatic dynamics. In fact, the
derivation of a memory\textbf{ }kernel involving the past evolution
of the local electronic states would be a formidable task in the Lagrangian
frame\emph{,} where the past behavior of the electronic state is determined
by the electronic Hamiltonian $H_{\text{el}}(\mathbf{x}_{\tau})$
evaluated \emph{along the trajectory} that brings the nuclear coordinates
to $\mathbf{x}_{t}$ at time $t$. 

This issue actually poses a fundamental interpretative problem when
extending \emph{classical} approaches to memory (as opposed to memoryless
or Markov) friction: how can the system found at $\mathbf{x}$ at
time $t$ experience a memory, in the electronic response, that depends
only on $H_{\text{el}}(\mathbf{x})$, if it never visited $\mathbf{x}$
at earlier times? The memory, if any, should involve the past dynamics,
hence $H_{\text{el}}(\mathbf{x})$ along the visited trajectory. The
reason why we \emph{do} get a coordinate-local memory (in spite of
the oddities it implies for the nuclear dynamics in the classical
limit) is, we believe, a manifestation of non-locality\textbf{ }of\textbf{
}quantum\textbf{ }mechanics, here arising in an apparently ``innocent''
setting: depending on the time-scale of the electron dynamics, the
representative point found at $\mathbf{x}$ at time $t$ may experience
the effect of a perturbation in the electronic state $\ket{u(\mathbf{x},t)}$
which was induced by \emph{different} representatives that visited
$\mathbf{x}$ in the past.

Having elaborated on these interpretative issues -- and made clear
that a re-derivation of the electronic friction picture in the Lagrangian
frame is a challenging task when memory effects are non-negligible
--- let us first recast the results of our previous work \citep{Martinazzo2022,Martinazzo2022a}
in the hydrodynamical setting developed in this work. These results
can be summarized as follows. At $T=0$ K, provided the ground-state
is non-degenerate, the whole $e-n$ state is pure and the quantum
nuclear dynamics is described by Eq. \ref{eq:HEF trajectory equation II}
with the total classical force $F_{k}$ provided by Eq. \ref{eq:dynamically corrected total force},
evaluated with the evolving electronic state $\ket{u}$. The latter
is assumed to be always close to the ground-state $\ket{u_{0}}$,
hence 
\[
F_{k}=F_{k}^{0}+\delta F_{k}
\]
where $F_{k}^{0}$ is the ground-state, dynamically corrected force
of Eq. \ref{eq:dynamically corrected total force} and $\delta F_{k}=2\Re\braket{\partial_{k}u_{0}|QH_{\text{el}}|\Delta u}$
to linear order in $\ket{\Delta u}=\ket{u}-\ket{u_{0}}$ (upon neglecting
terms that depend on the spatial derivatives of $\ket{\Delta u}$
and that arise from the geometric contribution $g_{ij}$). The electronic
fluctuation $\ket{\Delta u}$ can be obtained from linear response
theory using Eq. \ref{eq:electronic equation of motion} and it leads
to
\[
\delta F_{k}=-2\sum_{j}\Re\int_{0}^{\infty}\Gamma_{kj}(\tau)V^{j}(\mathbf{x},t-\tau)\,d\tau+F_{k}^{\text{geom}}
\]
where the first term is a \emph{pseudo}-friction contribution with
the kernel
\begin{equation}
\Gamma_{kj}(t)=\braket{\partial_{k}u_{0}|Q_{0}H'_{\text{el}}e^{-\frac{i}{\hbar}H'_{\text{el}}t}|\partial_{j}u_{0}}\label{eq:memory kernel}
\end{equation}
and the second is a geometric contribution
\begin{align}
F_{k}^{\text{geom}} & =2\hbar\Re\braket{\partial_{k}u_{0}|Ru_{0}}+\nonumber \\
 & -2\pi\hbar\braket{\partial_{k}u_{0}|Q_{0}H'_{\text{el}}\delta(H'_{\text{el}})|Ru_{0}}\label{eq:geometric contribution}
\end{align}
Here, $H_{\text{el}}'$ is the ground-state referenced, local Hamiltonian,
i.e., $H'_{\text{el}}=H{}_{\text{el}}(\mathbf{x})-E_{0}(\mathbf{x})$,
and $Q_{0}=\ket{u_{0}}\bra{u_{0}}$. In the memoryless (Markov) limit
only the integrated kernel matters and this can be recast as
\begin{equation}
\bar{\gamma}_{kj}=2\lim_{\epsilon\rightarrow0}\int_{0}^{\infty}e^{-\epsilon t}\Gamma_{kj}(t)\,dt=-2i\hbar q_{kj}^{0}+\gamma_{kj}\label{eq:Markovian friction}
\end{equation}
where $q_{kj}^{0}$ is the ground-state geometric tensor and 
\[
\gamma_{kj}=\pi\lim_{\omega\rightarrow0^{+}}\frac{\braket{u_{0}|(\partial_{k}H_{\text{el}})Q_{0}\delta(\hbar\omega-H'_{\text{el}})Q_{0}(\partial_{j}H_{\text{el}})|u_{0}}}{\omega}
\]
is the Markovian friction kernel. Hence, upon neglecting the second
term in Eq. \ref{eq:geometric contribution} (which is non-zero only
when time-reversal invariance is broken), the correction force in
the Markov limit reads as
\[
\delta F_{k}\approx F_{k}^{\text{friction}}+F_{k}^{\text{mag}}-\delta F_{k}^{\text{el}}
\]
where 
\begin{equation}
F_{k}^{\text{friction}}=-\Re\left(\sum_{j}\gamma_{kj}V^{j}\right)\label{eq:Markov-limit firctional force}
\end{equation}
is the electronic friction contribution, $F_{k}^{\text{mag}}=\sum_{j}v^{j}B_{kj}$
and $\delta F_{k}^{\text{el}}$ is the dynamical correction to the
\emph{pseudo}-electric force defined in Eq. \ref{eq:dynamical correction to electric force}.
In other words, the Markov-limit quantum hydrodynamical equation of
motion governing the nuclear dynamics is
\begin{equation}
\frac{d\pi_{k}}{dt}=F_{k}^{\text{ad}}+F_{k}^{\text{friction}}+(F_{q})_{k}\label{eq:momentum equation with electronic friction}
\end{equation}
where $F_{k}^{\text{ad}}$ is the ground-state \emph{adiabatic} force
of Eq. \ref{eq:total classical force in the adiabatic approximation}
and $(F_{q})_{k}$ is the quantum force of Eq. \ref{eq:quantum force}. 

Next, we consider electronic friction at finite temperature, using
the theory developed in Section \ref{subsec:Type-e-statistical-mixtures}
for type-$e$ statistical mixtures. This is the useful framework in
the mixed quantum-classical limit where the electronic friction regime
is most often considered, henceforth we shall start with re-deriving
the results recently obtained by Duo, Miao \& Subotnik (DMS) \citep{Dou2017}
in a mixed quantum-classical framework. To this end we neglect the
``quantum couplings'' and the geometric terms in the nuclear equation
of motion (Eq. \ref{eq:momentum equation type-e}) and truncate the
momentum moment hierarchy for the electrons at the zero-th order (see
Eq. \ref{eq:zero-moment electronic equation}), \emph{i.e.}, we consider
\begin{align*}
i\hbar\dot{\rho}_{\text{el}} & \approx[H_{\text{el}},\rho_{\text{el}}]
\end{align*}
\[
\frac{d\pi_{k}}{dt}\approx-\text{Tr}_{\text{el}}\left(\rho_{\text{el}}\partial_{k}H_{\text{el}}\right)
\]
where we have set $\rho_{\text{el}}\equiv\hat{\mathcal{M}}^{(0)}$
for clarity. This is clearly the mixed-state generalization of the
simple Ehrenfest method discussed in the previous Section. Next, we
apply linear response theory to the electron dynamics in the Eulerian
frame, treating the advection term as a perturbation. The result is
\begin{equation}
\rho_{\text{el}}(t)=\rho_{\text{el}}^{0}-\sum_{j}\int_{0}^{\infty}v^{j}(t-\tau)e^{-\frac{i}{\hbar}H_{\text{el}}\tau}\left(\partial_{j}\rho_{\text{el}}^{0}\right)e^{+\frac{i}{\hbar}H_{\text{el}}\tau}d\tau\label{eq:LRT result for rho (DMS-like)}
\end{equation}
where $\rho_{\text{el}}^{0}$ is the density operator that characterizes
the electronic system at steady-state. Eq. \ref{eq:LRT result for rho (DMS-like)}
can then be used to evaluate the force governing the nuclear dynamics
\begin{align*}
F_{k} & =-\text{Tr}_{\text{el}}\left(\rho_{\text{el}}\left(\partial_{k}H_{\text{el}}\right)\right)\\
 & \approx F_{k}^{0}-\sum_{j}\int_{0}^{\infty}W_{kj}(\tau)v^{j}(t-\tau)d\tau
\end{align*}
where $F_{k}^{0}=-\text{Tr}_{\text{el}}\left(\rho_{\text{el}}^{0}\left(\partial_{k}H_{\text{el}}\right)\right)$
is the average microscopic force acting on the nuclei at steady state
and the kernel reads as
\begin{align}
W_{kj}(\tau) & =-\text{Tr}_{\text{el}}\left[e^{+\frac{i}{\hbar}H_{\text{el}}\tau}\left(\partial_{k}H_{\text{el}}\right)e^{-\frac{i}{\hbar}H_{\text{el}}\tau}\left(\partial_{j}\rho_{\text{el}}^{0}\right)\right]\nonumber \\
 & =\text{Tr}_{\text{el}}\left[f_{k}(\tau)\left(\partial_{j}\rho_{\text{el}}^{0}\right)\right]\nonumber \\
 & =\text{Tr}_{\text{el}}\left[\delta f_{k}(\tau)\left(\partial_{j}\rho_{\text{el}}^{0}\right)\right]\label{eq:DMS kernel}
\end{align}
Here, we have introduced the (Heisenberg evolved) microscopic force
operator
\[
f_{k}(\tau)=-e^{+\frac{i}{\hbar}H_{\text{el}}\tau}\left(\partial_{k}H_{\text{el}}\right)e^{-\frac{i}{\hbar}H_{\text{el}}\tau}
\]
and its fluctuation 
\begin{equation}
\delta f_{k}(t)=f_{k}(t)-F_{k}^{\text{0}}\label{eq:force fluctuation}
\end{equation}
upon exploiting $\text{Tr}_{\text{el}}(\partial_{j}\rho_{\text{el}}^{0})\equiv0$.
In the Markov limit only the integrated kernel matters and we have
\[
F_{k}\approx F_{k}^{0}-\sum_{j}\gamma_{kj}^{\text{DMS}}v^{j}
\]
with 
\[
\gamma_{kj}^{\text{DMS}}=\lim_{\epsilon\rightarrow0^{+}}\int_{0}^{\infty}e^{-\epsilon t}W_{kj}(t)\,dt
\]
These are the results obtained by DMS in their mixed quantum-classical
theory of electronic friction \citep{Dou2017}. 

In our previous work \citep{Martinazzo2022,Martinazzo2022a} we showed
that in the $T=0$ K limit the DMS kernel reduces to the real part
of Eq. \ref{eq:Markovian friction} --- i.e., $\gamma_{kj}^{\text{DMS}}\rightarrow\Re\bar{\gamma}_{kj}$
at vanishing temperature --- and that $\bar{\gamma}_{kj}$ is actually
a \emph{pseudo}-friction kernel since it also contains a \emph{pseudo}-magnetic
contribution (see the quantum geometric contribution in the rightmost
hand side of Eq. \ref{eq:Markovian friction}). The same applies at
finite temperatures, where $W_{kj}$ and its integrated version $\gamma_{kj}^{\text{DMS}}$
are real by construction but not necessarily symmetric in their indexes,
unless time-reversal symmetry holds. Hence, in general, $\gamma_{kj}^{\text{DMS}}=\eta_{kj}-B_{kj}^{\text{DMS}}$
with $\eta_{kj}$ symmetric and $B_{kj}^{\text{DMS}}$antisymmetric,
and we have, in the Markov limit, 
\begin{equation}
F_{k}\approx F_{k}^{0}-\sum_{j}\eta_{kj}v^{j}+\sum_{j}B_{kj}^{\text{DMS}}v^{j}\label{eq:DMS Markov force}
\end{equation}
Here, only the second term represents a true frictional force, the
last one is a \emph{pseudo}-magnetic force. 

The above results hold for a general steady-state. If the electronic
system is in (local) canonical equilibrium additional, interesting
properties hold. First of all, the force $F_{k}^{0}$ becomes an equilibrium
force which is manifestly conservative,
\[
F_{k}^{0}=\frac{\sum_{n}(-\partial_{k}E_{n})e^{-\beta E_{n}}}{\sum_{n}e^{-\beta E_{n}}}\equiv\frac{1}{\beta}\partial_{k}\ln Z_{\text{el}}=-\partial_{k}\Phi_{\text{el}}
\]
where $n$ labels the eigenstates of $H_{\text{el}}$, $Z_{\text{el}}=\sum_{n}e^{-\beta E_{n}}$
is the (local) partition function of the electronic system and $\Phi_{\text{el}}=-k_{\text{B}}T\ln Z_{\beta}$
--- the corresponding (coordinate-dependent) Helmholtz free-energy
--- is the appropriate ``thermal'' potential that generates the
force. Secondly, the kernel $W_{kj}$ can be recast as a Kubo correlation
function
\begin{equation}
W_{kj}(t)=\beta\text{Tr}_{\text{el}}\left(\rho_{\text{el}}^{0}(\delta f_{j})_{\text{K}}^{\beta}\,\delta f_{k}(t)\right)\label{eq:Kubo correlation function}
\end{equation}
where, in general,
\[
A_{\text{K}}^{\beta}=\frac{1}{\beta}\int_{0}^{\beta}e^{+\alpha H}Ae^{-\alpha H}d\alpha
\]
is the Kubo transform of the operator $A$ at the inverse temperature
$\beta$ ($H\equiv H_{\text{el}}$ in our problem). This follows directly
from the definition of the kernel upon noticing the identity (see
Appendix \ref{app:Lie-Trotter})
\[
\partial_{j}\rho_{\text{el}}^{0}=\beta\rho_{\text{el}}^{0}(\delta f_{j})_{\text{K}}^{\beta}
\]

Eq. \ref{eq:Kubo correlation function} is a form of the\textbf{ }fluctuation-dissipation\textbf{
}theorem of the second kind relating the (time-symmetrized) memory
kernel and the correlation function of the ``random force'' $\delta f_{k}$,
and gives the Markovian kernel in Green-Kubo form 
\[
\gamma_{kj}^{\text{DMS}}=\beta\lim_{\text{\ensuremath{\epsilon}\ensuremath{\rightarrow0}}^{+}}\int_{0}^{\infty}e^{-\epsilon t}K_{\delta f_{k},\delta f_{j}}^{\beta}(t)dt
\]
where 
\[
K_{AB}^{\beta}(t)=\text{Tr}\left(\rho B_{K}^{\beta}A(t)\right)
\]
for $A=\delta f_{k}$ and $B=\delta f_{j}$. All this connects electronic
friction theory with linear-response irreversible thermodynamics,
a rather interesting issue that deserves consideration in a separate
manuscript. Here, we just notice that in thermal equilibrium the
\emph{pseudo}-magnetic field appearing in the Markov-limit expression
of Eq. \ref{eq:DMS Markov force} is given by $B_{kj}^{\text{DMS}}=-2\hbar\Im q_{kj}^{\text{av}}$,
where $q_{kj}^{\text{av}}$ is the quantum geometric tensor defined
in Eq. \ref{eq:average quantum geometric tensor}, averaged over the
canonical ensemble (see Appendix \ref{app:Pseudo-magnetic}). Likewise,
the ``corrected'' Markovian friction kernel reads, in thermal equilibrium,
as 
\begin{equation}
\eta_{kj}=\lim_{\omega\rightarrow0}\Re\gamma_{kj}^{\beta}(\omega)(1-e^{-\beta\hbar\omega})\label{eq:Markovian friction at finite T}
\end{equation}
where 
\begin{equation}
\gamma_{kj}^{\beta}(\omega)=\frac{\pi}{\omega}\sum_{n}p_{n}\braket{u_{n}|(\partial_{k}H_{\text{el}})Q_{n}\delta(\hbar\omega-H_{\text{el}}^{n})(\partial_{j}H_{\text{el}})|u_{n}}\label{eq:frequency depedent kernel at finite T}
\end{equation}
satisfies the detailed balance condition
\begin{equation}
\gamma_{kj}^{\beta}(\omega)=-e^{\beta\hbar\omega}\gamma_{jk}^{\beta}(-\omega)\label{eq:detailed balance}
\end{equation}
as shown in detail in Appendix \ref{app:Pseudo-magnetic}. Here, the
$p_{n}$'s are Boltzmann populations, $n$ labels again energy eigenstates,
$Q_{n}=1-\ket{u_{n}}\bra{u_{n}}$ and $H_{\text{el}}^{n}=H_{\text{el}}-E_{n}$.
Clearly, the difference with the $T=0$ K case is due to the electronic
de-excitation processes that can occur in the electronic bath at finite
temperature and that work by pumping energy into the nuclear system.
Detailed balance guarantees that the kernel $\Re\gamma_{kj}^{\beta}(\omega)$
remains positive definite for $\omega>0$, hence it correctly describes
friction. Finally, Appendix \ref{app:Finite-temperature-friction-kern}
shows that for independent electrons Eq. \ref{eq:Markovian friction at finite T}
becomes 
\begin{equation}
\eta_{kj}=\lim_{\omega\rightarrow0}\frac{\pi}{\omega}\Re\sum_{ab}D_{ab}^{k}D_{ba}^{j}(f(\epsilon_{a})-f(\epsilon_{b}))\delta(\hbar\omega-\Delta\epsilon_{ba})\label{eq:HGT finite-temperature}
\end{equation}
where $D_{ab}^{k}=\braket{\phi_{a}|\partial_{k}h|\phi_{b}}$, $\ket{\phi_{a}}$
($\epsilon_{a}$) are single-particle states (energies), $h$ is the
single-particle Hamiltonian, $\Delta\epsilon_{ba}=\epsilon_{b}-\epsilon_{a}$
and $f(\epsilon)$ is the Fermi occupation function at the given temperature.
Eq. \ref{eq:HGT finite-temperature} is the finite-temperature extension
of the original Head-Gordon and Tully expression \citep{Head-Gordon1995},
\[
\eta_{kj}=\pi\hbar\sum_{ab}D_{ab}^{k}D_{ba}^{j}\delta(\epsilon_{a}-\epsilon_{F})\delta(\epsilon_{b}-\epsilon_{F})
\]
to which it reduces in the limit $T\rightarrow0$ K (see Appendix
\ref{app:Finite-temperature-friction-kern}).

So far we have considered electronic friction theory in the context
of the simple Ehrenfest theory for mixed-states. Improvements are
possible (in a yet mixed quantum-classical framework) along the lines
of arguments used in Section \ref{subsec:Electronic-friction-at},
by either introducing geometric contributions or describing the \emph{e-n}
couplings quantally or both. The main difference with Section \ref{subsec:Electronic-friction-at}
is that now the hierarchy of electronic moment equations does not
terminate at the zero-th order (that of the conditional density operators),
rather requires the first moment equation to introduce corrections
(Eq. \ref{eq:first-moment electronic equation}). The latter can be
simplified to its free-evolution
\[
\frac{d\hat{\mathcal{M}}_{k}^{(1)}}{dt}\approx-\frac{i}{\hbar}[H_{\text{el}},\hat{\mathcal{M}}_{k}^{(1)}]
\]
and handled in linear response theory, consistently with zero-th moment
equation, giving rise to correction terms that account for the quantum
couplings. We discuss these improvements in a forthcoming paper.

\section{Summary and Conclusions\label{sec:Conclusions}}

This study has delved into the quantum dynamics of electron-nuclear
systems by leveraging the concept of exact factorization of the wavefunction.
Our primary objective was to reformulate the equations of motion for
both electrons and nuclei in a manner that eliminates the dependence
on the arbitrary phase of the electronic (and nuclear) wavefunction,
leaving only physical contributions.

In the case of pure states, we achieved this in an exact way by adopting
a quantum hydrodynamical description for the nuclei, coupled with
the dynamics of specific density operators for electrons. These density
operators are conditional, meaning that they depend on the nuclei
being in a particular configuration, and therefore move in tandem
with the fluid elements that describe the nuclear dynamics. This is
a setup often invoked in modern quantum approaches to the problem
where the dynamics is ``direct'' and the electronic problem is solved
``on-the-fly'' alongside the propagation of the nuclear wavefunction.
The advantage of the proposed approach is that it removes completely
the need of selecting smooth phases for the wavefunctions. Importantly,
the proposed approach establishes a natural connection between exact
quantum dynamics and various mixed quantum-classical methodologies,
laying the groundwork for their systematic enhancement. This is where
direct dynamical approaches are mostly employed and where a protocol
for systematic corrections is yet missing. 

The present theory further enables the extension of the exact factorization
approach to encompass statistical mixtures of states, where either
the nuclei or the electrons are in a inherently mixed state. The primary
adjustment in this scenario involves a hierarchical construction that
incorporates moments associated with the variables (either electrons
or nuclei) existing in a mixed state. These advancements establish
a formal connection to approximate non-adiabatic theories in condensed
phases, where nuclei are often classical and electrons are in a mixed
state. As an illustrative example, we explored the concept of electronic
friction at finite temperatures, in the quantum-classical limit where
it is commonly invoked. 

In concluding, we emphasize the role played by the exact factorization
of the wavefunction in the developments presented in this work. It
emerges as a key tool for investigating electron-nuclear systems,
forming the basis for unifying many different themes and views in
molecular dynamics.

\newpage{}

\appendix

\section{\label{app:Circulation} Circulation theorem}

The quantization condition on the circulation of the (Hamiltonian)
momentum field given in Eq. \ref{eq:circulation}, i.e., 
\[
\Gamma_{\gamma}=\sum_{k}\oint_{\gamma}dx^{k}p_{k}=2\pi\hbar\,n\ \ n\in\mathbb{Z}
\]
with $p_{k}\equiv\pi_{k}+\hbar A_{k}$, is necessary to single out
from the many solutions of the hydrodynamical equations of motion
(Eqs. \ref{eq:real Madelung (general)} and \ref{eq:continuity equation Lagrange (general)})
those that are compatible with the Schr\"{o}dinger equation we started
from. It guarantees the existence of a smooth scalar function $S(\mathbf{x})$,
that can be used to synthesize the wavefunction $\psi(\mathbf{x})=n(\mathbf{x})\exp(iS(\mathbf{x})/\hbar)$,
once the solutions $\{n(\mathbf{x}),\pi_{k}(\mathbf{x})\}$ of the
hydrodynamical problem have been obtained. This is the open-path line
integral along the curve $\gamma(\mathbf{x}_{0},\mathbf{x})$ joining
$\mathbf{x}$ to a reference point $\mathbf{x}_{0}$ 
\[
S(\mathbf{x})=\int_{\gamma(\mathbf{x}_{0},\mathbf{x})}dx^{k}(\pi_{k}+\hbar A_{k})
\]
that is in fact path-independent (modulo $2\pi\hbar$) under the quantization
condition of Eq. \ref{eq:circulation}. However, for consistency with
the dynamical evolution one has to check that this condition is preserved
during the dynamics, i.e. $d\Gamma_{\gamma}/dt=0$.

We first consider the simple case where the loop $\gamma$ is fixed
in space and
\[
\frac{d}{dt}\Gamma_{\gamma}=\sum_{k}\oint_{\gamma}dx^{k}\left(\frac{\partial\pi_{k}}{\partial t}+\hbar\frac{\partial A_{k}}{\partial t}\right)
\]
This is easily seen to vanish since, upon taking the real part of
Eq. \ref{eq:fundamental momentum equation}, i.e., 
\[
\frac{\partial\pi_{k}}{\partial t}+\hbar\frac{\partial A_{k}}{\partial t}=-\partial_{k}\Re\left(\frac{H\psi}{\psi}\right)
\]
one sees that the integrand is in fact purely longitudinal and therefore
gives a vanishing contribution to the closed-path line integral. The
only (implicit) assumption here is that in the flow of the probability
fluid no wavefunction node crosses the loop $\gamma$, otherwise $p_{k}$
would change abruptly and $\partial p_{k}/\partial t$ would be singular
at the crossing point (see Section \ref{subsec:Momentum-and-velocity-fields}).
If this condition is satisfied we thus have
\[
\sum_{k}\oint_{\gamma}dx^{k}p_{k}(t)=\sum_{k}\oint_{\gamma}dx^{k}p_{k}(t')
\]
for arbitrary times $t$ and $t'$. Let then be $\gamma'$ the loop
$\gamma$ that has been transported by the flow $v^{k}$ from $t$
to $t'$. We also have 
\[
\sum_{k}\oint_{\gamma}dx^{k}p_{k}(t')=\sum_{k}\oint_{\gamma'}dx^{k}p_{k}(t')
\]
since Eq. \ref{eq:circulation} holds for arbitrary loops and $\gamma$
and $\gamma'$ are connected by a continuous transformation that cannot
change the value of $n$. Hence, $\Gamma_{\gamma}$ is preserved along
closed paths that are in motion with the fluid elements. This is Kelvin's
circulation theorem for the many-body quantum hydrodynamical fluid
investigated in Sec. \ref{subsec:Many-particle-QHD-with}, \emph{i.e.},
that describing the nuclear dynamics in the exact factorization approach
of Sec. \ref{sec:Quantum-Hydrodynamics-of}. 

The final result can also be obtained with a direct calculation that
bypasses the need of a non-crossing condition. Indeed, when the curve
follows the fluid, the rate of change of circulation involves the
material derivative of the fields and contains contributions from
the moving path
\begin{align*}
\frac{d}{dt}\sum_{k}\oint_{\gamma}dx^{k}(\pi_{k}+\hbar A_{k}) & =\sum_{k}\oint_{\gamma}dx^{k}(\dot{\pi}_{k}+\hbar\dot{A}_{k})+\\
 & +\sum_{kj}\oint_{\gamma}dx^{k}(\pi_{j}+\hbar A_{j})\partial_{k}v^{j}
\end{align*}
Here, the second term on the r.h.s. can be rearranged as
\begin{align*}
\sum_{kj}\oint_{\gamma}dx^{k}(\pi_{j}+\hbar A_{j})\partial_{k}v^{j} & =\\
=\oint_{\gamma}\sum_{kj}dx^{k}\partial_{k}[(\pi_{j}+\hbar A_{j})v^{j}] & -\oint_{\gamma}\sum_{kj}dx^{k}v^{j}\partial_{k}[(\pi_{j}+\hbar A_{j})]\\
=-\oint_{\gamma}\sum_{kj}dx^{k}v^{j}\partial_{k}\pi_{j} & -\oint_{\gamma}\sum_{kj}dx^{k}v^{j}B_{kj}+\\
 & -\hbar\oint_{\gamma}\sum_{kj}dx^{k}v^{j}\partial_{j}A_{k}
\end{align*}
upon noticing that the first integral on the second line vanishes
identically. In second step we have introduced the magnetic field
(Eq. \ref{eq:E,B fields}) and singled out an advective term that
precisely cancels that coming from 
\[
\dot{A}_{k}=\partial_{t}A_{k}+\sum_{j}v^{j}\partial_{j}A_{k}
\]
Hence, we are left with 
\begin{align*}
\frac{d\Gamma_{\gamma}}{dt} & =\oint_{\gamma}\sum_{k}(\dot{\pi}_{k}+\hbar\partial_{t}A_{k})dx^{k}+\\
 & -\oint_{\gamma}\sum_{kj}dx^{k}v^{j}\partial_{k}\pi_{j}+\\
 & -\oint_{\gamma}\sum_{kj}dx^{k}v^{j}B_{kj}
\end{align*}
On the other hand, Eq. \ref{eq:real Madelung (general)} and Eq. \ref{eq:E,B fields}
give
\[
\dot{\pi}_{k}=\hbar(\partial_{k}A_{0}-\partial_{t}A_{k})+\sum_{j}v^{j}B_{kj}-\partial_{k}(\mathcal{V}+Q)
\]
hence
\[
\frac{d\Gamma_{\gamma}}{dt}=\oint_{\gamma}\sum_{k}dx^{k}\partial_{k}(\hbar A_{0}-\mathcal{V}-Q)\equiv0
\]
which is again Kelvin circulation theorem: circulation is conserved
along any dynamically evolved path --- hence continuous in time ---
and no problem arises from the possible presence of singularities
which, too, evolve during the dynamics. Note, however, that this results
does not conflict with the formation of vortexes. Vortexes can emerge
in pairs with opposite chirality, or they can manifest as closed vortex
lines originating from a single point \citep{Bialynicki-Birula2000}.
The distinction from a classical, inviscid fluid lies in the fact
that in QHD extended vortex structures (such as line vortexes in three
dimensions) are anticipated, whereas in a classical fluid, the concept
of a line vortex does not appear to be a plausible dynamical limit
\citep{Saffman1992}. 

In closing this Appendix we notice that the particle trajectories
define useful paths to synthesize the wavefunction \citep{RobertE.Wyatt2005}.
Indeed, according to the continuity equation in Lagrangian form, Eq.
\ref{eq:continuity equation Lagrange (general)}, we can write
\begin{equation}
n(\mathbf{x}_{t},t)=n(\mathbf{x}_{0},t_{0})\exp\left(-\int_{t_{0}}^{t}\left(\boldsymbol{\nabla}\mathbf{v}\right)(\mathbf{x}_{\tau},\tau)\,d\tau\right)\label{eq:density evolution along q-trajectories-1}
\end{equation}
where the divergence is evaluated along the trajectory $\mathbf{x}_{t}$
started from $\mathbf{x}_{0}$ at $t=t_{0}$. Likewise, we have 
\begin{equation}
S(\mathbf{x}_{t},t)=S(\mathbf{x}_{0},t_{0})+\int_{t_{0}}^{t}\mathcal{L}(\mathbf{x}_{\tau},\tau)\,d\tau\label{eq:action evolution along q-trajectories-1}
\end{equation}
where $\mathcal{L}$ is a \emph{pseudo}-electromagnetic Lagrangian
in which $\mathcal{V}-\hbar A_{0}+Q$ plays the role of potential
(see the Hamilton-Jacobi equation of Eq. \ref{eq:Hamilton-Jacobi}).
Hence, 
\begin{align}
\psi(\mathbf{x}_{t},t) & =\exp\left(-\frac{1}{2}\int_{t_{0}}^{t}\left(\boldsymbol{\nabla}\mathbf{v}\right)(\mathbf{x}_{\tau},\tau)\,d\tau\right)\label{eq:wavefunction synthesis-1}\\
 & \times\exp\left(\frac{i}{\hbar}\int_{t_{0}}^{t}\mathcal{L}(\mathbf{x}_{\tau},\tau)\,d\tau\right)\psi(\mathbf{x}_{0},t_{0})
\end{align}
builds the wavefunction amplitude along the trajectory \textbf{$\mathbf{x}_{t}$}
that solves the QHD equations of motion. Noteworthy, the ``propagator''
appearing here is determined by a \emph{single} trajectory, in striking
contrast with a Feymann's path integral where the propagator is the
sum of contributions from infinitely many \emph{paths}. This might
be useful for interpretative purposes but it is of limited help in
practice since, as emphasized in the main text, it is \emph{not} possible
to solve the QHD equations for a \emph{single} quantum trajectory,
unless the evolving density is known in advance. 

\section{\label{app:momentum moments}Wigner function and momentum moments}

Momentum moments represent an essential ingredient for the development
of a hydrodynamical picture of quantum dynamics beyond the pure state
case. With their help one can recast the evolution of a quantum system
in terms of a hierarchical set of equations for the momentum moments,
where each moment is seen to couple to the next (higher order) one.
For pure states the hierarchy can be truncated at the \emph{first}
moment equation --- second and higher moments of a pure state can
be expressed in terms of the zero-th and the first moment alone ---
but in general the whole structure of the hierarchy is needed and
often a \emph{closure} is pragmatically introduced to make the problem
manageable. The purpose of this Appendix is to extend such construction
to the case where the quantum system is subjected to \emph{gauge}
fields (in addition to the usual scalar potential typically considered)
which make the standard construction of limited help (since it is\emph{
gauge} dependent). Hence, we define here a new type of \emph{gauge}
invariant moments --- which we call mechanical\textbf{ }moment\textbf{
}moments (MMMs) because of their relation with the mechanical momentum
operator --- and show that their equation of motion take a simple,
physically sound form. We start with some general remarks about the
momentum moments and recast them in a form that suit better to the
presence of \emph{gauge} fields. Henceforth, we shall use the hat
on operators to denote the \emph{abstract} operators and and \emph{not}
their Schr\"{o}dinger representation. 

We recall that for a $D$ dimensional quantum system described by
a continuous position variable $\mathbf{x}\in\mathbb{R}^{D}$ the
Wigner function $\rho_{W}$ is defined as a Fourier transform of the
function
\[
W(\mathbf{q},\mathbf{r})=\left\langle \mathbf{q}-\frac{\mathbf{r}}{2}|\hat{\rho}|\mathbf{q}+\frac{\mathbf{r}}{2}\right\rangle =\rho\left(\mathbf{\mathbf{q}}-\frac{\mathbf{r}}{2},\mathbf{q}+\frac{\mathbf{r}}{2}\right)
\]
with respect to $\mathbf{r}$, at fixed $\mathbf{q}$. Here, $\hat{\rho}$
is the system density operator and $\rho(\mathbf{x},\mathbf{x}')=\braket{\mathbf{x}|\hat{\rho}|\mathbf{x}'}$
is the density matrix (hence, $\mathbf{r}\equiv\mathbf{x}'-\mathbf{x}$
the ``difference variable'' or ``relative position''). That is,
\[
\rho_{W}(\mathbf{q},\mathbf{p})=\int\frac{d^{D}\mathbf{r}}{(2\pi\hbar)^{D}}W(\mathbf{q},\mathbf{r})e^{i\frac{\mathbf{p}\mathbf{r}}{\hbar}}
\]
in such a way that $\rho_{W}(\mathbf{q},\mathbf{p})$ is a kind of
distribution function in phase space, $(\mathbf{q},\mathbf{p})\in\mathbb{R}^{2D}$.
Then, the momentum moments are the \emph{moments} of this ``distribution''
at fixed $\mathbf{q}$
\[
S_{i_{1}i_{2}..i_{D}}^{(n)}(\mathbf{q})=\int d^{D}\mathbf{p}\,p_{1}^{i_{1}}p_{2}^{i_{2}}..p_{D}^{i_{D}}\rho_{W}(\mathbf{q},\mathbf{p})
\]
(irrespective of the fact that $\rho_{W}$ is a true distribution
or not) and can be obtained from the $\mathbf{s}$-derivatives at
$\mathbf{s}=\mathbf{0}$ of the characteristic function 
\[
F_{\mathbf{q}}(\mathbf{s})=\braket{e^{i\mathbf{s}\mathbf{p}/\hbar}}_{\mathbf{q}}\equiv\int d^{D}\mathbf{p}e^{i\mathbf{s}\mathbf{p}/\hbar}\rho_{W}(\mathbf{q},\mathbf{s})\ \ \ \mathbf{s}\in\mathbb{R}^{D}
\]
where $\braket{..}$ denotes the average w.r.t. to the Wigner ``distribution''.
Specifically,
\[
F_{\mathbf{q}}(\mathbf{s})=\sum_{n=0}^{\infty}\left(\frac{i}{\hbar}\right)^{n}\frac{\braket{(\mathbf{s}\mathbf{p})^{n}}}{n!}
\]
where 
\[
\braket{(\mathbf{s}\mathbf{p})^{n}}=\sum_{i_{1}i_{2}..i_{D}}^{[n]}\frac{n!}{i_{1}!i_{2}!..i_{D}!}\braket{(s_{1}p_{1})^{i_{1}}(s_{2}p_{2})^{i_{2}}..(s_{S}p_{D})^{i_{D}}}
\]
and $[n]$ stands for the constraint $\sum_{k=1}^{D}i_{k}=n$. In
other words we have 
\[
F_{q}(\mathbf{s})=\sum_{i_{1}i_{2}..i_{D}}^{\infty}\left(\frac{i}{\hbar}\right)^{n}\frac{S_{i_{1}i_{2}..i_{D}}^{(n)}(\mathbf{q})}{i_{1}!i_{2}!..i_{D}!}s_{1}^{i_{1}}s_{2}^{i_{2}}..s_{D}^{i_{D}}\ \ \ \text{with}\ n=\sum_{k}i_{k}
\]
and then 
\begin{equation}
\left(\frac{i}{\hbar}\right)^{n}S_{i_{1}i_{2}..i_{D}}^{(n)}(\mathbf{q})=\frac{\partial^{n}}{\partial_{s_{1}}^{i_{1}}\partial_{s_{2}}^{i_{2}}..\partial_{s_{D}}^{i_{D}}}F_{\mathbf{q}}(\mathbf{s})\vert_{\mathbf{s}=\mathbf{0}}\label{eq:derivative expression}
\end{equation}
By definition, the characteristic function is the Fourier transform
(w.r.t. to $\mathbf{p}$) of the function $\rho_{W}(\mathbf{q},\mathbf{p})$,
which is itself the Fourier transform of $W(\mathbf{q},\mathbf{r})$.
Hence, 
\[
F_{\mathbf{q}}(\mathbf{s})\equiv\rho\left(\mathbf{\mathbf{q}}+\frac{\mathbf{s}}{2},\mathbf{\mathbf{q}}-\frac{\mathbf{s}}{2}\right)
\]
and the moments are seen to characterize the behavior of the density
matrix along the second diagonal of the $(\mathbf{x},\mathbf{x}')$
``plane'', for a fixed position $\mathbf{q}$ along the first (main)
diagonal. The characteristic function can be conveniently given in
terms of translational operators, namely
\[
F_{\mathbf{q}}(\mathbf{s})\equiv\braket{\mathbf{q}|T_{-\mathbf{s}/2}\hat{\rho}T_{-\mathbf{s}/2}|\mathbf{q}}=\lim_{\mathbf{x}\rightarrow\mathbf{q},\mathbf{x}'\rightarrow\mathbf{q}}\braket{\mathbf{x}|T_{-\mathbf{s}/2}\hat{\rho}T_{-\mathbf{s}/2}|\mathbf{x}'}
\]
where $T_{\mathbf{a}}=\exp(-i\mathbf{a}\hat{\mathbf{p}}/\hbar)$ is
the translation operator by $\mathbf{a}$, $T_{\mathbf{a}}\ket{\mathbf{x}}=\ket{\mathbf{x}+\mathbf{a}}$
(or equivalently, $\braket{\mathbf{x}|T_{\mathbf{a}}\psi}=\psi(\mathbf{x}-\mathbf{a})$).
The usefulness of the representation in the rightmost side of the
last equation becomes evident when expanding the exponentials
\[
\braket{\mathbf{x}|T_{-\mathbf{s}/2}\hat{\rho}T_{-\mathbf{s}/2}|\mathbf{x}'}=\sum_{n,m=0}^{\infty}\frac{1}{n!m!}\braket{\mathbf{x}|\left(\frac{i\mathbf{s}\hat{\mathbf{p}}}{2\hbar}\right)^{n}\hat{\rho}\left(\frac{i\mathbf{s}\hat{\mathbf{p}}}{2\hbar}\right)^{m}|\mathbf{x}'}
\]
and remembering that 
\[
\braket{\mathbf{x}|\hat{p}_{i}|\psi}=(-i\hbar)\partial_{i}\psi\ \ \braket{\psi|\hat{p}_{i}|\mathbf{x}'}=(i\hbar)\partial'_{i}\psi'^{*}
\]
where we have introduced the notation $\psi\equiv\psi(\mathbf{x})$
and $\psi'\equiv\psi(\mathbf{x}')$ for the wavefunction and, correspondingly,
$\partial_{i}\equiv\partial/\partial x_{i}$ and $\partial_{i}'\equiv\partial/\partial x_{i}'$.
It is then clear that
\begin{align}
\mathcal{P}^{(0)}(\mathbf{q}) & =\braket{\mathbf{q}|\hat{\rho}|\mathbf{q}}\equiv n(\mathbf{q})\nonumber \\
\mathcal{P}_{k}^{(1)}(\mathbf{q}) & =\frac{\braket{\mathbf{q}|[\hat{p}_{k},\hat{\rho}]_{+}|\mathbf{q}}}{2}\equiv S_{0..1_{k}..0}^{(1)}(\mathbf{q})\nonumber \\
\mathcal{P}_{jk}^{(1)}(\mathbf{q}) & =\frac{\braket{\mathbf{q}|[\hat{p}_{j}[\hat{p}_{k},\hat{\rho}]_{+}]_{+}|\mathbf{q}}}{2^{2}}\equiv S_{0..1_{j}..1_{k}..0}^{(1)}(\mathbf{q})\nonumber \\
.. & ..\nonumber \\
\mathcal{P}_{ijk..}^{(n)}(\mathbf{q}) & =\frac{\braket{\mathbf{q}|[\hat{p}_{i},[\hat{p}_{j},[\hat{p}_{k},..\hat{\rho}]_{+}]_{+}]_{+}|\mathbf{q}}}{2^{n}}\label{eq:anti-commutator (1)}
\end{align}
where $[..,..]_{+}$ is the anti-commutator. Here, we have introduced
new moments $\mathcal{P}_{ijk..}^{(n)}$ in which the subscripts denote
explicitly the order of the products, i.e. the momentum operators
appearing in the nested products are labeled with an index from right
to left, starting from the innermost product. This is useless in this
case since the $\hat{p}_{k}$'s commute with each other --- meaning
that the newly defined moments are fully symmetric w.r.t. their indexes
--- but it will become important soon when we generalize this construction
to non-commuting operator. This representation of moments in terms
of symmetrized product turns out to be useful in view of the following
property
\[
\braket{\mathbf{q}|[\hat{p}_{k},\hat{O}]_{+}|\mathbf{q}}=(p_{k}+p_{k}^{'*})O(\mathbf{x},\mathbf{x}')\vert_{\mathbf{x}=\mathbf{x}'=\mathbf{q}}
\]
which holds for arbitrary operators $\hat{O}$. Here $p_{k}=-i\hbar\partial_{k}$,
$p_{k}^{'}=-i\hbar\partial'_{k}$ are the Schr\"{o}dinger-representation
momentum operators and $O(\mathbf{x},\mathbf{x}')=\braket{\mathbf{x}|\hat{O}|\mathbf{x}'}$.
Indeed, it allows us to re-write the above moments in the form
\begin{align}
\mathcal{P}_{ijk..}^{(n)}(\mathbf{q}) & =\left(\frac{p_{i}+p_{i}^{'*}}{2}\right)\left(\frac{p_{j}+p_{j}^{'*}}{2}\right)\left(\frac{p_{k}+p_{k}^{'*}}{2}\right)...\nonumber \\
 & ..\,\,\rho(\mathbf{x},\mathbf{x}')\vert_{\mathbf{x}=\mathbf{x}'=\mathbf{q}}\label{eq:anti-commutator (2)}
\end{align}
which makes evident the symmetry of the moments (the $p_{k}$ 's commute
with each other and with the $p'_{j}$'s). Furthermore, the last equation
provides a useful expression for the moments
\begin{align*}
\mathcal{P}_{ijk..}^{(n)}(\mathbf{q}) & =\left(\frac{-i\hbar}{2}\right)^{n}\left(\partial_{i}-\partial'_{i}\right)\left(\partial_{j}-\partial'_{j}\right)\left(\partial_{k}-\partial'_{k}\right)...\\
 & ...\,\rho(\mathbf{x},\mathbf{x}')\vert_{\mathbf{x}=\mathbf{x}'=\mathbf{q}}
\end{align*}
or, equivalently,
\[
\mathcal{P}_{ijk..}^{(n)}(\mathbf{q})=\left(-i\hbar\right)^{n}\partial_{s_{i}}\partial_{s_{j}}\partial_{s_{k}}..\rho\left(\mathbf{q}+\frac{\mathbf{s}}{2},\mathbf{q}-\frac{\mathbf{s}}{2}\right)
\]
which is the same as Eq. \ref{eq:derivative expression}. As mentioned
above, these moments are dynamically coupled to each other. The (infinite)
set of equation of motion can be obtained in rather general form if
the system Hamiltonian takes the standard form
\[
H=\frac{\hat{p}^{2}}{2m}+\mathcal{V}
\]
where $\mathcal{V}$ is a scalar potential, see Refs. \citep{Johansen1998,Lill1989,Burghardt2002}. 

We now consider the case of a quantum mechanical system subjected
to \emph{gauge} fields, in addition to the scalar potential $\mathcal{V}$,
that is, we take our usual Hamiltonian 
\[
H=\sum_{ij}\frac{\xi^{ij}}{2}\hat{\pi}_{i}\hat{\pi}_{j}+\left(\mathcal{V}-\hbar A_{0}\right)
\]
with \footnote{The \emph{gauge} potentials  $A_{k}$, as well as $A_{0}$ are local,
\emph{abstract} operators but we shall omit the hat from them.} $\hat{\pi}_{k}=\hat{p}_{k}-\hbar A_{k}$. Furthermore, to distinguish
this case from the previous one we shall use $\hat{\sigma}$ in place
of $\hat{\rho}$ for the density operator and, correspondingly, $\sigma(\mathbf{x},\mathbf{x}')$
in place of $\rho(\mathbf{x},\mathbf{x}')$ for the density matrices.
Clearly, under such circumstances, the above momentum moments $\mathcal{P}_{ijk..}^{(n)}(\mathbf{q})$
are of limited help since they are \emph{gauge} dependent and \emph{cannot}
provide a \emph{gauge} invariant formulation of the dynamics. Progress
can be made by generalizing the anticommutator-based definition, Eq.
\ref{eq:anti-commutator (1)}, namely
\begin{align*}
\mathcal{M}^{(0)}(\mathbf{q}) & =\braket{\mathbf{q}|\hat{\sigma}|\mathbf{q}}\equiv\mathcal{P}^{(0)}(\mathbf{q})\\
\mathcal{M}_{k}^{(1)}(\mathbf{q}) & =\frac{\braket{\mathbf{q}|[\hat{\pi}_{k},\hat{\sigma}]_{+}|\mathbf{q}}}{2}\\
\mathcal{M}_{jk}^{(1)}(\mathbf{q}) & =\frac{\braket{\mathbf{q}|[\hat{\pi}_{j}[\hat{\pi}_{k},\hat{\sigma}]_{+}]_{+}|\mathbf{q}}}{2^{2}}\\
.. & ..\\
\mathcal{M}_{ijk..}^{(n)}(\mathbf{q}) & =\frac{\braket{\mathbf{q}|[\hat{\pi}_{i},[\hat{\pi}_{j},[\hat{\pi}_{k},..\hat{\sigma}]_{+}]_{+}]_{+}|\mathbf{q}}}{2^{n}}
\end{align*}
Equivalently, upon noticing 
\begin{align*}
\braket{\mathbf{x}|[\hat{\pi}_{k},\hat{O}]_{+}|\mathbf{x}'} & =\braket{\mathbf{x}|\hat{\pi}_{k}\hat{O}|\mathbf{x}'}+\braket{\mathbf{x}'|\hat{\pi}_{k}\hat{O}^{\dagger}|\mathbf{x}}^{*}\\
 & =(\pi_{k}+\pi_{k}^{'*})O(\mathbf{x},\mathbf{x}')
\end{align*}
we can write (cf. Eq. \ref{eq:anti-commutator (2)})
\begin{align*}
\mathcal{M}_{ijk..}^{(n)}(\mathbf{q}) & =\left(\frac{\pi_{i}+\pi_{i}^{'*}}{2}\right)\left(\frac{\pi_{j}+\pi_{j}^{'*}}{2}\right)\left(\frac{\pi_{k}+\pi_{k}^{'*}}{2}\right)...\\
 & ...\,\,\sigma(\mathbf{x},\mathbf{x}')\vert_{\mathbf{x}=\mathbf{x}'=\mathbf{q}}
\end{align*}
where now $\pi_{k}=-i\hbar\partial_{k}-\hbar A_{k}$, $\pi'_{k}=-i\hbar\partial'_{k}-\hbar A'_{k}$
and $A_{k}\equiv A_{k}(\mathbf{x})$, $A_{k}'\equiv A_{k}(\mathbf{x}')$. 

Since the $\pi_{k}$'s are \emph{gauge} covariant, these moments are,
by construction, \emph{gauge} invariant and they thus fulfill our
requirement. The price to be paid is that they are no longer symmetric
in their indexes since 
\[
[\pi_{j}+\pi_{j}^{'*},\pi_{k}+\pi_{k}^{'*}]=[\pi_{j},\pi_{k}]+[\pi_{j}^{'*},\pi_{k}^{'*}]\equiv-i\hbar(B_{jk}-B_{jk}^{'})
\]
where again we have primed (unprimed) quantities for functions of
$\mathbf{x}'$ ($\mathbf{x}$). For instance, a direct calculation
shows that 
\[
\mathcal{M}_{ijk}^{(3)}-\mathcal{M}_{ikj}^{(3)}=\left(-\frac{i\hbar}{2}\right)^{2}\partial_{i}B_{jk}(\mathbf{q})\,\mathcal{M}^{(0)}(\mathbf{q})
\]
\begin{align*}
\mathcal{M}_{mijk}^{(4)}-\mathcal{M}_{mikj}^{(4)} & =\left(-\frac{i\hbar}{2}\right)^{2}\left(\partial_{i}B_{jk}(\mathbf{q})\,\mathcal{M}_{m}^{(1)}(\mathbf{q})\right.+\\
 & \left.+\partial_{m}B_{jk}(\mathbf{q})\,\mathcal{M}_{i}^{(1)}\right)
\end{align*}
The exception is the symmetry in the two leftmost indexes 
\begin{align*}
\mathcal{M}_{ijk..}^{(n)}(\mathbf{q}) & =\left(\frac{\pi_{j}+\pi_{j}^{'*}}{2}\right)\left(\frac{\pi_{i}+\pi_{i}^{'*}}{2}\right)\left(\frac{\pi_{k}+\pi_{k}^{'*}}{2}\right)..\\
 & ..\,\sigma(\mathbf{x},\mathbf{x}')\vert_{\mathbf{x}=\mathbf{x}'=\mathbf{q}}+\\
 & -i\hbar\left(B_{ij}-B_{ij}^{'}\right)\left(\frac{\pi_{k}+\pi_{k}^{'*}}{2}\right)..\,\sigma(\mathbf{x},\mathbf{x}')\vert_{\mathbf{x}=\mathbf{x}'=\mathbf{q}}\\
 & \equiv\mathcal{M}_{jik..}^{(n)}(\mathbf{q})
\end{align*}
which holds for any moment since the second term disappears in the
limit $\mathbf{x}=\mathbf{x}'=\mathbf{q}$. We note that we \emph{could}
define moments which are yet \emph{gauge} invariant and symmetric
in their indexes, for instance 
\begin{align*}
\mathcal{N}_{ijk..}^{(n)}(\mathbf{q}) & =\left(\frac{p_{i}+p_{i}^{'*}}{2}-\hbar A_{i}(\mathbf{q})\right)\left(\frac{p_{j}+p_{j}^{'*}}{2}-\hbar A_{j}(\mathbf{q})\right)\\
 & \left(\frac{p_{k}+p_{k}^{'*}}{2}-\hbar A_{k}(\mathbf{q})\right)..\,\sigma(\mathbf{x},\mathbf{x}')\vert_{\mathbf{x}=\mathbf{x}'=\mathbf{q}}
\end{align*}
However, as it will become soon evident, only for the $\mathcal{M}_{ijk..}^{(n)}$'s
defined above the equations of motion take a simple form.

To obtain the dynamical equations for the moments we need the Liouville
- von Neumann (LvN) equation \footnote{Henceforth, the minus sign will be occasionally employed to denote
the commutator, $[A,B]_{-}\equiv[A,B]$.}
\[
i\hbar\partial_{t}\hat{\sigma}=[H,\hat{\sigma}]\equiv\sum_{ij}\frac{\xi^{ij}}{2}[\hat{\pi}_{i}[\hat{\pi}_{j},\hat{\sigma}]_{+}]+[\mathcal{V}-\hbar A_{0},\hat{\sigma}]
\]
the identities 
\begin{align*}
\braket{\mathbf{x}|[\hat{\phi},\hat{O}]_{\pm}|\mathbf{x}'}=\left(\phi\pm\phi'\right)O(\mathbf{x},\mathbf{x}')\\
\braket{\mathbf{x}|[\hat{\pi}_{k},\hat{O}]_{\pm}|\mathbf{x}'}=\left(\pi_{k}\pm\pi_{k}^{'*}\right)O(\mathbf{x},\mathbf{x}')
\end{align*}
(which hold for any operator $\hat{O}$ and an arbitrary local operator
$\hat{\phi}$), the commutation properties
\begin{align*}
[\pi_{j}+\pi_{j}^{'*},\pi_{k}\pm\pi_{k}^{'*}]=i\hbar(B_{jk}\mp B_{jk}^{'*})\\{}
[\pi_{j}+\pi_{j}^{'*},\phi\pm\phi^{'*}]=-i\hbar(\partial_{j}\phi\mp\partial_{j}\phi^{'*})
\end{align*}
and 
\[
(\partial_{k}+\partial'_{k})O(\mathbf{x},\mathbf{x}')\vert_{\mathbf{x}=\mathbf{x}'=\mathbf{q}}=\partial_{q_{k}}O(\mathbf{q},\mathbf{q})
\]
In the above, the kinetic energy contribution of the LvN has been
re-written in the convenient form 
\[
[T,\hat{\sigma}]\equiv\frac{1}{2}\sum_{ij}\xi^{ij}[\hat{\pi}_{i}[\hat{\pi}_{j},\hat{\sigma}]_{+}]
\]
that can be easily checked with a direct calculation.

The equation of motion for the $0^{\text{th}}$ moment can be obtained
readily upon taking the $\bra{\mathbf{q}}$ ..$\ket{\mathbf{q}}$
matrix element of LvN and noticing that 
\[
\braket{\mathbf{q}|[\mathcal{V}-\hbar A_{0},\hat{\sigma}]|\mathbf{q}}=0
\]
and that
\begin{align*}
\braket{\mathbf{q}|[\hat{\pi}_{i},[\hat{\pi}_{j},\hat{\sigma}]_{+}]|\mathbf{q}} & =\left(\pi_{i}-\pi_{i}^{'*}\right)\left(\pi_{j}+\pi_{j}^{'*}\right)\sigma(\mathbf{x},\mathbf{x}')\vert_{\mathbf{x}=\mathbf{x}'=\mathbf{q}}\\
 & =-i\hbar(\partial_{i}+\partial_{i}^{'})\left(\pi_{j}+\pi_{j}^{'*}\right)\sigma(\mathbf{x},\mathbf{x}')\vert_{\mathbf{x}=\mathbf{x}'=\mathbf{q}}\\
 & \equiv-2i\hbar\partial_{q_{i}}\mathcal{M}_{j}^{(1)}(\mathbf{q})
\end{align*}
It takes the form \footnote{To avoid confusion, here and in the following, in the calculations
we shall use the symbol $\partial_{q_{i}}$ for the derivative w.r.t.
$q^{i}$. However, once the equations have been put in final form,
i.e., with $\mathbf{q}$ in place of $\mathbf{x}$ and $\mathbf{x}'$,
it should be clear that $\partial_{i}f$ means the derivative w.r.t.
the $i^{\text{th}}$ variable of the function $f$, which is $q^{i}$
in our case.} 
\[
\frac{\partial\mathcal{M}^{(0)}}{\partial t}(\mathbf{q})=-\sum_{ij}\xi^{ij}\partial_{i}\mathcal{M}_{j}^{(1)}(\mathbf{q})
\]

The equation of motion for the first order moments follows similarly.
For, on the one hand, we have
\begin{align*}
\frac{\partial\mathcal{M}_{k}^{(1)}}{\partial t}(\mathbf{q}) & =\frac{\partial}{\partial t}\braket{\mathbf{q}|[\hat{\pi}_{k},\hat{\sigma}]_{+}|\mathbf{q}}=\\
 & =-\hbar\partial_{t}A_{k}\mathcal{M}^{(0)}(\mathbf{q})+\braket{\mathbf{q}|[\hat{\pi}_{k},\partial_{t}\hat{\sigma}]_{+}|\mathbf{q}}
\end{align*}
and, on the other hand,
\begin{align*}
\braket{\mathbf{q}|[\hat{\pi}_{k},\partial_{t}\hat{\sigma}]_{+}|\mathbf{q}} & =-\frac{i}{2\hbar}\sum_{ij}\frac{\xi^{ij}}{2}\braket{\mathbf{q}|[\hat{\pi}_{k},[\hat{\pi}_{i},[\hat{\pi}_{j},\hat{\sigma}]_{+}]]_{+}|\mathbf{q}}\\
 & =-\frac{i}{2\hbar}\braket{\mathbf{q}|[\hat{\pi}_{k},[\mathcal{V}-\hbar A_{0},\hat{\sigma}]]_{+}|\mathbf{q}}
\end{align*}
Here the first term reads as
\begin{align*}
 & -\frac{i}{2\hbar}\sum_{ij}\frac{\xi^{ij}}{2}(\pi_{k}+\pi_{k}^{'*})(\pi_{i}-\pi_{i}^{'*})(\pi_{j}+\pi_{j}^{'*})\sigma(\mathbf{x},\mathbf{x}')\vert_{\mathbf{x}=\mathbf{x}'=\mathbf{q}}=\\
 & =-\frac{i}{2\hbar}\sum_{ij}\frac{\xi^{ij}}{2}(\pi_{i}-\pi_{i}^{'*})(\pi_{k}+\pi_{k}^{'*})(\pi_{j}+\pi_{j}^{'*})\sigma(\mathbf{x},\mathbf{x}')\vert_{\mathbf{x}=\mathbf{x}'=\mathbf{q}}+\\
 & -\frac{i}{\hbar}\sum_{ij}\frac{\xi^{ij}}{2}\frac{i\hbar(B_{ki}+B'_{ki})}{2}(\pi_{j}+\pi_{j}^{'*})\sigma(\mathbf{x},\mathbf{x}')\vert_{\mathbf{x}=\mathbf{x}'=\mathbf{q}}=\\
 & \equiv-\sum_{ij}\xi^{ij}\partial_{q_{i}}\mathcal{M}_{kj}^{(2)}(\mathbf{q})+\sum_{ij}\xi^{ij}B_{ki}(\mathbf{q})\mathcal{M}_{j}^{(1)}(\mathbf{q})
\end{align*}
where we have used $[(\pi_{k}+\pi_{k}^{'*}),(\pi_{i}-\pi_{i}^{'*})]=i\hbar(B_{ki}+B'_{ki})$,
while for the second term we find
\begin{align*}
 & -\frac{i}{2\hbar}\braket{\mathbf{q}|[\hat{\pi}_{k},[\mathcal{V}-\hbar A_{0},\hat{\sigma}]]_{+}|\mathbf{q}}=\\
 & =-\frac{i}{2\hbar}(\pi_{k}+\pi_{k}^{'*})\left(\mathcal{V}-\hbar A_{0}-(\mathcal{V}'-\hbar A'_{0})\right)\sigma(\mathbf{x},\mathbf{x}')\vert_{\mathbf{x}=\mathbf{x}'=\mathbf{q}}=\\
 & \equiv-\frac{i}{2\hbar}(-i\hbar)\left(\partial_{k}(\mathcal{V}-\hbar A_{0})+\partial'_{k}(\mathcal{V}'-\hbar A'_{0})\right)\sigma(\mathbf{x},\mathbf{x}')\vert_{\mathbf{x}=\mathbf{x}'=\mathbf{q}}=\\
 & =-\partial_{q_{k}}(\mathcal{V}(\mathbf{q})-\hbar A_{0}(\mathbf{q}))\mathcal{M}^{(0)}(\mathbf{q})
\end{align*}
Hence, summing up we arrive at
\[
\frac{\partial\mathcal{M}_{k}^{(1)}}{\partial t}(\mathbf{q})=\mathcal{F}_{k}-\sum_{ij}\xi^{ij}\partial_{i}\mathcal{M}_{kj}^{(2)}(\mathbf{q})
\]
where 
\[
\mathcal{F}_{k}(\mathbf{q})=(-\partial_{k}\mathcal{V}(\mathbf{q})+E_{k})\mathcal{M}^{(0)}(\mathbf{q})+\sum_{ij}\xi^{ij}B_{ki}(\mathbf{q})\mathcal{M}_{j}^{(1)}(\mathbf{q})
\]
is the \emph{density} of classical forces, $E_{k}=\hbar(\partial_{k}A_{0}-\partial_{t}A_{k})$
being the \emph{pseudo}-electric field. 

The equation for the third moment can be derived straightforwardly
along steps similar to those followed above. It is only considerably
longer and thus here we only quote the final result
\begin{align*}
\frac{\partial\mathcal{M}_{km}^{(2)}}{\partial t}(\mathbf{q}) & =-\sum_{ij}\xi^{ij}\partial_{i}\mathcal{M}_{kmj}^{(3)}(\mathbf{q})+\\
 & +\sum_{ij}\xi^{ij}\left(B_{ki}(\mathbf{q})\mathcal{M}_{mj}^{(2)}(\mathbf{q})+B_{mi}(\mathbf{q})\mathcal{M}_{kj}^{(2)}(\mathbf{q})\right)+\\
 & \left(-\partial_{k}\mathcal{V}+E_{k}\right)\mathcal{M}_{m}^{(1)}(\mathbf{q})+\left(-\partial_{m}\mathcal{V}+E_{m}\right)\mathcal{M}_{k}^{(1)}(\mathbf{q})
\end{align*}
which shows the ``typical'' classical contribution, comprising the
intrinsic force and the \emph{gauge} forces, and the coupling to the
higher order moment. Please notice that this equation is symmetric
w.r.t. the indexes $k$ and $m$, as it should be since the moment
is yet of second order. However, the third moment $\mathcal{M}_{kmj}^{(3)}$
appearing in the first line of the r.h.s. of this equation \emph{cannot}
be replaced by its symmetric counterpart $\mathcal{N}_{kmj}^{(3)}$,
hence the latter cannot be used to build up a hierarchy of equations
of motion.

To summarize, the equations of motion for the first few moments are
\[
\frac{\partial\mathcal{M}^{(0)}}{\partial t}(\mathbf{q})=-\sum_{ij}\xi^{ij}\partial_{i}\mathcal{M}_{j}^{(1)}(\mathbf{q})
\]
\begin{align*}
\frac{\partial\mathcal{M}_{k}^{(1)}}{\partial t}(\mathbf{q}) & =-\sum_{ij}\xi^{ij}\partial_{i}\mathcal{M}_{kj}^{(2)}(\mathbf{q})+\\
 & +(-\partial_{k}\mathcal{V}(\mathbf{q})+E_{k})\mathcal{M}^{(0)}(\mathbf{q})+\\
 & +\sum_{ij}\xi^{ij}B_{ki}(\mathbf{q})\mathcal{M}_{j}^{(1)}(\mathbf{q})
\end{align*}
\begin{align*}
\frac{\partial\mathcal{M}_{km}^{(2)}}{\partial t}(\mathbf{q}) & =-\sum_{ij}\xi^{ij}\partial_{i}\mathcal{M}_{kmj}^{(3)}(\mathbf{q})+\\
 & +\left(-\partial_{k}\mathcal{V}+E_{k}\right)\mathcal{M}_{m}^{(1)}(\mathbf{q})+\\
 & +\left(-\partial_{m}\mathcal{V}+E_{m}\right)\mathcal{M}_{k}^{(1)}(\mathbf{q})+\\
 & +\sum_{ij}\xi^{ij}\left(B_{ki}(\mathbf{q})\mathcal{M}_{mj}^{(2)}(\mathbf{q})+B_{mi}(\mathbf{q})\mathcal{M}_{kj}^{(2)}(\mathbf{q})\right)
\end{align*}
\[
...
\]
In general, for the $n^{\text{th}}$ moment we have an ``upward''
coupling to the $(n+1)^{\text{th}}$ layer of moments that occurs
through a surface term containing the spatial derivatives of the $(n+1)^{\text{th}}$
moments, and both ``downward'' and ``horizontal'' couplings that
occur through ``classical'' forces. The rank-1 \emph{pseudo}-electric
field $E_{k}$ and the intrinsic force $-\partial_{k}\mathcal{V}$
link the $n^{\text{th}}$ layer of moments to the layer \emph{below}
it. The rank-2 \emph{pseudo-magnetic} field $B_{kj}$ increases the
order by one and thus describes ``horizontal'' coupling to moments
of the same order $n$ (starting from $n=1$). 

In the above hydrodynamic set of equations the first is continuity\textbf{
}equation, with $\mathcal{J}^{i}(\mathbf{q})=\sum_{j}\xi^{ij}\mathcal{M}_{j}^{(1)}(\mathbf{q})$
playing the role of density current. The second equation extends the
quantum\textbf{ }Navier-Stokes\textbf{ }equation given in the main
text for a pure state, namely
\[
\partial_{t}(n\pi_{k})=\mathcal{F}_{k}-\sum_{j}\partial_{j}(nv^{j}\pi_{k})+\sum_{j}\partial_{j}\sigma_{k}^{j}
\]
where $\mathcal{F}_{k}=nF_{k}$ is the density of classical forces.
Indeed, with the notation of the main text,
\begin{align*}
-\sum_{j}\partial_{j}(nv^{j}\pi_{k})+\sum_{j}\partial_{j}\sigma_{k}^{j}\equiv\\
-\sum_{ij}\xi^{ij}\partial_{j}\left(n\pi_{i}\pi_{k}+nw_{i}w_{k}-\frac{\hbar^{2}}{4}\partial_{i}\partial_{k}n\right)
\end{align*}
and the term between brackets on the r.h.s. is the moment $\mathcal{M}_{ki}^{(2)}(\mathbf{q})$
for a pure state. This can be checked with a simple calculation: setting
$\hat{\sigma}=\ket{\psi}\bra{\psi}$
\[
\mathcal{M}_{kj}^{(2)}(\mathbf{q})=\left(\frac{\pi_{k}+\pi_{k}^{'*}}{2}\right)\left(\frac{\pi_{j}+\pi_{j}^{'*}}{2}\right)\psi(\mathbf{x})\psi^{*}(\mathbf{x})\vert_{\mathbf{x}=\mathbf{x}'=\mathbf{q}}
\]
we recognize the complex-valued momentum fields of the main text (evaluated
at both $\mathbf{x}$ and $\mathbf{x}'$) 
\[
\left(\frac{\pi_{j}+\pi_{j}^{'*}}{2}\right)\psi(\mathbf{x})\psi^{*}(\mathbf{x})=\frac{\Pi_{j}+\Pi_{j}^{'*}}{2}\psi(\mathbf{x})\psi^{*}(\mathbf{x})
\]
Thus, with the help of the commutation properties
\[
[\pi_{k}+\pi_{k}^{'*},\Pi_{j}+\Pi_{j}^{'*}]=-i\hbar(\partial_{k}\Pi_{j}-\partial_{k}^{'}\Pi_{j}^{'*})
\]
(which reduce to $2\hbar\partial_{k}\Im\Pi_{j}=2\hbar\partial_{k}w_{j}\equiv-\hbar^{2}\partial_{k}\partial_{j}\ln n$
in the required limit $\mathbf{x}=\mathbf{x}'=\mathbf{q}$) we find
\begin{align*}
\mathcal{M}_{kj}^{(2)}(\mathbf{q}) & =\frac{\Pi_{j}+\Pi_{j}^{'*}}{2}\frac{\Pi_{k}+\Pi_{k}^{'*}}{2}\psi(\mathbf{x})\psi^{*}(\mathbf{x})\vert_{\mathbf{x}=\mathbf{x}'=\mathbf{q}}+\\
 & -\frac{\hbar^{2}}{4}n\,\partial_{k}\partial_{j}\ln n\\
 & \equiv\pi_{j}\pi_{k}n-\frac{\hbar^{2}}{4}n\,\partial_{k}\partial_{j}\ln n\\
 & =\pi_{j}\pi_{k}n-\frac{\hbar^{2}}{4}\left(\partial_{k}\partial_{j}n-\frac{\partial_{k}n\partial_{j}n}{n}\right)
\end{align*}
where now $\pi_{k}=\Re\Pi_{k}$ in the notation of the main text,
and we remember $-\hbar/2\partial_{k}\ln n\equiv w_{k}$. This analysis
is insightful, since in the last expression the first term is a ``uncorrelated''
term, while the term between brackets is right the one responsible
for the surface forces. Specifically, upon setting (in current notation),
\[
\mathcal{C}_{kj}^{(2)}(\mathbf{q})=\mathcal{M}_{kj}^{(2)}(\mathbf{q})-\frac{\mathcal{M}_{k}^{(1)}(\mathbf{q})\mathcal{M}_{j}^{(1)}(\mathbf{q})}{\mathcal{M}^{(0)}(\mathbf{q})}
\]
we have for a pure state
\[
-\sum_{ij}\xi^{ij}\partial_{i}\mathcal{C}_{kj}=F_{k}^{S}\equiv\sum_{i}\partial_{i}\sigma_{k}^{i}
\]
where $\sigma_{k}^{i}$ is the quantum stress tensor. This suggests
to rewrite the general first moment equation (i.e. that for arbitrary
mixed states) introducing the rank-2 tensor $\mathcal{C}_{kl}^{(2)}(\mathbf{q})$
defined as in the above equation. The result is 
\begin{align*}
\frac{d\pi_{k}}{dt}(\mathbf{q}) & =-\frac{1}{\mathcal{M}^{(0)}(\mathbf{q})}\sum_{ij}\xi^{ij}\partial_{i}\mathcal{C}_{kj}^{(2)}(\mathbf{q})+\\
 & +(-\partial_{k}\mathcal{V}(\mathbf{q})+E_{k})+\sum_{j}B_{kj}(\mathbf{q})v^{j}(\mathbf{q})
\end{align*}
where $\pi_{k}=\mathcal{M}_{k}^{(1)}/\mathcal{M}^{(0)}$ and $v^{j}=\sum_{i}\xi^{ji}\pi_{i}$
are the usual momentum and velocity fields, and $d/dt$ is the material
derivative. Written in this way the first moment equation describes
the Lagrangian-frame evolution of momentum in terms of classical ``bulk''
forces and a \emph{surface} force. Whether the system is in a pure
or in a mixed state it is the latter responsible for quantum effects. 

More generally, we argue that a similar decomposition is possible
for higher order moments. That is, we expect them to contain \emph{all}
quantum effects in a surface term that can be singled out by subtracting
the uncorrelated contributions from their definition. One can then
envisage different ``classical closures'' to the hydrodynamic hierarchy,
depending on the momentum order $n$ at which the surface term is
forced to be null. Classical mechanics would result upon setting $\mathcal{C}^{(2)}=0$
whereas $\mathcal{C}^{(n)}=0$ for $n>2$ would describe a classical
world with quantum corrections of the $(n-2)^{\text{th}}$ order. 

\section{\label{app:gauge invariant formulation of electron dynamics}\emph{Gauge}
invariant formulation of the pure-state electron dynamics}

We show in this Appendix that the electronic equation of motion of
the exact factorization approach, Eq. \ref{eq:electronic equation of motion (Euler)}
for the Eulerian frame or Eq. \ref{eq:electronic equation of motion (Lagrange)}
for the Lagrangian frame, is fully equivalent to Eq. \ref{eq:electronic equation final}
for the conditional electronic density operator $\rho_{\text{el}}=\ket{u}\bra{u}$.
We start writing 
\begin{align*}
i\hbar Q\ket{\dot{u}(\mathbf{x})}\bra{u(\mathbf{x})} & =QH_{\text{el}}(\mathbf{x}_{t})\ket{u(\mathbf{x})}\bra{u(\mathbf{x})}+\\
 & \hbar\sum_{j}u^{j}\ket{D_{j}u(\mathbf{x})}\bra{u(\mathbf{x})}+\\
 & -\hbar\ket{Ru(\mathbf{x})}\bra{u(\mathbf{x})}
\end{align*}
and
\begin{align*}
i\hbar\ket{u(\mathbf{x})}\bra{\dot{u}(\mathbf{x})}Q & =-\ket{u(\mathbf{x})}\bra{u(\mathbf{x})}H_{\text{el}}(\mathbf{x})Q\\
 & -\hbar\sum_{j}u^{j}\ket{u(\mathbf{x})}\bra{D_{j}u(\mathbf{x})}+\\
 & +\hbar\ket{u(\mathbf{x})}\bra{Ru(\mathbf{x})}
\end{align*}
where the dot denotes the material derivative. Summing the two equations
and noticing that
\[
Q\ket{\dot{u}(\mathbf{x})}\bra{u(\mathbf{x})}+\ket{u(\mathbf{x})}\bra{\dot{u}(\mathbf{x})}Q\equiv\dot{\rho}_{\text{el}}
\]
(since $\Re\braket{u|\dot{u}}=0$) and that 
\[
QH_{\text{el}}\rho_{\text{el}}-\rho_{\text{el}}H_{\text{el}}Q\equiv[H,\rho_{\text{el}}]
\]
we arrive at 
\begin{align*}
i\hbar\dot{\rho}_{\text{el}} & =[H_{\text{el}},\rho_{\text{el}}]+\\
 & +\hbar\sum_{j}u^{j}\left(\ket{D_{j}u(\mathbf{x})}\bra{u(\mathbf{x})}-\ket{u(\mathbf{x})}\bra{D_{j}u(\mathbf{x})}\right)+\\
 & -\frac{\hbar^{2}}{2}\sum_{ij}\xi^{ij}\left(Q\ket{D_{i}D_{j}u}\bra{u}-\ket{u}\bra{D_{i}D_{j}u}Q\right)
\end{align*}
where the effective Liouville super-operator is the sum of three \emph{gauge}
invariant contributions. Here, the third term can be shown to be the
derivative of contributions that appear in the second term, namely
\[
-\frac{\hbar^{2}}{2}\sum_{ij}\xi^{ij}\partial_{i}\left(\ket{D_{j}u(\mathbf{x})}\bra{u(\mathbf{x})}-\ket{u(\mathbf{x})}\bra{D_{j}u(\mathbf{x})}\right)
\]
and these (\emph{gauge} invariant) contributions (i.e. the terms between
brackets in the last equation) turn out to be
\[
\ket{D_{j}u(\mathbf{x})}\bra{u(\mathbf{x})}-\ket{u(\mathbf{x})}\bra{D_{j}u(\mathbf{x})}\equiv[\partial_{j}\rho_{\text{el}},\rho_{\text{el}}]
\]
To prove these identities we first notice that 
\begin{align*}
Q\ket{D_{i}D_{j}u}\bra{u}-\ket{u}\bra{D_{i}D_{j}u}Q & =\\
\left(\,Q\ket{\partial_{i}D_{j}u}\bra{u}-\ket{u}\bra{\partial_{i}D_{j}u}Q\,\right) & +iA_{i}\left(\ket{D_{j}u}\bra{u}+\ket{u}\bra{D_{j}u}\right)
\end{align*}
where $Q\ket{D_{j}u}=\ket{D_{j}u}$. Then, we move the derivatives
in the first term out of the kets and the bras 
\begin{align*}
Q\ket{\partial_{i}D_{j}u}\bra{u}-\ket{u}\bra{\partial_{i}D_{j}u}Q & =\\
\partial_{i}\left(\ket{D_{j}u}\bra{u}-\ket{u}\bra{D_{j}u}\right) & +\\
-(\partial_{i}Q)\ket{D_{j}u}\bra{u}+\ket{u}\bra{D_{j}u}(\partial_{i}Q) & +\\
-\ket{D_{j}u}\bra{\partial_{i}u}+\ket{\partial_{i}u}\bra{D_{j}u}
\end{align*}
and notice that $-\partial_{i}Q=\partial_{i}P=\ket{\partial_{i}u}\bra{u}+\ket{u}\bra{\partial_{i}u}$
to write
\begin{align*}
-(\partial_{j}Q)\ket{D_{j}u}\bra{u}+\ket{u}\bra{D_{j}u}(\partial_{j}Q) & =\\
\rho_{\text{el}}\left(\braket{\partial_{i}u|D_{j}u}-\braket{D_{j}u|\partial_{i}u}\right) & =\\
\rho_{\text{el}}\left(\braket{D_{i}u|D_{j}u}-\braket{D_{j}u|D_{i}u}\right)
\end{align*}
since $\braket{u|D_{j}u}\equiv0$. On the other hand, we also have
\begin{align*}
-\ket{D_{j}u}\bra{\partial_{i}u}+\ket{\partial_{i}u}\bra{D_{j}u} & =\\
-\ket{D_{j}u}\bra{D_{i}u}+\ket{D_{i}u}\bra{D_{j}u} & +\\
-iA_{i}\left(\ket{D_{j}u}\bra{u}+\ket{u}\bra{D_{j}u}\right)
\end{align*}
hence, summing up, we obtain
\begin{align*}
Q\ket{D_{i}D_{j}u}\bra{u}-\ket{u}\bra{D_{i}D_{j}u}Q & =\\
\partial_{i}\left(\ket{D_{j}u}\bra{u}-\ket{u}\bra{D_{j}u}\right) & +\\
-\ket{D_{j}u}\bra{D_{i}u}+\ket{D_{i}u}\bra{D_{j}u}
\end{align*}
where the last two terms cancel each other once multiplied by the
(symmetric) inverse mass tensor $\xi^{ij}$ and summed over $i$ and
$j$. This proves the first identity. As for the second identity a
direct calculation gives
\begin{align*}
[\partial_{j}\rho_{\text{el}},\rho_{\text{el}}] & =\partial_{j}\rho_{\text{el}}\ket{u}\bra{u}-\ket{u}\bra{u}\partial_{j}\rho_{\text{el}}=\\
 & =\ket{\partial_{j}u}\bra{u}-\ket{u}\bra{\partial_{j}u}+\\
 & +\ket{u}\braket{\partial_{j}u|u}\bra{u}-\ket{u}\braket{u|\partial_{j}u}\bra{u}\\
 & \equiv\ket{D_{j}u}\bra{u}-\ket{u}\bra{D_{j}u}
\end{align*}
since $\braket{\partial_{j}u|u}=\braket{u|\partial_{j}u}^{*}=iA_{j}$
and $\braket{u|\partial_{j}u}=-iA_{j}$. Noteworthy, since in fact
we have 
\[
\partial_{j}\rho_{\text{el}}\equiv\ket{D_{j}u}\bra{u}+\ket{u}\bra{D_{j}u}
\]
we find that the \emph{gauge} invariant dyad $\ket{D_{j}u}\bra{u}$
can be written solely in terms of $\rho_{\text{el}}$ and its derivatives
\[
\ket{D_{j}u}\bra{u}=\frac{1}{2}\left(\partial_{j}\rho_{\text{el}}+[\partial_{j}\rho_{\text{el}},\rho_{\text{el}}]\right)
\]
and similarly for its adjoint. 

As a second step we prove that the equation of motion
\begin{align*}
i\hbar\dot{\rho}_{\text{el}} & =[H_{\text{el}},\rho_{\text{el}}]+\hbar\sum_{j}u^{j}[\partial_{j}\rho_{\text{el}},\rho_{\text{el}}]+\\
 & -\frac{\hbar^{2}}{2}\sum_{ij}\xi^{ij}\partial_{i}\left([\partial_{j}\rho_{\text{el}},\rho_{\text{el}}]\right)
\end{align*}
is trace and purity conserving since, as mentioned in the main text,
this suffices to show its equivalence to the original equation, Eq.
\ref{eq:electronic equation of motion (Lagrange)}, for the state
vector. The trace conserving property is immediate since on the r.h.s.
there appear only commutators, and these are traceless by virtue of
the trace properties. As for purity, we need to prove that 
\[
\text{Tr}\left(\rho_{\text{el}}\dot{\rho}_{\text{el}}\right)=0
\]
Left multiplication by $\rho_{\text{el}}$ gives three terms, the
first two of which are easily seen to vanish
\begin{align*}
\text{Tr}\left(\rho_{\text{el}}[A,\rho_{\text{el}}]\right) & =\text{Tr}\left(\rho_{\text{el}}A\rho_{\text{el}}\right)-\text{Tr}\left(\rho_{\text{el}}\rho_{\text{el}}A\right)\\
 & =\text{Tr}\left(\rho_{\text{el}}\rho_{\text{el}}A\right)-\text{Tr}\left(\rho_{\text{el}}\rho_{\text{el}}A\right)
\end{align*}
by cycling under the trace operation. As for the third contribution,
we notice that 
\[
\rho_{\text{el}}\partial_{i}\left([\partial_{j}\rho_{\text{el}},\rho_{\text{el}}]\right)=\partial_{i}\left(\rho_{\text{el}}[\partial_{j}\rho_{\text{el}},\rho_{\text{el}}]\right)-\left(\partial_{i}\rho_{\text{el}}\right)[\partial_{j}\rho_{\text{el}},\rho_{\text{el}}]
\]
and then upon tracing 
\[
\text{Tr}\left(\rho_{\text{el}}\partial_{i}\left([\partial_{j}\rho_{\text{el}},\rho_{\text{el}}]\right)\right)=-\text{Tr}\left(\left(\partial_{i}\rho_{\text{el}}\right)[\partial_{j}\rho_{\text{el}},\rho_{\text{el}}]\right)
\]
since the first contribution vanishes as above. The remaining term
gives the contributions
\begin{align*}
\frac{\hbar^{2}}{2}\sum_{ij}\xi^{ij}\text{Tr}\left(\left(\partial_{i}\rho_{\text{el}}\right)\left(\partial_{j}\rho_{\text{el}}\right)\rho_{\text{el}}\right)+\\
-\frac{\hbar^{2}}{2}\sum_{ij}\xi^{ij}\text{Tr}\left(\left(\partial_{j}\rho_{\text{el}}\right)\left(\partial_{i}\rho_{\text{el}}\right)\rho_{\text{el}}\right) & \equiv0
\end{align*}
as is evident upon expanding the commutator, cycling under the trace
and remembering that $\xi^{ij}=\xi^{ji}$. This shows that if $\rho_{\text{el}}$
represents a pure state at initial time, it will remain a pure state
at any later time when evolving according to Eq. \ref{eq:electronic equation final}. 

Finally, in closing this Appendix we show that the second and the
third term on the r.h.s. of the equation of motion can be compound
into a single term
\begin{align*}
\hbar\sum_{j}u^{j}[\partial_{j}\rho_{\text{el}},\rho_{\text{el}}]-\frac{\hbar^{2}}{2}\sum_{ij}\xi^{ij}\partial_{i}\left([\partial_{j}\rho_{\text{el}},\rho_{\text{el}}]\right) & =\\
\frac{\hbar^{2}}{2n}\sum_{ij}\xi^{ij}\partial_{j}\left(n[\rho_{\text{el}},\partial_{j}\rho_{\text{el}}]\right)
\end{align*}
and, interestingly, the latter can be recast as an electron-nuclear\textbf{
}correction\textbf{ }to\textbf{ }the\textbf{ }Hamiltonian\textbf{
}that\textbf{ }defines\textbf{ }the\textbf{ }Liouvillian, namely
\[
i\hbar\dot{\rho}_{\text{el}}=[H_{\text{el}}+\delta H_{\text{en}},\rho_{\text{el}}]
\]
where 
\[
\delta H_{\text{en}}=-\frac{\hbar^{2}}{2n}\sum_{ij}\xi^{ij}\partial_{i}(n\partial_{j}\rho_{\text{el}})
\]
The first step is simple, and requires just to recognize $u^{j}=\sum_{i}\xi^{ij}w_{i}=-\frac{\hbar}{2}\sum_{i}\xi^{ij}\partial_{i}\ln\,n$.
As for the second result, we notice that 
\begin{align*}
-\frac{\hbar^{2}}{2n}\sum_{ij}\xi^{ij}[\partial_{i}(n\partial_{j}\rho_{\text{el}}),\rho_{\text{el}}]=\\
=-\frac{\hbar^{2}}{2n}\sum_{ij}\xi^{ij}[\partial_{i}(n\partial_{j}\rho_{\text{el}}\rho_{\text{el}})-\partial_{i}(\rho_{\text{el}}n\partial_{j}\rho_{\text{el}})+\\
-n(\partial_{j}\rho_{\text{el}})(\partial_{i}\rho_{\text{el}})+n(\partial_{i}\rho_{\text{el}})(\partial_{j}\rho_{\text{el}})]\\
\equiv\frac{\hbar^{2}}{2n}\sum_{ij}\xi^{ij}\partial_{i}(n[\rho_{\text{el}},\partial_{j}\rho_{\text{el}}])
\end{align*}
since the terms in the last line of the second step give no contributions
to the sum. 

\section{\label{app:electronic momentum moments}Electronic moments}

In the exact factorization of the electron-nuclear wavefunction describing
a pure state, the doubly parametrized electronic dyad
\[
\hat{\mathcal{W}}(\mathbf{x},\mathbf{x}')=\ket{u(\mathbf{x})}\bra{u(\mathbf{x}')}
\]
plays a key role, in that it determines through its diagonal elements
the conditional electronic density operator of the system and, in
general, through its trace, the \emph{true} nuclear density matrix
\[
\rho_{n}(\mathbf{x},\mathbf{x}')=\psi(\mathbf{x})\psi^{*}(\mathbf{x}')\braket{u(\mathbf{x}')|u(\mathbf{x})}=\sigma(\mathbf{x},\mathbf{x}')\text{Tr}_{e}\hat{\mathcal{W}}(\mathbf{x},\mathbf{x}')
\]
where $\sigma(\mathbf{x},\mathbf{x}')=\psi(\mathbf{x})\psi^{*}(\mathbf{x}')$
is the \emph{apparent} nuclear density matrix. The purpose of this
Appendix is to generalize this concept to the case where a \emph{single}
nuclear wavefunction can be yet singled out from an ensemble, whereas
the electronic state is in an intrinsically mixed state. That is,
the total density operator of the system is taken of the form (in
the Schr\"{o}dinger representation for the nuclei)\footnote{The notation here adopted should make clear that $\hat{\rho}(\mathbf{x},\mathbf{x}')$
is a \emph{matrix} in nuclear coordinates and an \emph{operator} in
the electronic ones. The hat emphasizes right this latter aspect.}
\[
\hat{\rho}(\mathbf{x},\mathbf{x}')=\psi(\mathbf{x})\psi^{*}(\mathbf{x}')\hat{\mathcal{W}}(\mathbf{x},\mathbf{x}')
\]
where $\hat{\mathcal{W}}$ is the ensemble average of the operator
defined above. In other words, given a total density operator $\hat{\rho}(\mathbf{x},\mathbf{x}')$
\emph{and} a nuclear wavefunction $\psi(\mathbf{x})$ (hence, the
``pure-state'' $\sigma(\mathbf{x},\mathbf{x}')=\psi(\mathbf{x})\psi^{*}(\mathbf{x}')$)
we introduce a ``conditional electronic operator'' according to
\[
\hat{\mathcal{W}}(\mathbf{x},\mathbf{x}')=\frac{\hat{\rho}(\mathbf{x},\mathbf{x}')}{\sigma(\mathbf{x},\mathbf{x}')}
\]
for any $\mathbf{x}$ and $\mathbf{x}'$ where the wavefunction does
not vanish. Note that our assumption about the existence of a single
nuclear wavefunction implies that the electronic states involved in
the ensemble (call them $\ket{u_{\alpha}(\mathbf{x})}$, with $\alpha=1,2,..$)
have only an overall \emph{gauge} degree of freedom, i.e. their phases
cannot be changed independently of each other. Rather, a \emph{single}
set of \emph{gauge} fields exist, $A_{k}$, irrespective of $\alpha$. 

Next, we define moments of these conditional electronic operators
(which we call electronic moments) by analogy with the momentum moments
introduced in Appendix C, that is, 
\[
\hat{\mathcal{M}}^{(0)}(\mathbf{q})=\hat{\mathcal{W}}(\mathbf{q},\mathbf{q})
\]
\[
\hat{\mathcal{M}}_{k}^{(1)}(\mathbf{q})=\frac{\mu_{k}+\mu_{k}^{'*}}{2}\hat{\mathcal{W}}(\mathbf{x},\mathbf{x}')\vert_{\mathbf{x}=\mathbf{x}'=\mathbf{q}}
\]
\[
\hat{\mathcal{M}}_{kj}^{(2)}(\mathbf{q})=\frac{\mu_{k}+\mu_{k}^{'*}}{2}\frac{\mu_{j}+\mu_{j}^{'*}}{2}\hat{\mathcal{W}}(\mathbf{x},\mathbf{x}')\vert_{\mathbf{x}=\mathbf{x}'=\mathbf{q}}
\]
\[
...
\]
where $\mu_{k}=-i\hbar\partial_{k}+\hbar A_{k}\equiv-i\hbar D_{k}$
and $D_{k}=\partial_{k}+iA_{k}$ is the \emph{gauge} covariant derivative,
using the \emph{gauge} potential common to the ensemble elements.
By definition, these moments are \emph{gauge} invariant with respect
to the \emph{gauge} transformations mentioned above (notice the sign
change of the \emph{gauge} field when compared to the mechanical momenta
$\pi_{k}$'s, i.e., $\mu_{k}\equiv-\pi_{k}^{*}$ ). For a pure state,
for instance, we find 
\[
\hat{\mathcal{M}}^{(0)}(\mathbf{q})=\ket{u(\mathbf{q})}\bra{u(\mathbf{q})}=\rho_{\text{el}}(\mathbf{q})
\]
\[
\hat{\mathcal{M}}_{k}^{(1)}(\mathbf{q})=\left(-\frac{i\hbar}{2}\right)\left(\ket{D_{k}u(\mathbf{q})}\bra{u(\mathbf{q})}-\ket{u(\mathbf{q})}\bra{D_{k}u(\mathbf{q})}\right)
\]
\begin{align*}
\hat{\mathcal{M}}_{kj}^{(2)}(\mathbf{q}) & =\frac{\hbar^{2}}{4}\left(\ket{D_{k}u(\mathbf{q})}\bra{D_{j}u(\mathbf{q})}+\ket{D_{j}u(\mathbf{q})}\bra{D_{k}u(\mathbf{q})}\right)+\\
 & -\frac{\hbar^{2}}{4}\left(\ket{D_{k}D_{j}u(\mathbf{q})}\bra{u(\mathbf{q})}+\ket{u(\mathbf{q})}\bra{D_{k}D_{j}u(\mathbf{q})}\right)
\end{align*}
in the same notation used in the main text. 

Then, we proceed in deriving the equations of motion for these moments.
We start from the Lagrangian-frame pure-state result
\begin{align*}
i\hbar\dot{\rho}_{\text{el}} & =[H_{\text{el}},\rho_{\text{el}}]+\\
 & +\hbar\sum_{j}u^{j}\left(\ket{D_{j}u(\mathbf{x})}\bra{u(\mathbf{x})}-\ket{u(\mathbf{x})}\bra{D_{j}u(\mathbf{x})}\right)+\\
 & -\frac{\hbar^{2}}{2}\sum_{ij}\xi^{ij}\partial_{i}\left(\ket{D_{j}u}\bra{u}-\ket{u}\bra{D_{j}u}\right)
\end{align*}
and re-interpret it in terms of moments, 
\begin{align*}
\frac{d\hat{\mathcal{M}}^{(0)}}{dt}(\mathbf{q}) & =-\frac{i}{\hbar}[H_{\text{el}}(\mathbf{q}),\hat{\mathcal{M}}^{(0)}(\mathbf{q})]+\\
 & +\frac{2}{\hbar}\left(\sum_{j}u^{j}\hat{\mathcal{M}}_{j}^{(1)}(\mathbf{q})-\frac{\hbar}{2}\sum_{ij}\xi^{ij}\partial_{i}\hat{\mathcal{M}}_{j}^{(1)}(\mathbf{q})\right)
\end{align*}
Here, $u^{j}=\sum_{i}\xi^{ij}w_{i}=-\frac{\hbar}{2}\sum_{i}\xi^{ij}\partial_{i}\ln n$,
thus 
\begin{align*}
i\hbar\frac{d\hat{\mathcal{M}}^{(0)}}{dt}(\mathbf{q}) & =[H_{\text{el}}(\mathbf{q}),\hat{\mathcal{M}}^{(0)}(\mathbf{q})]-\frac{i\hbar}{n}\sum_{ij}\xi^{ij}\partial_{i}\left(n\hat{\mathcal{M}}_{j}^{(1)}\right)
\end{align*}
This equation \emph{survives} to the kind of ensemble average we are
interested in (being linear in the $\mathcal{\hat{M}}$'s ), hence
it represents the equation of motion of our zero-th order electronic
moment (the conditional density operator for electrons generalized
to arbitrary mixed states). The key difference with a pure state is
that only for the latter the first order moment appearing on the right
hand side can be expressed in terms of the zero-th order one and its
spatial derivative, namely 
\[
\hat{\mathcal{M}}_{j}^{(1)}=i\frac{\hbar}{2}[\hat{\mathcal{M}}^{(0)},\partial_{j}\hat{\mathcal{M}}^{(0)}]\ \ \ \text{(pure state)}
\]
Henceforth, we shall proceed by looking at the pure state case and
seeking a form of the equations of motion that suits (``survives'')
to the kind of ensemble average mentioned above. 

The dynamical equation of the first order moments follows from the
observation that 
\begin{align*}
\frac{\partial\hat{\mathcal{M}}_{k}^{(1)}}{\partial t}(\mathbf{q}) & =\left(\partial_{t}+\partial_{t}^{'}\right)\frac{\mu_{k}+\mu_{k}^{'*}}{2}\ket{u}\bra{u'}\vert_{\mathbf{x}=\mathbf{x}'=\mathbf{q},t=t'}\\
 & \equiv\hbar\partial_{t}A_{k}\ket{u}\bra{u}+\\
 & +\frac{\mu_{k}+\mu_{k}^{'*}}{2}\left(\partial_{t}+\partial_{t}^{'}\right)\ket{u}\bra{u'}\vert_{\mathbf{x}=\mathbf{x}'=\mathbf{q},t=t'}
\end{align*}
since $[\partial_{t}+\partial'_{t},\mu_{k}+\mu_{k}^{'*}]=\hbar\partial_{t}A_{k}+\hbar\partial'_{t}A'_{k}$,
where now primed and unprimed quantities refer to $(\mathbf{x'},t')$
and $(\mathbf{x},t)$, respectively. On the other hand,
\begin{align*}
\left(\partial_{t}+\partial_{t}^{'}\right)\ket{u}\bra{u'} & =\ket{D_{0}u}\bra{u'}+\ket{u}\bra{D'_{0}u'}+\\
 & -iA_{0}\ket{u}\bra{u'}+iA'_{0}\ket{u}\bra{u'}
\end{align*}
where $D_{0}=\partial_{t}+iA_{0}$ is the \emph{gauge} covariant time-derivative.
Now, upon using $[\mu_{k}+\mu_{k}^{'*},A_{0}-A'_{0}]=-i\hbar\left(\partial_{k}A_{0}+\partial'_{k}A'_{0}\right)$,
in the limit $\mathbf{x}=\mathbf{x}'=\mathbf{q}$ and $t=t'$ we obtain
\begin{align}
\frac{\partial\hat{\mathcal{M}}_{k}^{(1)}}{\partial t}(\mathbf{q}) & =-E_{k}\hat{\mathcal{M}}^{(0)}(\mathbf{q})+\nonumber \\
 & +\frac{\pi_{k}+\pi_{k}^{'*}}{2}\left(\ket{D_{0}u}\bra{u'}+\ket{u}\bra{D'_{0}u'}\right)\vert_{\mathbf{x}=\mathbf{x}'=\mathbf{q},t=t'}\label{eq:first moment time-derivative}
\end{align}
where $E_{k}$ is the \emph{pseudo}-electric field. We then use the
electronic equation of the exact factorization in \emph{gauge} covariant
form (as explained in the main text, see Eq. \ref{eq:electronic equation (gauge covariant derivatives)})
\[
i\hbar\ket{D_{0}u}=(H_{\text{el}}-\mathcal{V})\ket{u}-i\hbar\sum_{j}V^{j}\ket{D_{j}u}-\frac{\hbar^{2}}{2}\sum_{ij}\xi^{ij}\ket{D_{i}D_{j}u}
\]
and consider first the simplified situation where we neglect the term
which is of second order in the spatial derivatives of the electronic
state (third term on the r.h.s.). With the replacement $V^{j}=v^{j}+iu^{j}$
and
\[
\ket{D_{0}u}\approx-\frac{i}{\hbar}(H_{\text{el}}-\mathcal{V})\ket{u}-\sum_{j}v^{j}\ket{D_{j}u}-i\sum_{j}u^{j}\ket{D_{j}u}
\]
the r.h.s. of Eq. \ref{eq:first moment time-derivative} is seen to
contain three terms. The first is the Halmitonian driving
\begin{align*}
\text{I} & =\frac{1}{2}(\ket{D_{k}u}\bra{u'}(H_{\text{el}}^{'}-\mathcal{V}')-\ket{u}\bra{D_{k}^{'}(H_{\text{el}}^{'}-\mathcal{V}')u'}+\\
 & -\ket{D_{k}(H_{\text{el}}-\mathcal{V})u}\bra{u'}+(H_{\text{el}}-\mathcal{V})\ket{u}\bra{D'_{k}u'})
\end{align*}
where
\[
\ket{D_{k}(H_{\text{el}}-\mathcal{V})u}=(H_{\text{el}}-\mathcal{V})\ket{D_{k}u}+\partial_{k}(H_{\text{el}}-\mathcal{V})\,\ket{u}
\]
hence
\begin{align*}
\text{I} & =\frac{1}{2}([\ket{D_{k}u}\bra{u'}-\ket{u}\bra{D_{k}^{'}u'}](H_{\text{el}}^{'}-\mathcal{V}')+\\
 & -(H_{\text{el}}-\mathcal{V})[\ket{D_{k}u}\bra{u'}-\ket{u}\bra{D_{k}^{'}u'}]\\
 & -\partial_{k}(H_{\text{el}}-\mathcal{V})\ket{u}\bra{u'}-\ket{u}\bra{u'}\partial'_{k}(H'_{\text{el}}-\mathcal{V}'))
\end{align*}
and, in the limit $\mathbf{x}'=\mathbf{x}$,
\[
\text{I}=-\frac{i}{\hbar}[H_{\text{el}},\hat{\mathcal{M}}_{k}^{(1)}]-\frac{1}{2}[\partial_{k}(H_{\text{el}}-\mathcal{V}),\hat{\mathcal{M}}^{(0)}]_{+}
\]
where we have introduced the moments. The second term involves the
real velocity field
\[
\text{II}=\frac{i\hbar}{2}(D_{k}-D_{k}^{'*})\sum_{j}(v^{j}\ket{D_{j}u}\bra{u'}+v{}^{'j}\ket{u}\bra{D_{j}^{'}u'})
\]
and can be re-arranged as follows
\begin{align*}
\text{II} & =\frac{i\hbar}{2}\sum_{j}v^{j}(\ket{D_{k}D_{j}u}\bra{u}+\ket{D_{k}u}\bra{D_{j}u}+\\
 & -\ket{D_{j}u}\bra{D_{k}u}-\ket{u}\bra{D_{k}D_{j}u})+\\
 & +\frac{i\hbar}{2}\sum_{j}\partial_{k}v^{j}\,(\ket{D_{j}u}\bra{u}-\ket{u}\bra{D_{j}u})
\end{align*}
in the required limit $\mathbf{x}'=\mathbf{x}$. Here, the second
derivatives are conveniently swapped (by introducing the commutator
$\hbar[D_{k},D_{j}]\equiv iB_{kj}$) and the terms of the first sum
presented as 
\[
\frac{2i}{\hbar}\partial_{j}\hat{\mathcal{M}}_{k}^{(1)}+\frac{2i}{\hbar}B_{kj}\hat{\mathcal{M}}^{(0)}
\]
As a result
\[
\text{II}=-\sum_{j}v^{j}\partial_{j}\hat{\mathcal{M}}_{k}^{(1)}+\sum_{j}v^{j}B_{kj}\hat{\mathcal{M}}^{(0)}-\sum_{j}(\partial_{k}v^{j})\hat{\mathcal{M}}_{j}^{(1)}
\]
Finally, the third contribution (involving the imaginary part of the
velocity field) reads as 
\[
\text{III}=-\frac{\hbar}{2}(D_{k}-D_{k}^{'*})\sum_{j}(u^{j}\ket{D_{j}u}\bra{u'}-u{}^{'j}\ket{u}\bra{D_{j}^{'}u'})
\]
and can be manipulated in the form
\[
-\sum_{ij}\xi^{ij}(\partial_{i}\ln\,n)\hat{\mathcal{M}}_{kj}^{(2)}+\frac{i\hbar}{2}\sum_{ij}\xi^{ij}(\partial_{k}\partial_{i}\ln\,n)\hat{\mathcal{M}}_{j}^{(1)}
\]
upon remembering the second moment expression \ref{eq:M_2 (pure-state)}.
Summing up the contributions we find
\begin{align*}
\left(\frac{\partial}{\partial t}+\sum_{j}v^{j}\partial_{j}\right)\hat{\mathcal{M}}_{k}^{(1)}(\mathbf{q})\approx\\
-\left(E_{k}+\sum_{j}B_{kj}v^{j}\right)\hat{\mathcal{M}}^{(0)}(\mathbf{q})+\\
-\frac{i}{\hbar}[H_{\text{el}},\hat{\mathcal{M}}_{k}^{(1)}]-\frac{1}{2}[\partial_{k}(H_{\text{el}}-\mathcal{V}),\hat{\mathcal{M}}^{(0)}]_{+}+\\
-\sum_{j}(\partial_{k}v^{j})\hat{\mathcal{M}}_{j}^{(1)}+\\
-\sum_{ij}\xi^{ij}(\partial_{i}\ln\,n)\hat{\mathcal{M}}_{kj}^{(2)}+\frac{i\hbar}{2}\sum_{ij}\xi^{ij}(\partial_{k}\partial_{i}\ln\,n)\hat{\mathcal{M}}_{j}^{(1)}
\end{align*}
thereby identifying the total derivative of the Lagrangian-frame dynamics. 

If we considered the neglected term (second order in the spatial derivatives)
we would have the additional contribution
\begin{align*}
\frac{i\hbar}{2}\frac{\mu_{k}+\mu_{k}^{'*}}{2}\sum_{ij}\xi^{ij}\left(\ket{D_{i}D_{j}u}\bra{u'}\right.+\\
\left.-\ket{u}\bra{D'_{i}D'_{j}u'}\right)\vert_{\mathbf{x}=\mathbf{x}'=\mathbf{q}} & =\frac{\hbar^{2}}{4}\sum_{ij}\xi^{ij}L_{kij}
\end{align*}
where now 
\[
L_{kij}=(D_{k}-D_{k}^{'*})(\ket{D_{i}D_{j}u}\bra{u}-\ket{u}\bra{D'_{i}D'_{j}u'})
\]
The latter can be re-arranged as follows
\begin{align*}
L_{kij} & =\frac{(D_{k}-D_{k}^{'*})}{2}[(D_{i}-D_{i}^{'*})(D_{j}+D_{j}^{'*})+\\
 & +(D_{i}+D_{i}^{'*})(D_{j}-D_{j}^{'*})]\ket{u}\bra{u'}
\end{align*}
with the usual limit implicit, and the sum of derivatives can be moved
to the leftmost position -- upon introducing the appropriate commutators
-- while the moments can be easily identified, since they only involve
differences of derivatives. The result is 
\begin{align*}
L_{kij} & =\frac{1}{2}\left(\frac{2i}{\hbar}\right)^{2}\{\partial_{j}\hat{\mathcal{M}}_{ki}^{(2)}+\partial_{i}\hat{\mathcal{M}}_{kj}^{(2)}+\\
 & +B_{kj}\hat{\mathcal{M}}_{i}^{(1)}+B_{ij}\hat{\mathcal{M}}_{k}^{(1)}+B_{ki}\hat{\mathcal{M}}_{j}^{(1)}\}
\end{align*}
and thus the additional term turns out to be 
\begin{align*}
-\sum_{ij}\frac{\xi^{ij}}{2}\{\partial_{j}\hat{\mathcal{M}}_{ki}^{(2)}+\partial_{i}\hat{\mathcal{M}}_{kj}^{(2)}+B_{kj}\hat{\mathcal{M}}_{i}^{(1)}+B_{ki}\hat{\mathcal{M}}_{j}^{(1)}\}\equiv\\
=-\sum_{ij}\xi^{ij}\{\partial_{i}\hat{\mathcal{M}}_{kj}^{(2)}+B_{kj}\hat{\mathcal{M}}_{i}^{(1)}\}
\end{align*}
although the form on the l.h.s. is more appropriate for generalizations
to higher orders. Overall, the complete equation of motion for the
first moment reads as
\begin{align*}
\frac{d\hat{\mathcal{M}}_{k}^{(1)}}{dt} & =-(E_{k}+\sum_{j}B_{kj}v^{j}-\partial_{k}\mathcal{V})\hat{\mathcal{M}}^{(0)}\\
 & -\frac{i}{\hbar}[H_{\text{el}},\hat{\mathcal{M}}_{k}^{(1)}]-\frac{1}{2}\left[\partial_{k}H_{\text{el}},\mathcal{M}^{(0)}\right]_{+}+\\
 & -\sum_{j}(\partial_{k}v^{j})\hat{\mathcal{M}}_{j}^{(1)}+\\
 & -\sum_{ij}\xi^{ij}(\partial_{i}\ln\,n)\mathcal{M}_{kj}^{(2)}+\frac{i\hbar}{2}\sum_{ij}\xi^{ij}(\partial_{k}\partial_{i}\ln\,n)\hat{\mathcal{M}}_{j}^{(1)}\\
 & -\sum_{ij}\xi^{ij}\{\partial_{i}\hat{\mathcal{M}}_{kj}^{(2)}+B_{kj}\hat{\mathcal{M}}_{i}^{(1)}\}
\end{align*}
This concludes our derivation of the first two moment equations. We
stress once again that these equations of motion are most naturally
formulated in the Lagrangian frame, which is in motion on the nuclear
trajectories.

\section{Lie-Trotter identity\label{app:Lie-Trotter}}

The Lie-Trotter identity is 
\begin{equation}
\frac{d}{d\lambda}\left(e^{A(\lambda)}\right)=\int_{0}^{1}dt\ e^{tA(\lambda)}\frac{dA}{d\lambda}e^{(1-t)A(\lambda)}\label{eq:Lie-Trotter}
\end{equation}
and can be proved using the operator identity

\[
\frac{d}{dt}\left(e^{tA(\lambda')}e^{(1-t)A(\lambda)}\right)=e^{tA(\lambda')}\left(A(\lambda')-A(\lambda)\right)e^{(1-t)A(\lambda)}
\]
upon integrating over $t\in[0,1]$ to write
\[
e^{A(\lambda')}-e^{A(\lambda)}=\int_{0}^{1}dt\ e^{tA(\lambda')}\left(A(\lambda')-A(\lambda)\right)e^{(1-t)A(\lambda)}
\]
Indeed, setting $\lambda'=\lambda+\Delta\lambda$ and taking the limit
of the incremental ratio 
\begin{align*}
\lim_{\Delta\lambda\rightarrow0}\frac{e^{A(\lambda')}-e^{A(\lambda)}}{\Delta\lambda}=\\
\int_{0}^{1}dt\ \lim_{\Delta\lambda\rightarrow0}\left(e^{tA(\lambda')}\frac{A(\lambda')-A(\lambda)}{\Delta\lambda}\right)e^{(1-t)A(\lambda)}
\end{align*}
one obtains Eq. \ref{eq:Lie-Trotter}. Eq. \ref{eq:Lie-Trotter} is
particularly useful under a trace operation, 
\begin{align*}
\frac{d}{d\lambda}\text{Tr}\left(e^{A(\lambda)}\right) & =\int_{0}^{1}dt\ \text{Tr}\left(e^{tA(\lambda)}\frac{dA}{d\lambda}e^{(1-t)A(\lambda)}\right)\\
 & \equiv\text{Tr}\left(e^{A(\lambda)}\frac{dA}{d\lambda}\right)
\end{align*}

For the purposes of the developments described in the main text, the
Lie-Trotter identity takes the form
\begin{align*}
\frac{d}{d\lambda}\left(e^{-\beta H(\lambda)}\right) & =-\int_{0}^{\beta}d\alpha\ e^{-\alpha H(\lambda)}\frac{dH}{d\lambda}e^{-(\beta-\alpha)H(\lambda)}\\
 & =-\int_{0}^{\beta}d\alpha\ e^{-(\beta-\alpha)H(\lambda)}\frac{dH}{d\lambda}(\lambda)e^{-\alpha H(\lambda)}
\end{align*}
where $\beta$ is the inverse temperature. This leads one to consider
the Kubo transform $A_{\text{K}}^{\beta}$ of an operator $A$
\[
A_{\text{K}}^{\beta}=\frac{1}{\beta}\int_{0}^{\beta}d\alpha\ e^{\alpha H}Ae^{-\alpha H}
\]
and its Bolztmannized version, the ``Kubo-Boltzmann transform''
$A_{\text{KB}}^{\beta}$ defined by
\[
A_{\text{KB}}^{\beta}=e^{-\beta H}A_{\text{K}}^{\beta}=(A_{\text{K}}^{\beta})^{\dagger}e^{-\beta H},
\]
hence to re-write the Lie-Trotter identity as
\[
\frac{d}{d\lambda}\left(e^{-\beta H(\lambda)}\right)=-\beta\left(\frac{dH}{d\lambda}\right)_{\text{KB}}^{\beta}
\]
In the main text we have used this expression for $\lambda=x^{k}$
and set then $-dH/d\lambda=f_{k}$ to write
\[
\partial_{j}e^{-\beta H}=\beta e^{-\beta H}(f_{j})_{\text{K}}^{\beta}
\]
Finally, using the canonical density operator $\rho^{0}=e^{-\beta H}/Z$,
where $Z=\text{Tr}e^{-\beta H}$ is the partition function, we find
\[
\partial_{j}\rho^{0}=\beta\rho^{0}(\delta f_{j})_{\text{K}}^{\beta}
\]
where $\delta f_{j}=f_{j}-F_{j}^{0}$ and $F_{j}^{0}=\beta^{-1}\partial_{j}\ln\,Z$
is the ensemble-average of the microscopic force $f_{j}$. 

\section{Pseudo-magnetic contribution to the finite-temperature friction kernel\label{app:Pseudo-magnetic}}

In this Appendix we provide an expression, which is valid in thermal
equilibrium, for the \emph{pseudo}-magnetic contribution contained
in the DMS friction kernel of Eq. \ref{eq:DMS kernel}. We start from
\[
W_{kj}(t)=\text{Tr}_{\text{el}}(\delta f_{k}(t)\partial_{j}\rho_{\text{el}}^{0})
\]
and write the spatial derivative of the density operator at equilibrium
as
\begin{align*}
\partial_{j}\rho_{\text{el}}^{0} & =\sum_{n}\partial_{j}p_{n}\ket{u_{n}}\bra{u_{n}}+\\
 & +\sum_{n}p_{n}\ket{D_{j}^{n}u_{n}}\bra{u_{n}}+\sum_{n}p_{n}\ket{u_{n}}\bra{D_{j}^{n}u_{n}}
\end{align*}
where $p_{n}$ are Boltzmann populations and the $D_{j}^{n}=\partial_{j}+iA_{j}^{n}$
are the covariant derivatives defined with the state-specific Berry
connection $A_{j}^{n}=i\braket{u_{n}|\partial_{j}u_{n}}$ (see Section
\ref{subsec:Type-e-statistical-mixtures} and the relationship with
the ensemble connection, Eq. \ref{eq:ensemble connection}). Thus,
\[
W_{kj}(t)=I_{kj}(t)+\Gamma_{kj}(t)+\Gamma_{kj}^{*}(t)
\]
where the first term is 
\[
I_{kj}(t)=\sum_{n}\braket{u_{n}|\delta f_{k}(t)|u_{n}}\partial_{j}p_{n}\equiv I_{kj}(0)
\]
and $\Gamma_{kj}(t)$ is the finite-temperature generalization of
the kernel of Eq. \ref{eq:memory kernel},
\[
\Gamma_{kj}(t)=\sum_{n}p_{n}\braket{u_{n}|\delta f_{k}(t)|D_{j}^{n}u_{n}}
\]
as becomes evident upon noticing that
\begin{align*}
\Gamma_{kj}(t) & \equiv\sum_{n}p_{n}\braket{u_{n}|(-\partial_{k}H_{\text{el}})e^{-\frac{i}{\hbar}(H_{\text{el}}-E_{n})t}Q_{n}|u_{n}}\\
 & =\sum_{n}p_{n}\braket{\partial_{k}u_{n}|Q_{n}H_{\text{el}}^{n}e^{-\frac{i}{\hbar}H_{\text{el}}^{n}t}|u_{n}}
\end{align*}
with $Q_{n}=1-\ket{u_{n}}\bra{u_{n}}$ and $H_{\text{el}}^{n}:=H_{\text{el}}-E_{n}$.
The identities above follow because $Q_{n}\ket{u_{n}}\equiv\ket{D_{j}^{n}u_{n}}$,
$Q_{n}$ commutes with any function of $H_{\text{el}}$ and 
\[
Q_{n}\partial_{k}H_{\text{el}}\ket{u_{n}}=-H_{\text{el}}^{n}Q_{n}\ket{\partial_{k}u_{n}}
\]

We first address the $I_{kj}$ contribution, which explicitly reads
as a sort of cross-correlation matrix
\[
I_{kj}=\beta\sum_{n}p_{n}\braket{u_{n}|\delta f_{k}|u_{n}}\braket{u_{n}|\delta f_{j}|u_{n}}
\]
as can be seen upon using $\partial_{j}p_{n}=\beta p_{n}\braket{u_{n}|\delta f_{j}(t)|u_{n}}$.
This term is vanishingly small if the microscopic forces in the energy
eigenstates are close to each other, $f_{k}^{n}=\braket{u_{n}|f_{k}|u_{n}}\approx F_{k}^{0}$.
In any case, it is symmetric in its indexes, hence it does not affect
our conclusions on the \emph{pseudo-}magnetic term. Secondly, we consider
the (causal) Fourier transform of $\Gamma_{kj}(t)$ defined as
\[
\bar{\gamma}_{kj}(\omega)=2\lim_{\epsilon\rightarrow0^{+}}\int_{0}^{\infty}e^{-\epsilon t}e^{i\omega t}\Gamma_{kj}(t)\,dt
\]
and address the Markovian regime as the $\omega\rightarrow0$ limit
of this expression. With arguments similar to those given in Ref.
\citep{Martinazzo2022a} (Section IV.C) we find
\begin{align*}
\bar{\gamma}_{kj}(\omega) & =-2i\hbar\sum_{n}p_{n}\braket{u_{n}|(\partial_{k}H_{\text{el}})Q_{n}G^{+}(E_{n}+\hbar\omega)|\partial_{j}u_{n}}\\
 & =-2i\hbar q_{kj}^{\beta}(\omega)+2\gamma_{kj}^{\beta}(\omega)
\end{align*}
Here, $G^{+}(E)$ is the retarded Green operator of the (many-body)
electronic Hamiltonian, 
\begin{align*}
G^{+}(E) & =\lim_{\epsilon\rightarrow0^{+}}(E+i\epsilon-H_{\text{el}})^{-1}\\
 & =\frac{1}{i\hbar}\lim_{\epsilon\rightarrow0^{+}}\int_{0}^{\infty}e^{-\epsilon t}e^{-\frac{i}{\hbar}(H_{\text{el}}-E)t}\,dt
\end{align*}
and the second equality follows upon splitting it into real (or principal)
part and imaginary one, namely with the help of
\[
G^{+}(E)=G^{P}(E)-i\pi\delta(E-H_{\text{el}})
\]
The kernels thus introduced are the ensemble averaged generalization
of the kernel introduced in Ref. \citep{Martinazzo2022a} (Section
IV.C) and represent a frequency-dependent quantum geometric tensor
\[
q_{kj}^{\beta}(\omega)=\sum_{n}p_{n}\braket{u_{n}|(\partial_{k}H_{\text{el}})Q_{n}G^{+}(E_{n}+\hbar\omega)|\partial_{j}u_{n}}
\]
and the true, frequency-dependent friction tensor
\begin{align*}
\gamma_{kj}^{\beta}(\omega) & =-\pi\hbar\sum_{n}p_{n}\braket{u_{n}|(\partial_{k}H_{\text{el}})Q_{n}\delta(\hbar\omega-H_{\text{el}}^{n})|\partial_{j}u_{n}}\\
 & =\frac{\pi}{\omega}\sum_{n}p_{n}\braket{u_{n}|(\partial_{k}H_{\text{el}})Q_{n}\delta(\hbar\omega-H_{\text{el}}^{n})(\partial_{j}H_{\text{el}})|u_{n}}\\
 & =\frac{\pi}{\omega}\sum_{m,n}^{m\neq n}p_{n}(f_{k})_{nm}(f_{j})_{mn}\delta(E_{n}+\hbar\omega-E_{m})
\end{align*}
where for the second equality we have used $\delta(E_{n}+\hbar\omega-H_{\text{el}})Q_{n}\ket{\partial_{j}u_{n}}=-\frac{1}{\hbar\omega}\delta(E_{n}+\hbar\omega-H_{\text{el}})Q_{n}(\partial_{j}H_{\text{el}})\ket{u_{n}}$,
and for the third one we have expanded $Q_{n}$ in the energy basis
and introduced the matrix elements of the force $(f_{k})_{nm}=\braket{u_{n}|(-\partial_{k}H_{\text{el}})|u_{m}}$.
In the Markov limit we have
\begin{align*}
q_{kj}^{\beta}(\omega) & \rightarrow q_{kj}^{\beta}=\sum_{n}p_{n}\braket{\partial_{k}u_{n}|Q_{n}|\partial_{j}u_{n}}\\
 & =\sum_{n}p_{n}\braket{D_{k}^{n}u_{n}|D_{j}^{n}u_{n}}
\end{align*}
which amounts to the averaged quantum geometric tensor of Eq. \ref{eq:average quantum geometric tensor},
\[
q_{kj}^{\text{av}}=\sum_{n}p_{n}\braket{D_{k}u_{n}|D_{j}u_{n}}
\]
to within a minor correction in its symmetric part, 
\[
q_{kj}^{\beta}=q_{kj}^{\text{av}}+\sum_{n}p_{n}\delta A_{k}^{n}\delta A_{j}^{n}
\]
where $\delta A_{k}^{n}=A_{k}^{n}-A_{k}$. Hence, $\Im q_{kj}^{\beta}=\Im q_{kj}^{\text{av}}$
and the \emph{pseudo}-magnetic field contribution is precisely that
described by the ensemble averaged geometric tensor. 

As for the Markovian friction, the physical friction kernel $\eta_{kj}$
is given by the real part of 
\[
\gamma_{kj}=\lim_{\omega\rightarrow0}\frac{2\gamma_{kj}^{\beta}(\omega)+2\gamma_{kj}^{\beta}(-\omega)}{2}
\]
as extensively discussed in Ref. \citep{Martinazzo2022a} (see Appendix
E). In the previous work we found $\gamma_{kj}=\lim_{\omega\rightarrow0}\gamma_{kj}(|\omega|)$
since $\gamma_{kj}(-|\omega|)\equiv0$ at $T=0$ K where only the
ground-state is populated. Now this has to be replaced by 
\[
\eta_{kj}=\lim_{\omega\rightarrow0}\Re\gamma_{kj}^{\beta}(\omega)(1-e^{-\beta\hbar\omega})=\hbar\beta\lim_{\omega\rightarrow0}\omega\,\Re\gamma_{kj}^{\beta}(\omega)
\]
for any temperature but the smallest one, and the zero-frequency limit
needs to be carefully addressed. This result follows from the detailed-balance
condition on the kernel, a kind of Kubo-Martin-Schwinger relationship
\[
\gamma_{kj}^{\beta}(\omega)=-e^{\beta\hbar\omega}\gamma_{jk}^{\beta}(-\omega)
\]
that can be proved with a direct calculation in the energy basis 
\begin{align*}
\gamma_{kj}^{\beta}(\omega) & =\frac{\pi}{\omega Z}\sum_{m,n}^{m\neq n}e^{-\beta E_{n}}(f_{k})_{nm}(f_{j})_{mn}\delta(E_{n}+\hbar\omega-E_{m})\\
 & =\frac{\pi}{\omega Z}\sum_{m,n}^{m\neq n}e^{-\beta E_{m}}e^{+\beta\hbar\omega}(f_{k})_{nm}(f_{j})_{mn}\delta(E_{n}+\hbar\omega-E_{m})\\
 & =-e^{\beta\hbar\omega}\gamma_{jk}^{\beta}(-\omega)
\end{align*}
(in the last line we have swapped the indexes $n,m$). Notice that
$\gamma_{kj}$ is symmetric in its real part and anti-symmetric in
its imaginary one. Clearly, the difference with the $T=0$ K case
is due to the electronic de-excitation processes that can occur in
the electronic bath at finite temperature and that work by pumping
energy into the nuclear system. Detailed balance guarantees that the
kernel $\Re\gamma_{kj}(\omega)$ remains positive definite for $\omega>0$,
hence it correctly describes friction.

\section{Independent electrons\label{app:Finite-temperature-friction-kern}}

In this Appendix we focus on independent electrons, and address the
issue of obtaining the finite-temperature Markovian friction kernel
$\eta_{kj}$ for them. We start from the results of Section \ref{app:Pseudo-magnetic}
in the form $\eta_{kj}=\lim_{\omega\rightarrow0}\Re\eta_{kj}(\omega)$
where 
\[
\eta_{kj}(\omega)=\gamma_{kj}^{\beta}(\omega)(1-e^{-\beta\hbar\omega})
\]
and 
\[
\gamma_{kj}^{\beta}(\omega)=\frac{\pi}{\omega}\sum_{n}p_{n}\braket{u_{n}|(\partial_{k}H_{\text{el}})Q_{n}\delta(\hbar\omega-H_{\text{el}}^{n})(\partial_{j}H_{\text{el}})|u_{n}}
\]
Here, $n$ labels energy eigenstates, $p_{n}$ are Boltzmann populations,
$Q_{n}=1-\ket{u_{n}}\bra{u_{n}}$ and $H_{\text{el}}^{n}=E_{n}-H_{\text{el}}$.
For $N$ independent electrons the many-body states are Hartree-Fock
``determinants'' built with the single-particle states $\ket{\phi_{a}}$
of the monoelectronic Hamiltonian $h$, and the sum over states can
be recast as
\begin{align*}
\gamma_{kj}^{\beta}(\omega) & =\frac{\pi}{\omega}\sum_{I}\sum_{a\notin I}\sum_{b\notin I}p_{Ia}\braket{\phi_{a}|\partial_{k}h|\phi_{b}}\braket{\phi_{b}|\partial_{j}h|\phi_{a}}\times\\
 & \times\delta(\hbar\omega-\Delta\epsilon_{ba})
\end{align*}
where $I$ is a collection of $N-1$ particle labels, $\epsilon_{a}$
are single-particle energies and $\Delta\epsilon_{ba}=\epsilon_{b}-\epsilon_{a}$.
In obtaining the above expression we have exploited Slater rules and
included only single-particle excitations, i.e., $Ia\rightarrow Ib$.
This automatically accounts for the projector $Q_{n}$ without further
constraining the single-particle labels {[}$a\neq b$ in unnecessary
in the above expression unless $\omega\equiv0${]}. In the thermodynamic
limit we have
\[
\sum_{I}\sum_{a\notin I}\sum_{b\notin I}p_{Ia}\{..\}=\sum_{a,b}f(\epsilon_{a})(1-f(\epsilon_{b}))\{..\}
\]
where $f(\epsilon)=(e^{\beta(\epsilon-\mu)}+1)^{-1}$ is the Fermi
occupation function at the (inverse) temperature $\beta$ and chemical
potential $\mu$, while, on the other hand, we have the identity
\[
f(\epsilon_{a})(1-f(\epsilon_{b}))\left(e^{\beta(\epsilon_{a}-\epsilon_{b})}-1\right)\equiv f(\epsilon_{b})-f(\epsilon_{a})
\]
which easily follows from the definition of $f$. Hence, introducing
$D_{ab}^{k}=\braket{\phi_{a}|\partial_{k}h|\phi_{b}}$, we find
\[
\gamma_{kj}^{\beta}(\omega)=\frac{\pi}{\omega}\sum_{a,b}D_{ab}^{k}D_{ba}^{j}\frac{f(\epsilon_{b})-f(\epsilon_{a})}{e^{-\beta\hbar\omega}-1}\delta(\hbar\omega-\Delta\epsilon_{ba})
\]
and 
\[
\eta_{kj}(\omega)=\frac{\pi}{\omega}\sum_{a,b}D_{ab}^{k}D_{ba}^{j}(f(\epsilon_{a})-f(\epsilon_{b}))\delta(\hbar\omega-\Delta\epsilon_{ba})
\]
The latter kernel satisfies $\eta_{kj}(\omega)^{*}=\eta_{kj}(-\omega)=\eta_{jk}(\omega)$,
which makes its real part symmetric in its indexes, and even in $\omega$.
The Markovian friction kernel follows unambiguously as a zero frequency
limit,
\[
\eta_{kj}=\lim_{\omega\rightarrow0}\frac{\pi}{\omega}\Re\sum_{a,b}D_{ab}^{k}D_{ba}^{j}(f(\epsilon_{a})-f(\epsilon_{b}))\delta(\hbar\omega-\Delta\epsilon_{ba})
\]
or, equivalently,
\begin{align*}
\eta_{kj} & =\lim_{\omega\rightarrow0}\pi\hbar\Re\sum_{a,b}D_{ab}^{k}D_{ba}^{j}\left(-\frac{f(\epsilon_{a}+\hbar\omega)-f(\epsilon_{a})}{\hbar\omega}\right)\times\\
 & \times\delta(\hbar\omega-\Delta\epsilon_{ba})
\end{align*}
The latter form is particularly useful for taking the $T=0$ K limit,
where 
\[
\left(-\frac{f(\epsilon_{a}+\hbar\omega)-f(\epsilon_{a})}{\hbar\omega}\right)\rightarrow\delta(\epsilon_{a}-\mu)
\]
and $\mu\rightarrow\epsilon_{F}$, the Fermi energy. Hence, 
\[
\eta_{kj}=\pi\hbar\sum_{a,b}D_{ab}^{k}D_{ba}^{j}\delta(\epsilon_{a}-\epsilon_{F})\delta(\epsilon_{b}-\epsilon_{F})
\]
since the sum becomes real in this limit. This is the Head-Gordon
Tully result \citep{Head-Gordon1995}, who were the first to derive
an electronic friction kernel by working at zero temperature in the
independent electron approximation.


%
 \pagebreak
\end{document}